\documentclass[aps,rmp,twocolumn,floatfix,superscriptaddress]{revtex4}

\usepackage{amsmath}
\usepackage{amssymb}
\usepackage{graphicx}
\usepackage{dcolumn}
\usepackage{bm}
\usepackage[latin1]{inputenc}
\usepackage[mathscr]{eucal}
\usepackage{epsfig}
\usepackage{rotating}

\renewcommand{\vr}{{\mathbf r}}
\newcommand{\vp}{{\mathbf p}}
\newcommand{\vv}{{\mathbf v}}

\renewcommand{\Re}{\text{Re}}

\newcommand{\DT}[1][]{\mathcal{D}_{T#1}}
\newcommand{\DL}[1][]{\mathcal{D}_{L#1}}
\newcommand{\T}[1][]{\mathcal{T}_{#1}^{\rm an}}

\newcommand{\jS}[1][]{j_{S#1}}

\newcommand{\jL}[1][]{j^{L#1}}
\newcommand{\jT}[1][]{j^{T#1}}

\newcommand{\fL}[1][]{f^{L#1}}
\newcommand{\fT}[1][]{f^{T#1}}

\newcommand{\Tr}{\text{Tr}}

\newcommand{\GA}{\hat{G}^A}
\newcommand{\GR}{\hat{G}^R}

\newcommand{\Lor}{L_0}
\newcommand{\energy}{E}
\newcommand{\jQ}{\dot{Q}}
\newcommand{\cur}{I}
\newcommand{\tprob}{{\mathcal T}}
\newcommand{\Vol}{{\mathcal V}}

\newcommand{\tx}{\tau_1}

\newcommand{\tnz}{\hat{\tau}_3}
\newcommand{\tnv}{\hat{1}}
\def\Tr{\textrm{Tr}~}

\newcommand{\tmpnote}[1]%
   {\begingroup{\it (FIXME: #1)}\endgroup}

   \newcommand{\comment}[1]%
       {\marginpar{\tiny C: #1}}

\newcommand{\tmpnoteb}[1]%
   {\begingroup{\it (FIXME: #1)}\endgroup}

\renewcommand{\tmpnote}[1]{}
\renewcommand{\tmpnoteb}[1]{}
\renewcommand{\comment}[1]{}

\begin{document}
\title{Thermal properties in mesoscopics: physics and applications from thermometry to
refrigeration}

\author{Francesco Giazotto}
\email{F.Giazotto@sns.it} \affiliation{Low Temperature Laboratory,
Helsinki University of Technology, P.O. Box 2200, FIN-02015 HUT,
Finland}
 \affiliation{NEST-INFM and Scuola
Normale Superiore, I-56126 Pisa, Italy}

\author{Tero T. Heikkil\"a}
\email{Tero.T.Heikkila@hut.fi} \affiliation{Low Temperature
Laboratory, Helsinki University of Technology, P.O. Box 2200,
FIN-02015 HUT, Finland}

\affiliation{Department of Physics and Astronomy, University of
Basel, Klingelbergstr. 82, CH-4056 Basel, Switzerland}

\author{Arttu Luukanen}
\affiliation{National Institute of Standards and Technology,
Quantum Electrical Metrology Division, 325 Broadway, Boulder CO
80305 USA}

\affiliation{Millimetre-wave Laboratory of Finland - MilliLab, VTT,
P. O. Box 1000, FIN-02044 VTT, Finland}

\author{Alexander M. Savin}

\affiliation{Low Temperature Laboratory, Helsinki University of
Technology, P.O. Box 2200, FIN-02015 HUT, Finland}

\author{Jukka P. Pekola}
\affiliation{Low Temperature Laboratory, Helsinki University of
Technology, P.O. Box 2200, FIN-02015 HUT, Finland}

\date{\today}

\begin{abstract}
This review presents an overview of the thermal properties of
mesoscopic structures. The discussion is based on the concept of
electron energy distribution, and, in particular, on controlling and
probing it. The temperature of an electron gas is determined by this
distribution: refrigeration is equivalent to narrowing it, and
thermometry is probing its convolution with a function
characterizing the measuring device. Temperature exists, strictly
speaking, only in quasiequilibrium in which the distribution follows
the Fermi-Dirac form. Interesting nonequilibrium deviations can
occur due to slow relaxation rates of the electrons, e.g., among
themselves or with lattice phonons. Observation and applications of
nonequilibrium phenomena are also discussed. The focus in this paper
is at low temperatures, primarily below 4 K, where physical
phenomena on mesoscopic scales and hybrid combinations of various
types of materials, e.g., superconductors, normal metals,
insulators, and doped semiconductors, open up a rich variety of
device concepts. This review starts with an introduction to
theoretical concepts and experimental results on thermal properties
of mesoscopic structures. Then thermometry and refrigeration are
examined with an emphasis on experiments. An immediate application
of solid-state refrigeration and thermometry is in ultrasensitive
radiation detection, which is discussed in depth. This review
concludes with a summary of pertinent fabrication methods of
presented devices.
\end{abstract}

\maketitle
 \tableofcontents

\section{Introduction}
\label{sec:introduction} Solid state mesoscopic electronic systems
provide a micro-laboratory to realize experiments on low
temperature physics, to study quantum phenomena such as
fundamental relaxation mechanisms in solids, and a way to create
advanced cryogenic devices. In a broad sense, mesoscopic here
refers to micro- and nanostructures, whose size falls in between
atomic and macroscopic scales. The central concept of this Review
is the energy distribution of mesoscopic electron systems, which
in thermal equilibrium (Fermi-Dirac distribution) determines the
temperature of the electron gas. The non-Fermi distributions are
discussed in depth, since they are often encountered and utilized
in mesoscopic structures and devices. This Review aims to discuss
the progress mainly during the past decade on how electron
distributions can be controlled, measured and made use of in
various device concepts. When appropriate, earlier developments
are reviewed as well. The central devices and concepts to be
discussed are electronic refrigerators, thermometers, radiation
detectors, and distribution-controlled transistors. Typically the
working principles or resolution of these detectors rely on
phenomena that show up only at cryogenic temperatures, i.e., at
temperatures of the order of a few kelvin and below. A practical
threshold in terms of temperature is set by liquefaction of helium
at 4.2 K. This also sets the emphasis in this Review: the devices
and principles working mostly at temperatures above 4.2 K are at
times mentioned only for reference.

Section \ref{sec:thermalproperties} of this Review introduces
formally the central ingredients; the relevant theoretical results
are either derived or given there. We also review some of the new
developments concerning the thermoelectric effects in mesoscopic
systems. Although the theoretical analysis of the effects in the
later sections is based on Sec.~\ref{sec:thermalproperties}, the
main messages can be understood without reading it in detail.
Section \ref{sec:thermometry} explains how the electronic
temperature is typically measured and what is required of an
electronic thermometer. Accurate and fast thermometers can be
utilized for thermal radiation detection as explained in
Sec.~\ref{sec:thermaldetectors}, which reviews such detectors. The
resolution of these devices is ultimately limited by the thermal
noise, which can be lowered by refrigeration. In
Sec.~\ref{sec:ecool}, we show how the electron temperature can be
lowered via electronic means, and discuss the direct applications
of this refrigeration. Section \ref{sec:devicefabrication}
explains the main methods used in the fabrication of mesoscopic
electronic devices, and in Sec.~\ref{sec:future} we briefly
discuss some of the main open questions in the field and the
prospects of practical instruments based on electronic
refrigeration and using the peculiar out-of-equilibrium energy
distributions.

\begin{figure}[tb]
\centering
\includegraphics[width=\columnwidth]{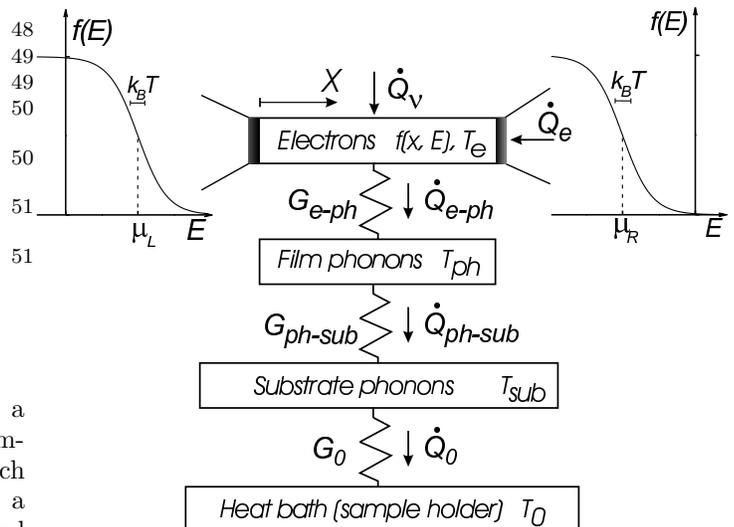}
\caption{Schematic picture of the system considered in this
review. We describe an electron system in a diffusive wire,
connected to two normal-metal or superconducting reservoirs via
contacts of resistance $R_N$. The reservoirs are further connected
to the macroscopic measurement apparatus (see
Subs.~\ref{subs:reservoir}). The heat flows and thermal
conductances between the studied electron system and the external
driving, the electromagnetic environment, and the phonons in the
lattice (Subs.~\ref{subs:colli} and \ref{subs:limits}) are
indicated with the arrows. The description of phonons in the
lattice can further be divided in the film phonons, substrate
phonons and finally the heat bath on which the substrate resides
(Subs.~\ref{subs:phonons}). If the system is used as a radiation
detector, it also couples to the radiation field, typically via
some matching circuit (Sec.~\ref{sec:thermaldetectors}). }
\label{fig:setup}
\end{figure}

\section{Thermal properties of mesoscopic scale hybrid structures at sub-kelvin
temperatures} \label{sec:thermalproperties}

The schematic picture of a setup studied in typical experiments
described in this Review is shown in Fig.~\ref{fig:setup}. The
main object is a diffusive metal or heavily-doped semiconductor
wire connected to large electrodes acting as reservoirs where
electrons thermalize quickly to the surroundings. The electrons in
the wire interact between themselves, and are coupled to the
phonons in the film and to the radiation and the fluctuations in
the electromagnetic environment. The temperature $T_{ph}$ of the
film phonons can, in a non-equilibrium situation, differ from that
of the substrate phonons, $T_{\rm sub}$ and this can even differ
from the phonon temperature $T_0$ in the sample holder that is
externally cooled. Under the applied voltage, the energy
distribution function $f(E)$ of electrons depends on each of these
couplings, and on the state (e.g., superconducting or normal) of
the various parts of the system. In certain cases detailed below,
$f(E)$ is a Fermi function
\begin{equation}
f_{\rm eq}(\energy;T_e,\mu)= \frac{1}{\exp[(\energy-\mu)/(k_B
T_e)]+1}, \label{eq:fermifunction}
\end{equation}
characterized by an electron temperature $T_e$ and potential
$\mu$. One of the main goals of this review is to explain how
$T_e$, and in some cases also $T_{ph}$, can be driven even much
below the lattice temperature $T_0$, and how this low $T_e$ can be
exploited to improve the sensitivity of applications relying on
the electronic degrees of freedom. We also detail some of the
out-of-equilibrium effects, where $f(E)$ is not of the form of
Eq.~\eqref{eq:fermifunction}. In some setups, the specific form of
$f(E)$ can be utilized for novel physical phenomena.

Throughout the Review, we concentrate on wires whose dimensions are
large enough to fall in the quasiclassical diffusive limit. This
means that the Fermi wavelength $\lambda_F$, elastic mean free path
$\ell_{el}$ and the length of the wire $L$ have to satisfy
$\lambda_F \ll \ell_{el} \ll L$. In this regime, the electron energy
distribution function is well defined, and its space dependence can
be described by a diffusion equation
(Eq.~\eqref{eq:diffusiveboltzmann}). In most parts of the Review, we
assume the capacitances $C$ of the contacts large enough, such that
the charging energy $E_C=e^2/2C$ is less than any of the relevant
energy scales and can thus be ignored.

Our approach is to describe the electron energy distribution
function $f(\vr,\energy)$ at a given position $\vr$ of the sample
and then relate this function to the charge and heat currents and
their noise. In typical metal structures in the absence of
superconductivity, phase-coherent effects are weak and often it is
enough to rely on a semiclassical description. In this case,
$f(\vr,\energy)$ is described by a diffusion equation, as
discussed in Subs.~\ref{subs:diffusionequation}. The electron
reservoirs impose boundary conditions for the distribution
functions, specified in Subs.~\ref{subs:bc}. The presence of
inelastic scattering due to electron-electron interaction, phonons
or the electromagnetic environment can be described by source and
sink terms in the diffusion equation, specified by the collision
integrals and discussed in Subs.~\ref{subs:colli}. In the limit
when these scattering effects are strong, the distribution
function tends to a Fermi function $f_{\rm
eq}(E;T_e(\vr),\mu(\vr))$ throughout the wire, with a position
dependent potential $\mu(\vr)$ and temperature $T_e(\vr)$. In this
quasiequilibrium case, detailed in Subs.~\ref{subs:limits}, it
thus suffices to find these two quantities. Finally, with the
knowledge of $f(\vr;\energy)$, one can obtain the observable
currents and their noise as described in
Subs.~\ref{subs:observables}.

In many cases, it is not enough to only describe the electrons
inside the mesoscopic wire, assuming that the surroundings are
totally unaffected by the changes in this electron system. If the
phonons in the film are not well coupled to a large phonon bath,
their temperature is influenced by the coupling to the electrons.
In this case, it is important to describe the phonon heating or
cooling in detail (see Subs.~\ref{subs:phonons}). Often also the
electron reservoirs may get heated due to an applied bias voltage,
which has to be taken into account in the boundary conditions.
This heating is discussed in Subs.~\ref{subs:reservoir}.

At the temperature range considered in this Review, many metals
undergo a transition to the superconducting state
\cite{tinkham:96}. This gives rise to several new phenomena that
can be exploited, for example, for thermometry (see
Sec.~\ref{sec:thermometry}), for radiation detection
(Sec.~\ref{sec:thermaldetectors}) and for electron cooling
(Sec.~\ref{sec:ecool}). The presence of superconductivity modifies
both the diffusion equation (inside normal-metal wires through the
proximity effect, see Subs.~\ref{subs:diffusionequation}) and
especially the boundary conditions (Subs.~\ref{subs:bc}). Also the
relations between the observable currents and the distribution
functions are modified (Subs.~\ref{subs:observables}).

Once the basic equations for finding $f(\vr,\energy)$ are
outlined, we detail its behavior in different types of
normal-metal -- superconductor heterostructures in
Subs.~\ref{subs:examples}.

\subsection{Boltzmann equation in a diffusive wire}
\label{subs:diffusionequation}

The semiclassical Boltzmann equation
\cite{ashcroftmermin,smithjensen} describes the average number of
particles, $f(\vr,\vp) d\vr d\vp/(2\pi)^3$, in the element
$\{d\vr,d\vp\}$ around the point $\{\vr,\vp\}$ in the
six-dimensional position-momentum space. The kinetic equation for
$f(\vr,\vp)$ is a continuity equation for particle flow,
\begin{equation}
\left(\frac{\partial}{\partial t}+\vv \cdot \partial_{\vr} +
e{\mathbf E} \cdot
\partial_{\vp} \right) f(\vr,\vp;t) = I_{\rm el}[f] + I_{\rm
in}[f]. \label{eq:fullboltzmann}
\end{equation}
Here ${\mathbf E}$ is the electric field driving the charged
particles and the elastic and inelastic collision integrals
$I_{\rm el}$ and $I_{\rm in}$, functionals of $f$, act as source
and sink terms. They illustrate the fact that scattering breaks
translation symmetries in space and time --- the total particle
numbers expressed through the momentum integrals of $f$ still
remain conserved.

In the metallic diffusive limit, Eq.~(\ref{eq:fullboltzmann}) may
be simplified as follows
\cite{nagaev:103,sukhorukov:13054,rammer}. The electric field term
can be absorbed in the space derivative by the substitution
$\energy = \varepsilon_\vp + \mu(\vr)$ in the argument of the
distribution function, such that $\energy$ describes both the
kinetic $\varepsilon_\vp$ and the potential energy $\mu$ of the
electron. Therefore, we are only left with the full
${\vr}$-dependent derivative $\vv \cdot \nabla f = \vv \cdot
\partial_\vr f + e{\mathbf E} \cdot
\partial_\vp f$ on the left-hand side of Eq.~(\ref{eq:fullboltzmann}). In the diffusive regime, one
may concentrate on length scales much larger than the mean free path
$\ell_{el}$. There, the particles quickly lose the memory of the
direction of their initial momentum, and the distribution functions
become nearly isotropic in the direction of $\vv$. Therefore, we may
expand the distribution function $f$ in the two lowest spherical
harmonics in the dependence on $\hat{\mathbf v}\equiv\vv/v$,
$f(\hat{\mathbf v})=f_0+\hat{\mathbf v}  \cdot {\mathbf \delta f}$,
and make the relaxation-time approximation to the elastic collision
integral with the elastic scattering time $\tau$, i.e., $I_{\rm
el}=-\hat{\mathbf v} \cdot {\mathbf \delta f}/\tau$. In the limit
where the time dependence of the distribution function takes place
in a much slower scale than $\tau$, this yields the diffusion
equation with a source term,
\begin{equation}
(\partial_t - D\nabla^2_\vr) f_0(\vr;\energy,t) = I_{\rm in}[f_0].
\label{eq:diffusiveboltzmann}
\end{equation}
Here we assume that the particles move with the Fermi velocity,
i.e., $\vv=v_F\hat{\mathbf v}$. As a result, their diffusive
motion is characterized by the diffusion constant $D =
v_F^2\tau/3$. In what follows, we will mainly concentrate on the
static limit, i.e., assume $\partial_t f_0(\vr;\energy,t)=0$ and
lift the subscript 0 from the angle-independent part $f_0$ of the
distribution function.

Equation \eqref{eq:diffusiveboltzmann} can also be derived
rigorously from the microscopic theory through the use of the
quasiclassical Keldysh formalism \cite{rammer:323}. With such an
approach, one can also take into account superconducting effects,
such as Andreev reflection \cite{andreev:1228} and the proximity
effect \cite{belzig:1251}. In the diffusive limit, one obtains the
Usadel equation \cite{usadel:507}, which in the static case is
\begin{equation}
\frac{D}{\sigma A}\nabla\cdot \check{I}= \left[-i\energy
\check{\tau}_3 + \check{\Delta}(\vr) + \check{\Sigma}_{\rm
in}(\vr),\check{G}(\vr;\energy)\right]. \label{eq:usadel}
\end{equation}
Here $\check{G}(\vr;\energy)$ is the isotropic part of the Keldysh
Green's function in the Keldysh $\otimes$ Nambu space, $A$ and
$\sigma$ are the cross section and the normal-state conductivity
of the wire, $\check{\tau}_3=\hat{1} \otimes \hat{\tau}_3$ is the
third Pauli matrix in Nambu space, $\check{\Delta}=\hat{1} \otimes
\hat{\Delta}$ is the pair potential matrix, and
$\check{\Sigma}_{\rm in}$ describes the inelastic scattering that
is not contained in $\check{\Delta}$. Usadel equation describes
the matrix current $\check{I}=\sigma A\check{G} \nabla \check{G}$
\cite{nazarov:1221}, whose components integrated over the energy
yield the physical currents. In the Keldysh space, $\check{G}$ is
of the form
\begin{equation*}
\check{G}= \begin{pmatrix} \hat{G}^R & \hat{G}^K \\ 0 & \hat{G}^A
                 \end{pmatrix},
\end{equation*}
where each component is a $2\times 2$ matrix in Nambu
particle-hole space. Equation \eqref{eq:usadel} has to be
augmented with a normalization condition $\check{G}^2=1$. This
implies $(\hat{G}^{R/A})^2=1$ and allows a parametrization
$\hat{G}^K=\hat{G}^R \hat{h}-\hat{h} \hat{G}^A$, where $\hat{h}$
is a distribution function matrix with two free parameters. The
equations for the retarded/advanced functions $\hat{G}^{R/A}$
((1,1) and (2,2) -Keldysh components of Eq.~\eqref{eq:usadel})
describe the behavior of the pairing amplitude. The solutions to
these equations yield the coefficients for the kinetic equations,
i.e., the (1,2) or the Keldysh part of Eq.~\eqref{eq:usadel}. This
describes the symmetric and antisymmetric parts of the energy
distribution function with respect to the chemical potential of
the superconductors. The latter is assumed everywhere equal to
allow a time-independent description. A common choice is a
diagonal $\hat{h}=f^L+f^T \hat{\tau_3}$ \cite{schmid:207}, where
$f^L(\energy)=f(-\energy)-f(\energy)$ is the antisymmetric and
$f^T(\energy)=1-f(\energy)-f(-\energy)$ the symmetric part of the
energy distribution function $f(\energy)$. With this choice,
inside the normal metals where $\hat{\Delta}=0$, we get two
kinetic equations of the form \cite{belzig:1251,virtanen:401}
\begin{subequations}
  \label{eq:kinetic}
  \begin{align}
    D\nabla\cdot \jL &= \Sigma_{in}^L, &
    \jL = \sigma A(\DL\nabla\fL - \T\nabla\fT + \jS\fT)
    \,,
    \label{eq:kinetic1}
    \\
    D\nabla\cdot \jT &= \Sigma_{in}^T, &
    \jT = \sigma A(\DT\nabla\fT + \T\nabla\fL + \jS\fL)
    \,.
    \label{eq:kinetic2}
  \end{align}
\end{subequations}
Here $\jL \equiv \frac{1}{4} \Tr[(\tx\otimes\tnv)\,\check{I}\,]$
describes the spectral energy current, and $\jT \equiv \frac{1}{4}
\Tr[(\tx\otimes\tnz)\,\check{I}\,] \,$ the spectral charge
current. The inelastic effects are described by the collision
integrals $\Sigma_{in}^L \equiv \frac{1}{4}
\Tr[(\tx\otimes\tnv)[\check{\Sigma}_{in},\check{G}]]$ and
$\Sigma_{in}^T \equiv \frac{1}{4}
\Tr[(\tx\otimes\tnz)[\check{\Sigma}_{in},\check{G}]]$. The kinetic
coefficients are
  \begin{align*}
    \DL &\equiv \frac{1}{4} \Tr[1\!-\!\GR\GA]
    \\
    \DT &\equiv \frac{1}{4} \Tr[1\!-\!\GR\tnz\GA\tnz]
    \\
    \T  &\equiv \frac{1}{4} \Tr[\GA\GR\tnz]
    \\
    \jS &\equiv \frac{1}{4} \Tr[(\GR\nabla\GR -
    \GA\nabla\GA)\tnz].
  \end{align*}
Here, $\DL$ and $\DT$ are the spectral energy and charge diffusion
coefficients, and $\jS$ is the spectral density of the
supercurrent-carrying states \cite{heikkila:184513}. The
cross-term $\T$ is usually small but not completely negligible. In
a normal-metal wire in the absence of a proximity effect,
$\GR=\hat{\tau}_3$ and $\GA=-\hat{\tau_3}$. Then we obtain
$\DL=\DT=1$, $\T=\jS=0$ and the kinetic equations
\eqref{eq:kinetic2} reduce to Eq.~\eqref{eq:diffusiveboltzmann} in
the static limit.

\subsection{Boundary conditions}
\label{subs:bc} The quasiclassical kinetic equations cannot
directly describe constrictions whose size is of the order of the
Fermi wavelength, such as tunnel junctions or quantum point
contacts. However, such contacts can be described by the boundary
conditions derived by \textcite{nazarov:1221},
\begin{equation}\label{eq:nazarovbc}
\check{I}_L = \check{I}_R =
Z(\frac{1}{2}[\check{G}_L,\check{G}_R]_+)[\check{G}_R,\check{G}_L]_-,
\end{equation}
where
\begin{equation*}
Z(x)=\frac{2e^2}{h} \sum_n \frac{\tprob_n}{2+\tprob_n (x-1)}.
\end{equation*}
Here, $[\check{G}_L,\check{G}_R]_{\pm} \equiv \check{G}_L
\check{G}_R \pm \check{G}_R \check{G}_L$, and $\check{I}_{L(R)}$ and
$\check{G}_{L(R)}$ are the matrix current and the Green's function
at the left (right) of the constriction, evaluated at the interface
and flowing towards the right. The constriction is described by a
set $\{\tprob_n\}$ of transmission eigenvalues through the function
$Z(x)$. For large constrictions, it is typically enough to transform
the sum over the eigenvalues to an integral over the transmission
probabilities $\tprob$, weighted by their probability distribution
$\rho(\tprob)$. In the case of a tunnel barrier, $\tprob_n \ll 1$,
and thus $Z(x)=e^2 \sum_n \tprob_n/h \equiv G_N/2$. For a ballistic
contact $\tprob_n \equiv 1$ and $Z(x)=G_N/(x+1)$. For other types of
contacts, it is typically useful to find the observable for
arbitrary $\tprob$ and weight it with $\rho(\tprob)$, e.g.,
$\rho(\tprob)=\hbar \pi G_N/[(2e^2)\tprob\sqrt{1-\tprob}]$ for a
diffusive contact \cite{nazarov:134}, $\rho(\tprob)=\hbar G_N/[e^2
\tprob^{3/2}\sqrt{1-\tprob}]$ for a dirty interface
\cite{schep:3015} or $\rho(\tprob)=2\hbar
G_N/[e^2\sqrt{\tprob(1-\tprob)}]$ for a chaotic cavity
\cite{baranger:142}. This way, the observables can be related to the
normal-state conductance $G_N$ of the junction.

Equation \eqref{eq:nazarovbc} yields a boundary condition both for
the "spectral" functions $\hat{G}^{R/A}$ and for the distribution
functions. In the absence of superconductivity, we simply have
$\hat{G}^{R/A}=\pm \hat{\tau}_3$, and the boundary condition for
the distribution functions becomes independent of the type of the
constriction,
\begin{equation}
j^{L/T}=G_N(f^{L/T}_R-f^{L/T}_L).
\end{equation}
In this case, the two currents can be included in the same
function by defining $j(\pm \energy)=(j^L(\energy) \pm
j^T(\energy))/2$. This yields the spectral current through the
constriction
\begin{equation}
j(\energy)=G_N(f_L(\energy)-f_R(\energy)), \label{eq:incohbc}
\end{equation}
where $f_{L/R}$ is the energy distribution function on the
left/right side of the constriction.

Another interesting yet tractable case is the one where a
superconducting reservoir (on the "left" of the junction) is
connected to a normal metal (on the "right") and the proximity
effect into the latter can be ignored. The latter is true, for
example, if we are interested in the distribution function at
energies far exceeding the Thouless energy of the normal-metal
wire, or in the presence of strong depairing. In this case, the
spectral energy and charge currents are
\begin{subequations}
\begin{align}
j^L&=\frac{2e^2}{h} \sum_{n} \tprob_n M_L(\tprob_n)
\theta(\bar{\energy})(f^L_R-f^L_L)\\
j^T&=\frac{2e^2}{h} \sum_n \tprob_n[M_T^{1}(\tprob_n)
\theta(-\bar{\energy})+M_T^{2}(\tprob_n)\theta(\bar{\energy})]f^T_R.
\end{align}
\label{eq:incohbcns}
\end{subequations}
Here $\bar{\energy}=|E|-\Delta$ and $\theta(\energy)$ is the
Heaviside step function, and the energy-dependent coefficients are
\begin{subequations}
\begin{align}\label{eq:cohfactors}
M_L(\tprob)&=\frac{2\left((2-\tprob)|\energy|\Omega+\tprob
\Omega^2\right)}{\left((2-\tprob)\Omega+\tprob
|\energy|\right)^2}\\
M_T^{1}(\tprob)&=\frac{2\tprob \Delta^2}{4
(\tprob-1)\energy^2+(\tprob-2)^2\Delta^2}\\
M_T^{2}(\tprob)&=\frac{2|\energy|}{(2-\tprob)\Omega+\tprob
|\energy|}
\end{align}
\end{subequations}
Here we defined $\Omega\equiv \sqrt{\energy^2-\Delta^2}$. In the
tunneling limit $\tprob \ll 1$, we get
\begin{equation}
j^{L/T}=G_N N_S(\energy)(f^{L/T}_R-f^{L/T}_L),
\label{eq:tunnellimitbc}
\end{equation}
where
\begin{equation}
N_S(\energy)=\left|\Re\left[\frac{\energy + i
\Gamma}{\sqrt{(\energy+i \Gamma)^2-\Delta^2}}\right]\right|
\overset{\Gamma \rightarrow 0}{\rightarrow}
\theta(\bar{\energy})|\energy|/\Omega \label{eq:sdos}
\end{equation}
is the reduced superconducting density of states (DOS). The first
form of Eq.~\eqref{eq:sdos} assumes a finite pair-breaking rate
$\Gamma$, which turns out to be relevant in some cases discussed
in Sec.~\ref{sec:ecool}.C.1 Unless specified otherwise, we assume
that the superconductors are of the conventional weak-coupling
type and the superconducting energy gap $\Delta$ at $T=0$ is
related to the critical temperature $T_c$ by $\Delta \approx 1.764
k_B T_c$ \cite{tinkham:96}.

\subsection{Collision integrals for inelastic scattering}
\label{subs:colli}

The collision integral $I_{\rm in}$ in
Eq.~\eqref{eq:diffusiveboltzmann} is due to electron--electron,
electron--phonon interaction and the interaction with the photons
in the electromagnetic environment.

\subsubsection{Electron-electron scattering}
For the electron--electron interaction, the collision integral is
of the form
\begin{equation} \label{eq:collee}
{\cal I}_{\rm coll}^{\rm e-e}=\kappa_{e-e}^{(d)} \int d\omega
d\energy' \omega^{\alpha}\left(\tilde{I}_{\rm coll}^{\rm
in}(\omega,\energy,\energy')-\tilde{I}_{\rm coll}^{\rm
out}(\omega,\energy,\energy')\right),
\end{equation}
where $\alpha$ and $\kappa_{e-e}^{(d)}$ depend on the type of
scattering and the "in" and "out" collisions are
\begin{subequations}
\begin{align}
\tilde{I}_{\rm coll}^{\rm in} &=
[1-f(\energy)][1-f(\energy')]f(\energy-\hbar\omega)f(\energy'+\hbar\omega)\\
\tilde{I}_{\rm coll}^{\rm out} &=
f(\energy)f(\energy')[1-f(\energy-\hbar\omega)][1-f(\energy'+\hbar\omega)].
\end{align}
\end{subequations}
Electron-electron scattering can be either due to the direct
Coulomb interaction \cite{altshuleraronov}, or mediated through
magnetic impurities which can flip their spin in a scattering
process \cite{kaminski:2400} or other types of impurities with
internal dynamics. In practice, both of these effects contribute
to the energy relaxation \cite{anthore:076806,pierre:437}.
Assuming the electron--electron interaction is local on the scale
of the variations in the distribution function, the direct
interaction yields \cite{altshuleraronov} Eq.~\eqref{eq:collee}
with $\alpha=-2+d/2$ for a $d$-dimensional wire. In a diffusive
wire, the effective dimensionality of the wire is determined by
comparing the dimensions to the energy-dependent length
$L_{\energy} \equiv \sqrt{\hbar D/\energy}$. The prefactor
$\kappa_{e-e}^{(d)}$ for a $d$-dimensional sample is
\begin{subequations}
\begin{align}
\kappa_{e-e}^{(1)} &=\frac{1}{\pi \sqrt{2D} \hbar^2 \nu_F A},\quad
\text{\cite{huard:599}}\\
\kappa_{e-e}^{(2)} &= \frac{1}{8E_F \tau}, \quad \text{\cite{rammer:323}}\\
\kappa_{e-e}^{(3)} &=\frac{1}{8 \pi^2\sqrt{2} \hbar^2 \nu_F
D^{3/2}}, \quad \text{\cite{rammer}}
\end{align}
\end{subequations}
where $\nu_F=\nu(E_F)$ is the density of states at the Fermi
energy $E_F$ and $A$ is the wire cross-section.

In the case of relaxation due to magnetic impurities, one expects
\cite{kaminski:2400} $\alpha=-2$ and
$\kappa_{e-e}=\frac{\pi}{2}\frac{c_m}{\hbar \nu_F} S
(S+1)\left[\ln\left(\frac{eV}{k_B T_K}\right)\right]^{-4}$. Here
$c_m$ is the concentration, $S$ is the spin, and $T_K$ is the
Kondo temperature of the magnetic impurities responsible for the
scattering. This form is valid for $T > T_K$. For a more detailed
account of the magnetic-impurity effects, see
\cite{goppert:033301,goppert:193301,goppert:195328,kroha:176803,ujsaghy:256805}
and the references therein.

For $d=3$, and for small deviations $\delta f$ from equilibrium, the
collision integral can be approximated \cite{rammer} by $-\delta
f/\tau_{e-e}$, where $\tau_{e-e} =
3\sqrt{3\pi}(\sqrt{8}-1)\zeta(3/2) (k_B T)^{3/2}/(16 k_F \ell_{el}
\sqrt{\hbar \tau} E_F)$ is the relaxation time ($\zeta(3/2) \approx
2.612$), $\tau=\ell_{el}/v_F$ is the elastic scattering time and
$k_F$ is the Fermi momentum. In the case when $\alpha < -1/2$, the
usual relaxation-time approach does not work for the
electron--electron interaction as the expression for the relaxation
time has an infrared divergence \cite{altshuleraronov,rammer}.
Therefore, one has to solve the full Boltzmann equation with the
proper collision integrals. To obtain an estimate for the length
scale at which the electron--electron interaction is effective, we
can proceed differently. Introducing dimensionless position
$\tilde{x}\equiv x/L$ and energy variables $\tilde{\energy'} \equiv
\energy'/\energy^*$ and $\tilde{\omega} \equiv \hbar
\omega/\energy^*$, we get
\begin{equation*}
\partial_{\tilde{x}}^2 f = -K_{e-e} \tilde{I}_{\rm coll}.
\end{equation*}
Here the dimensionless integral $\tilde{I}_{\rm coll}$
characterizes the deviation in the shape of the distribution
function from the Fermi function and $K_{e-e}$ depends on the
specific system. For a quasi-1d wire with bare Coulomb
interaction,
\begin{equation}
K_{e-e}=\frac{1}{\sqrt{2}} \frac{R_D}{R_K}
\sqrt{\frac{\energy^*}{\energy_T}}, \label{eq:colleepref}
\end{equation}
where $R_K=h/(2e^2)$, $R_D=L/(\sigma A)$ is the resistance of the
wire and $\energy_T=\hbar D/L^2$ is the Thouless energy. In the
case when the wire terminates in a point contact with resistance
$R_T$, the resistance $R_D$ in Eq.~\eqref{eq:colleepref} should be
replaced with the total resistance $R_D + R_T$
\cite{pekola:056804}. Typically the energy scale characterizing
the deviation from (quasi)equilibrium is $\energy^* = eV$. At $eV
\gg k_B T$, electron--electron collisions start to be effective
when $K_{e-e} \approx 1$. This yields a length scale
\begin{equation}
\ell^*_{e-e} = \sqrt{R_K A \sigma\sqrt{\frac{2\hbar D}{eV}}},
\end{equation}
where $A$ is the cross section of the wire and $\sigma$ its
conductivity. Using a wire with resistance $R_D = 10$ $\Omega$ and
$\energy_T \approx 10$ $\mu$eV for $L=1$ $\mu$m (close to the
values in \cite{huard:599}), and a voltage $V=100$ $\mu$V, we get
$\ell^*_{e-e} \approx 24$ $\mu$m. Increasing the temperature,
$\tilde{I}_{\rm coll}$ becomes smaller, and this effective length
also decreases. The experimental results of \textcite{huard:599}
indicate at least an order of magnitude larger $\kappa_{e-e}$ and
thus smaller $\ell^*_{e-e}$ than predicted by this theory. At
present, the reasons for this discrepancy have not been found.

\subsubsection{Electron--phonon scattering}
\label{subs:eph} Another source of inelastic scattering is due to
phonons, for which the collision integral is of the form
\cite{rammer,wellstood:5942}
\begin{equation}
{\cal I}_{\rm coll}^{\rm e-ph}=2\pi \int_0^\infty d\omega \alpha^2
F(\omega) \left[\tilde{I}_{\rm coll}^{\rm
in}(\energy,\omega)-\tilde{I}_{\rm coll}^{\rm
out}(\energy,\omega)\right].
\end{equation}
Here
\begin{equation}
\begin{split}
\tilde{I}_{\rm coll}^{\rm
in}(\energy,\omega)=&f(\energy+\hbar\omega)[1-f(\energy)][1+n_{\rm
ph}(\omega)]\\&+[1-f(\energy)] f(\energy-\hbar \omega) n_{\rm ph}(\omega))\\
\tilde{I}_{\rm coll}^{\rm
out}(\energy,\omega)=&f(\energy)[1-f(\energy-\hbar\omega)][1+n_{\rm
ph}(\omega)]\\&+f(\energy) [1-f(\energy+\hbar\omega)] n_{\rm
ph}(\omega).
\end{split}
\end{equation}
The kernel $\alpha^2 F(\omega)$ (the {\it Eliashberg function}) of
the interaction depends on the type of considered phonons
(longitudinal or transverse), on the relation between the phonon
wavelength $\lambda_{ph}$ and the electron mean free path
$\ell_{el}$, on the dimensionality of the electron and phonon system
\cite{sergeev:136602}, and on the characteristics of the Fermi
surface \cite{prunnila:206602}. At sub-kelvin temperatures and low
voltages, the optical phonons can be neglected, and one can only
concentrate on the acoustic phonons. In what follows, we also
neglect phonon quantization effects which may be important in
restricted geometries. Moreover, the phonon distribution function
$n_{\rm ph}(\omega)$ is considered to be in (quasi)equilibrium,
i.e., described by a Bose distribution function $n_{\rm
ph}(\omega)=n_{\rm eq}(\omega)\equiv [\exp(\hbar \omega/(k_B
T))-1]^{-1}$ (for phonon relaxation processes, see
Subs.~\ref{subs:phonons}).

When the phonon temperature $T_{ph}$ is much lower than the Debye
temperature $T_D$, the phonon dispersion relation is linear and one
can estimate the phonon wavelength using $\lambda_{ph} = hv_S/k_B
T_{ph}$. For typical metals, the speed of sound is $v_S \sim 3 \dots
5$ km/s, which yields a wavelength $\lambda_{ph} \sim 100\dots 200$
nm at $T_{ph}=1$ K and $\lambda_{ph} \sim 1 \dots 2$ $\mu$m at
$T_{ph}=100$ mK. In the clean limit $\lambda_{ph} \ll \ell_{el}$,
approximating the electron-phonon coupling with a scalar deformation
potential, only the longitudinal phonons are coupled to the
electrons. In this case for $\omega \ll k_B T_D/\hbar, k_F v_S$
\cite{wellstood:5942,rammer}
\begin{equation}
\alpha^2 F(\omega)=\frac{|M|^2}{4\pi^2\hbar^2 v_S^3 \nu_F}
\omega^2, \label{eq:eliashbergfunction}
\end{equation}
where $|M|^2$ is the square of the matrix element for the
deformation potential. Generally this is inversely proportional to
the mass density of the ions, but its precise microscopic form
depends on the details of the lattice structure. Therefore it is
useful to present $|M|$ in terms of a separately measurable
quantity, e.g., the prefactor $\Sigma$ of the power $P=\Sigma \Vol
T^5$ dissipated to the lattice of volume $\Vol$ in the
quasiequilibrium limit (see Table \ref{table:Sigma} and
Subs.~\ref{subs:limits}): $|M|^2=\pi \hbar^5 v_S^3 \Sigma/(12
\zeta(5) k_B^5)$, where $\zeta(5) \approx 1.0369$.

\begin{table}
 \centering
\begin{tabular}{|p{0.85cm}|p{1.22cm}|p{1.22 cm}|p{1.4cm}|p{3.26cm}|}
\firsthline

  & $T_{e}$ (mK) &  $T_{ph} $ (mK) & $\Sigma$ (W m$^{-3}$K$^{-5})$    & Measured in \\

 \hline

 Ag &  50...400 &   50...400 &   0.5 10$^{9}$    &     \cite{steinbach:3806} \\

 \hline

 Al&  35...130  &35 & 0.2 10$^{9}$ & \cite{kautz:2386}\\
  & 200...300  & 200  & 0.3 10$^{9}$&\cite{meschke:1119}  \\

\hline

 Au &  80...1200 &   80...1000 &   2.4 10$^{9}$    &     \cite{echternach:10339} \\

\hline

  AuCu &  50...120  &   20...120  &   2.4 10$^{9}$    &     \cite{wellstood:2599}\\

 \hline

  Cu &  25...800 &   25...320 &   2.0 10$^{9}$    &     \cite{roukes:422} \\

  &100...500  &280...400 & 0.9...4 10$^{9}$ & \cite{leivo:1996}\\

 & 50...200 & 50...150  & 2.0 10$^{9}$&\cite{meschke:1119}  \\

\hline

 Mo & 980   & 80...980    & 0.9 10$^{9}$ & \cite{savin:2005}\\

 \hline

 n$^{++}$Si & 120...400 & 175...400  & 0.1 10$^{9}$ & \cite{savin:1471}\\
 & 173...450  & 173  & 0.04 10$^{9}$ & \cite{prunnila:773}\\
 & 320...410  & 320...410  & 0.1 10$^{9}$ & \cite{buonomo:7784}\\

\hline

 Ti & 300...800  & 500...800  & 1.3 10$^{9}$  &\cite{manninen:3020}\\

 \hline

\end{tabular}
\caption{Measured electron-phonon coupling constant $\Sigma$ for
different materials. The second and third columns indicate the
temperature ranges (electron and phonon temperatures,
respectively) of the measurements.} \label{table:Sigma}
\end{table}

In the dirty limit $\lambda_{ph} \gg \ell_{el}$, the power of
$\omega$ in the Eliashberg function can be either 1 or 3, for the
cases of static or vibrating disorder, respectively. For further
details about the dirty limit, we refer to
\cite{belitz:2513,sergeev:6041,rammer:1352}.

The relaxation rate for electron-phonon scattering is given by
$1/\tau_{e-ph}=-\{\delta I_{e-ph}[f(\energy)]/\delta
f(\energy)\}|_{f(\energy)=f_{\rm eq}(\energy)}$, where $f_{\rm
eq}(\energy)$ now is a Fermi function at the lattice temperature.
With this definition at $\energy=\energy_F$, \cite{rammer}
\begin{equation}
\frac{1}{\tau_{e-ph}}=4\pi \int_0^\infty d\omega \frac{\alpha^2
F(\omega)}{\sinh\left(\frac{\hbar \omega}{k_B T}\right)}.
\end{equation}
Thus, in the clean case for $k_B T \ll 2 \hbar k_F v_S$ we obtain
$1/\tau_{e-ph}=\alpha T^3$, $\alpha=7 \zeta(3) \Sigma/(24\zeta(5)
k_B^2 \nu_F) \approx 0.34 \Sigma/(k_B^2 \nu_F)$. With typical values
for Cu, $\Sigma=2 \cdot 10^9$ ${\rm W K}^{-5}{\rm m}^{-3}$ and
$\nu_F = 1.6 \cdot 10^{47}$ J$^{-1}$m$^{-3}$, we get $\tau_{e-ph} =
45$ ns at $T=1$ K. Assuming $\lambda_F \ll \ell_{el} \ll
\ell_{e-ph}$, the electron--phonon scattering length is
$\ell_{e-ph}=\sqrt{D\tau_{e-ph}}$. For the above values and a
typical diffusion constant $D=0.01 $ m$^2$/s, $\ell_{e-ph} \approx
21$ $\mu$m at $T=1$ K and $\ell_{e-ph} \approx 670$ $\mu$m at
$T=100$ mK.

In the disordered limit $\lambda_{ph} \gg l_{el}$, the temperature
dependence of the electron-phonon scattering rate is expected to
follow either the $T^{2}$ or $T^4$ laws, depending on the nature of
the disorder \cite{sergeev:6041}.

It seems that although most of the experiments are done in the
limit where the phonon wavelength at least slightly exceeds the
electron mean free path, in majority of the cases the results have
fitted to the clean-limit expressions, i.e., the scattering rate
$\propto T_e^3$ and the heat current flowing into the phonon
system $\propto T_e^5$, see Eq.~\eqref{eq:phononpower} below (for
an exception, see \cite{karvonen:012302}). Finding the correct
exponent is not straightforward, as the film phonons are also
typically affected by the measurement, and because of the reduced
dimensionality of the phonon system (see
Subs.~\ref{subs:phonons}). In this Review, we concentrate on the
clean-limit expressions.

\subsection{Quasiequilibrium limit}
\label{subs:limits}

The shape of the distribution function at a given position of the
wire strongly depends on how the inelastic scattering length
$l_{\rm in}$ compares to the length $L$ of the wire. For $L \ll
l_{\rm in}$ ({\it nonequilibrium limit}), we may neglect the
inelastic scattering altogether. In this case, the distribution
function is a solution to either Eqs.~\eqref{eq:kinetic} or
Eq.~\eqref{eq:diffusiveboltzmann}, where the collision
integrals/self energies for inelastic scattering can be neglected.
As a result, the shape of the electron distribution functions
inside the wire at a finite bias voltage $eV \gtrsim k_B T$ may
strongly deviate from a Fermi distribution
\cite{pothier:3490,pierre:1078,heslinga:5157,heikkila:100502,pekola:056804,giazotto:137001}.
The nonequilibrium shape shows up in most of the observable
properties of the system, including the $I-V$ characteristics, the
current noise or the supercurrent. In general, it can only be
neglected in the $I-V$ characteristics if the charge transport
process is energy independent as in the case of purely
normal-metal samples. Even in this case the form of $f(\energy)$
can be observed in the current noise.

The kinetic equations can be greatly simplified in the limit where
$l_{\rm in}$ for one type of scattering is much smaller than $L$. In
the {\it quasiequilibrium} limit, the energy relaxation length due
to electron--electron scattering is much shorter than the wire,
$\ell_{e-e} \ll L$ \cite{nagaev:4740}. In this case, the local
distribution function is a Fermi function characterized by the
temperature $T_e(\vr,t)$ and potential $\mu(\vr,t)$. Mathematically,
this can be seen by considering the Boltzmann equation
\eqref{eq:diffusiveboltzmann} with the electron--electron collision
integral, Eq.~\eqref{eq:collee} in the limit where the prefactor of
the latter becomes very large. As the left-hand side of
Eq.~\eqref{eq:diffusiveboltzmann} is not strongly dependent on the
form of $f(\energy,\vr)$ as a function of energy, the equation can
only be satisfied if the collision integral without the prefactor
becomes small. It can be easily shown that the latter vanishes for
$f(\energy)=f_{\rm eq}(\energy)$. Thus, the deviations from the
Fermi-function shape will be at most of the order of $l_{\rm in}/L$,
and can be neglected in the quasiequilibrium limit.

In this limit, we are still left with two unknowns, $T_e(\vr,t)$
and $\mu(\vr,t)$. Substituting $f(\energy,\vr)=f_{\rm
eq}(\energy;T_e(\vr,t),\mu(\vr,t))$ in
Eq.~\eqref{eq:diffusiveboltzmann} yields
\begin{equation*}
\begin{split}
&(\partial_t - D\nabla^2) f=(\partial_t - D \nabla^2) (T_e
\partial_{T_e} f + \mu \partial_\mu f)-\\&D[(\nabla \mu)^2 \partial_\mu^2 f)+(\nabla T_e)^2
\partial_{T_e}^2 f+ 2 \nabla T_e \nabla \mu \partial_{T_e} \partial_\mu f] = I_{\rm coll}[f],
\end{split}
\end{equation*}
where $I_{\rm coll}[f]$ contains the other types of inelastic
scatterings, e.g., those with the phonons. In the right hand side
of the upper line, the differential operators $\partial_t$ and
$\nabla$ act only on $T_e$ and $\mu$. Integrating this over the
energy and multiplying by $\nu_F \energy$ and then integrating
over $\energy$ yields
\begin{align}
(\partial_t-D\nabla^2) \mu(\vr) &=0, \label{eq:quasieqpot}\\
C_e(\vr,t) \partial_t T_e - \nabla(\kappa(\vr,t) &\nabla T_e) -
\sigma (\nabla \mu/e)^2 = \tilde{I}_{\rm
coll}.\label{eq:quasieqtemp}
\end{align}
We assumed that the energy integral over $I_{\rm coll}[f]$
vanishes.\footnote{In the diffusive limit where the inelastic
scattering rates are lower than $1/\tau$, this is related to the
particle number conservation and is thus generally valid.} Here
$C_e(\vr,t)=\pi^2 \nu_F k_B^2 T_e(\vr,t)/3$ is the electron heat
capacity, $\sigma=D\nu_F e^2$ is the Drude conductivity,
$\kappa(\vr,t)=\sigma \Lor T_e(\vr,t)$ is the electron heat
conductivity, $\Lor=\pi^2 k_B^2/(3e^2) \approx 2.45 \cdot 10^{-8}$
${\rm W} \Omega {\rm K}^{-2}$ is the Lorenz number and
$\tilde{I}_{\rm coll}(T_e,\mu)$ contains the power per unit volume
emitted or absorbed by other excitations, such as phonons or the
electromagnetic radiation field. The last term on the left hand
side of Eq.~\eqref{eq:quasieqtemp} describes the Joule heating due
to the applied voltage. In what follows, we write the volume
explicitly in the collision integral by averaging over a small
volume $\Vol$ around the point $\vr$ where $T(\vr)$ is
approximately constant, thus defining $P_{\rm coll}(\vr) \equiv
\Vol \tilde{I}_{\rm coll}$.

For the electron-phonon scattering \cite{wellstood:5942}, $P_{\rm
coll}$ reads in the clean case (see also Table \ref{table:Sigma})
\begin{equation}
P_{\rm coll}^{e-ph} =\Sigma \Vol (T_e(\vr)^5-T_{ph}(\vr)^5).
\label{eq:phononpower}
\end{equation}
For the dirty limit specified below
Eq.~\eqref{eq:eliashbergfunction}, the Eliashberg functions
scaling with $\omega^n$ translate into temperature dependences
scaling as $T^{n+3}$, i.e., $T^{4}$ and $T^6$ \cite{sergeev:6041}.

The electrons can also be heated due to the thermal noise in their
electromagnetic environment unless proper filtering is realized to
prevent this heating. If one aims to detect the electromagnetic
environment as discussed in Sec.~\ref{sec:thermaldetectors}, this
discussion can of course be turned around to find the optimal
coupling to the radiation to be observed. A model for such
coupling in the quasiequilibrium limit was considered by
\textcite{schmidt:045901}, who obtained an expression for the
emitted/absorbed power due to the external noise in the form
\begin{equation}\label{eq:empower}
P_{\rm coll}^{\rm e-em}=r\frac{k_B^2 \pi^2}{6h}(T_e^2-T_\gamma^2).
\end{equation}
Here $r=4 R_e R_\gamma/(R_e+R_\gamma)^2$ is the coupling constant,
$T_\gamma$ is the (noise) temperature of the environment, and
$R_e$ and $R_\gamma$ are the resistances characterizing the
thermal noise in the electron system and the environment,
respectively. This expression assumes a frequency independent
environment in the relevant frequency range. For some examples on
the frequency dependence, we refer to \textcite{schmidt:045901}.

\subsection{Observables}
\label{subs:observables}

\subsubsection{Currents}

In the nonequilibrium diffusive limit, the charge current in a
normal-metal wire in the absence of a proximity effect or any such
interference effects as weak localization is obtained from the
local distribution function by
\begin{equation}
\cur=-eA \int_{-\infty}^\infty dE D(E) \nu(E) \nabla f(x;E)
\label{eq:curinawire}
\end{equation}
and the heat current from a reservoir with potential $\mu$ is
\begin{equation}
\jQ= -A\int_{-\infty}^\infty dE (E-\mu) D(E) \nu(E) \nabla f(x;E).
\label{eq:heatcurinawire}
\end{equation}
Here we included the possible energy dependence of the diffusion
constant $D(E)$ and of the density of states $\nu(E)$, due to the
energy dependence of the elastic scattering time, or due to the
nonlinearities in the quasiparticle dispersion relation. If the
Kondo effect \cite{vavilov:075119} can be neglected, such effects
are very small in good metals at temperatures of the order of 1 K
or less.

In the quasiequilibrium limit, assuming $D(E)=$const. and
$\nu(E)=\nu_F$, Eqs.~(\ref{eq:curinawire},\ref{eq:heatcurinawire})
can be simplified to
\begin{subequations}\label{eq:quasieqcurs}
\begin{align}
\cur&=-\sigma A \nabla \mu/e\\
\jQ&=-\kappa A \nabla T.
\end{align}
\end{subequations}

When a diffusive wire of resistance $R_D$ is connected to a
reservoir through a point contact characterized by the
transmission eigenvalues $\{\tprob_n\}$, the final distribution
function is obtained after solving the Boltzmann equation
\eqref{eq:diffusiveboltzmann} or Eqs.~\eqref{eq:kinetic} with the
boundary conditions given by Eq.~\eqref{eq:nazarovbc},
\eqref{eq:incohbc}, or by \eqref{eq:incohbcns}. However, when
$R_D$ is much less than the normal-state resistance $1/G_N$ of the
contact, we can ignore the wire and obtain the full current by a
direct integration over Eq.~\eqref{eq:incohbc} or
Eq.~\eqref{eq:incohbcns}. For example, for NIS or SIS tunnel
junctions, the expressions for the charge and heat currents from
the left side of the junction become
\begin{align}
\cur&=\frac{1}{e R_T} \int dE N_L(\tilde{\energy})
N_R(E)[f_L(E)-f_R(E)] \label{eq:tunnelcurrent}\\
\jQ&=\frac{1}{e^2 R_T} \int dE \tilde{\energy} N_L(\tilde{E})
N_R(E)[f_L(E)-f_R(E)]. \label{eq:tunnelheatcurrent}
\end{align}
Here $\tilde{\energy}=\energy-eV$, $R_T=1/G_N$,
$N_{L/R}(E)=\nu(E)/\nu_F$ is the reduced density of states for the
left/right wire, $N(E)=1$ for a normal metal and $N(E)=N_S(E)$ for
a superconductor. Furthermore, if the two wires constitute
reservoirs, $f_{L/R}$ are Fermi functions with potentials
$\mu_L=-eV$, $\mu_R=0$. The resulting NIS or SIS charge current is
a sensitive probe of temperature and can hence be used for
thermometry or radiation detection, as explained in
Secs.~\ref{subs:tjthermometry} and \ref{sec:thermaldetectors},
respectively. Moreover, analysis of the heat current
Eq.~\eqref{eq:tunnelheatcurrent} shows that the electrons can in
certain situations be cooled in NIS/SIS structures, as discussed
in Sec.~\ref{sec:SINIS}.

In the presence of a proximity effect, the equations for the
charge and thermal currents in the quasiclassical limit, i.e.,
ignoring the energy dependence of the diffusion constant 
and the normal-metal density of states, 
are
\begin{subequations}
\begin{align}
\cur&=\int dE j^T\\
\jQ&=\int dE (E j^L - \mu j^T). \label{eq:pecurs}
\end{align}
\end{subequations}
Here $\mu$ is the potential of the reservoir from which the heat
current is calculated. As in Eq.~\eqref{eq:kinetic}, these
currents can be separated into quasiparticle, anomalous, and
supercurrent parts.

\subsubsection{Noise}
\label{subsubs:noise}

Often one can express the zero-frequency current noise in terms of
the local distribution function \cite{blanter:1}. The noise is
characterized by the correlator
\begin{equation*}
S=2\int_{-\infty}^\infty dt' \langle \delta \hat{I}(t+t')
\delta\hat{I}(t)\rangle,
\end{equation*}
where $\delta \hat{I}=\hat{I}-\langle \hat{I} \rangle$ and
$\hat{I}$ is the current operator. In a stationary system $S$ is
independent of $t$. In a normal-metal wire of length $L$ in the
nonequilibrium limit, the current noise $S$ can be expressed as
\cite{nagaev:103}
\begin{equation}
S=\frac{4eD \nu_F A}{L^2}\int_0^L dx \int_{-\infty}^\infty dE
f(E,x)[1-f(E,x)].
\end{equation}
In the quasiequilibrium regime, this equation simplifies to
\cite{nagaev:4740}
\begin{equation}
S=\frac{4eD \nu_F A k_B}{L^2} \int_0^L dx T_e(x).
\end{equation}

When the resistance of a point contact dominates that of the wire,
the noise power can be expressed through \cite{blanter:1}
\begin{equation*}
\begin{split}
S_{NN}=&\frac{4e^2}{h}\sum_n \int dE \{\tprob_n
\left[f_L(1-f_L)+f_R(1-f_R)\right]\\&+\tprob_n(1-\tprob_n)(f_L-f_R)^2\}
\end{split}
\end{equation*}
for a normal-metal contact and \cite{dejong:16070}
\begin{equation*}
\begin{split}
S_{NS}=\frac{2e^2}{h}\sum_n \int dE
\bigg\{&\frac{\tprob_n^2}{(2-\tprob_n)^2}
2f_L(\energy)[1-f_L(\energy)]
\\&+ \frac{16
\tprob_n^2(1-\tprob_n)}{(2-\tprob_n)^4}(f^T_L)^2\bigg\}
\end{split}
\end{equation*}
for an incoherent NS contact at $E < \Delta$. Here $f^{L/R}$ are
the distribution functions in the left/right
(normal/superconducting in the latter case) side of the contact
and $f^T_L(E)=1-f_L(E)-f_L(-E)$, the symmetric part w.r.t.~the S
potential. If the scattering probabilities are independent of
energy and if the two sides are reservoirs with the same
temperature, these expressions simplify to
\begin{subequations}\label{eq:noise}
\begin{align}
S_{NN}&=\frac{2e^2}{h}\sum_n \left[2k_B T \tprob_n^2 + eV
\coth\left(v\right)
\tprob_n(1-\tprob_n)\right]\label{eq:nnnoise}\\
\begin{split}
S_{NS}&=\frac{2e^2}{h}\sum_n
\bigg[\frac{\tprob_n^2}{(2-\tprob_n)^2}2k_B T \\&+\frac{16
\tprob_n^2(1-\tprob_n)}{(2-\tprob_n)^4}\left(2eV
\coth\left(v\right)-4 k_B T\right)\bigg],
\end{split}\label{eq:nsnoise}
\end{align}
\end{subequations}
where $v=eV/(2 k_B T)$. For $eV=0$ (thermal Johnson-Nyquist noise),
$S_{NN/S}=4k_B T G_{N/S}$, where $G_N R_K= \sum_n {\mathcal T}_n$
and $G_S R_K = \sum_n {\mathcal T}_n^2/(2-{\mathcal T}_n)^2$ are the
conductances of the point contact in the NN and NS cases,
respectively. In the opposite limit $eV \gg k_B T$ (shot noise), one
obtains $S=2eF_{N/S}I_{N/S}$, where $I_{N/S}=G_{N/S}V$ is the
average current through the junction, and $F_{N/S}$ are the Fano
factors: $F_N G_N R_K=\sum_n {\mathcal T}_n (1-{\mathcal T}_n)$ and
$F_S G_S R_K = \sum_n 16 {\mathcal T}_n^2 (1-{\mathcal
T}_n)/(2-{\mathcal T}_n)^4$.

In the presence of the superconducting proximity effect, the
expression for the noise becomes more complicated
\cite{houzet:107004}. In general, it can be found by employing the
counting-field technique developed by Nazarov and coworkers, see
\cite{nazarov:196801} and the references therein. This technique
can also be applied to study the full counting statistics of the
transmitted currents through a given sample within a given
measurement time \cite{nazarovbook}.

Apart from the charge current, also the heat current in electric
circuits fluctuates. For example, the zero-frequency heat current
noise from the "left" of a tunnel contact biased with voltage $V$
is given by
\begin{equation}
\begin{split}
S_Q=&\frac{2}{e^2 R_T}\int d\energy \energy^2
N_L(\energy-eV)N_R(\energy)\\&[f_R(1-f_L)+f_L(1-f_R)].
\end{split}
\end{equation}
At low voltages $V \ll k_B T/e$, the heat current noise obeys the
fluctuation-dissipation result $S_Q = 4 k_B T^2 G_{\rm th}$, where
$G_{\rm th}$ is the thermal conductance. This quantity is related
to the noise equivalent power (NEP) discussed in the literature of
thermal detectors by $S_Q=$NEP$^2$. The total NEP contains
contributions not only from the electrical heat current noise, but
also from other sources, such as the direct charge current noise
and electron-phonon coupling. A detailed discussion of various NEP
sources is presented in Sec.~\ref{sec:thermaldetectors}.

Another important quantity is the cross-correlator between the
current and heat current fluctuations. At zero frequency, this is
given by
\begin{equation}
\begin{split}
S_{IQ}=&-\frac{2}{e R_T}\int d\energy \energy
N_L(\energy-eV)N_R(\energy)\\&[f_R(1-f_L)+f_L(1-f_R)].
\end{split}
\end{equation}
These types of fluctuations have to be taken into account for
example when analyzing the NEP of bolometers \cite{golubev:6464}.
Recently, also the general statistics of the heat current
fluctuations have been theoretically addressed by
\textcite{kindermann:155334}.

\begin{figure}[tb]
\centering
\includegraphics[width=0.48\columnwidth]{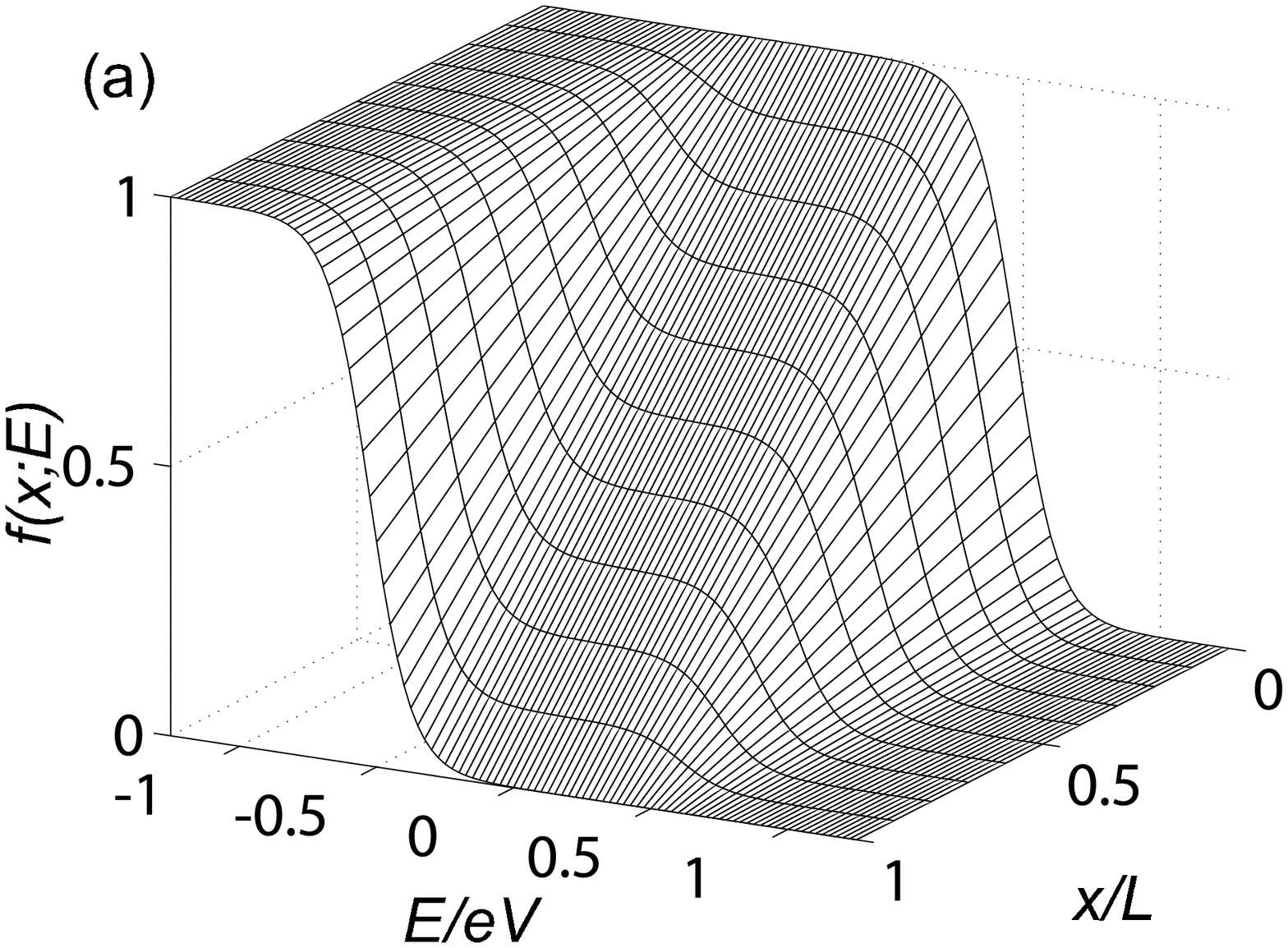}
\includegraphics[width=0.48\columnwidth]{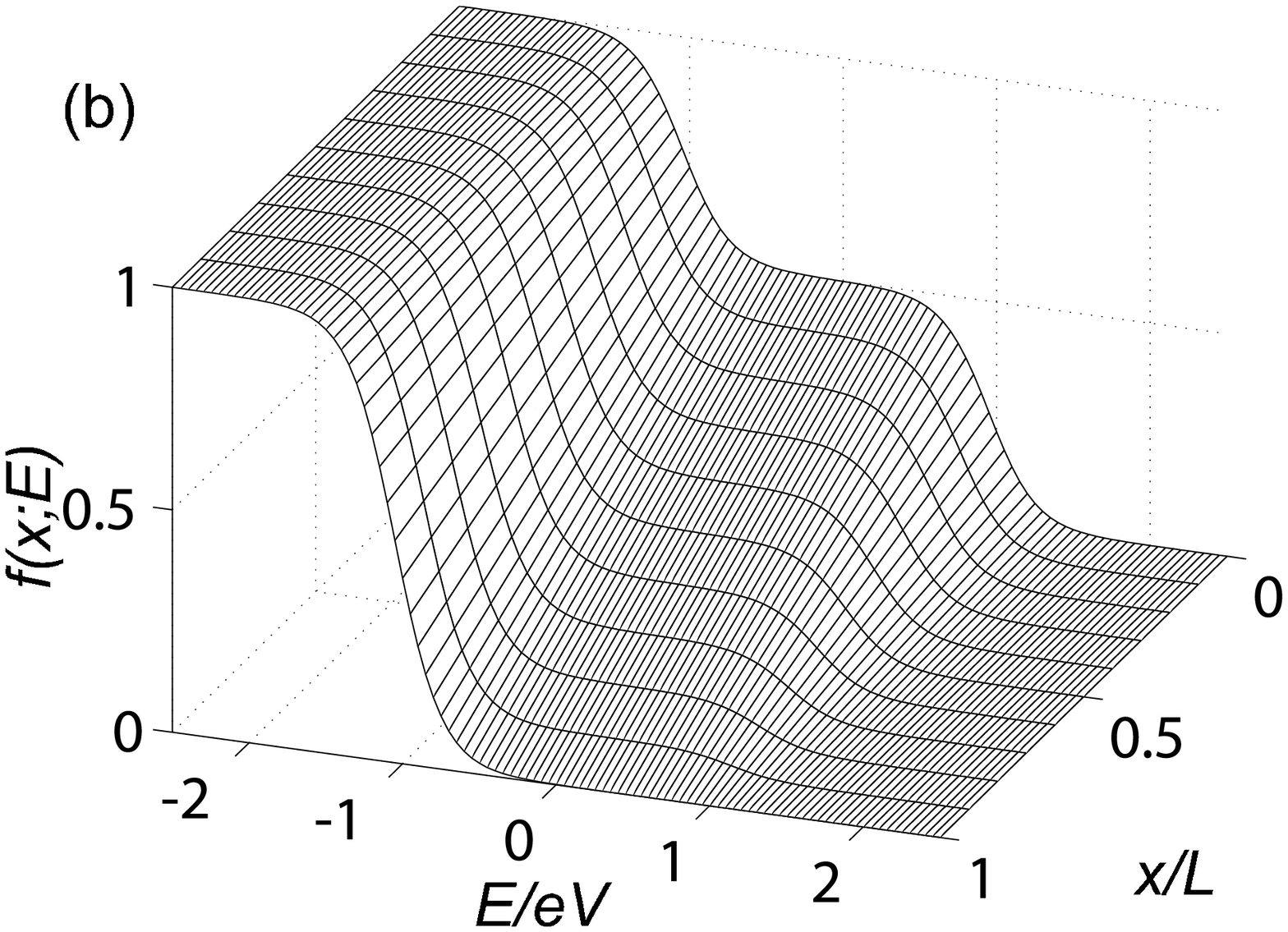}
\caption{Nonequilibrium quasiparticle energy distribution function
in a diffusive normal-metal wire in the absence of inelastic
scattering: (a) wire placed between two normal-metal reservoirs
and (b) wire placed between a normal-metal ($x=L$) and a
superconducting ($x=0$) reservoir. In the latter picture, we
assume $\hbar D/L^2 \ll eV \ll \Delta$, such that the proximity
effect and the states above the gap can be neglected. In both
pictures, the lattice temperature was fixed to $k_B T=eV/10$.}
\label{fig:noneqdistfuncs}
\end{figure}

\subsection{Examples on different systems}
\label{subs:examples}

Below, we detail the solutions to the kinetic equations,
\eqref{eq:diffusiveboltzmann} or \eqref{eq:kinetic} in some
example systems. The aim is first to provide an understanding of
the general behavior of the distribution function in these
systems, but also to show the generic features, such as the
electron cooling in NIS junctions.

\subsubsection{Normal-metal wire between normal-metal reservoirs}
The simplest example is a quasi-one-dimensional normal-metal
diffusive wire of resistance $R_D=L/(A \sigma)$, connected to two
normal-metal reservoirs by clean contacts. In the full
nonequilibrium limit, we find (see Fig.~\ref{fig:noneqdistfuncs}
(a); the coordinate $x$ follows Fig.~\ref{fig:setup})
\begin{equation}
f(E,x)=f_L(E)+[f_R(E)-f_L(E)]\frac{x}{L},
\end{equation}
where $f_L$ and $f_R$ are the (Fermi) distribution functions in
the left and right reservoirs with temperatures $T_L$ and $T_R$
and potentials $\mu_L$ and $\mu_R=\mu_L+eV$, respectively. The
resulting two-step form illustrated in
Fig.~\ref{fig:noneqdistfuncs}a was first measured by
\textcite{pothier:3490}.

\begin{figure}[tb]
\centering
\includegraphics[width=0.5\columnwidth]{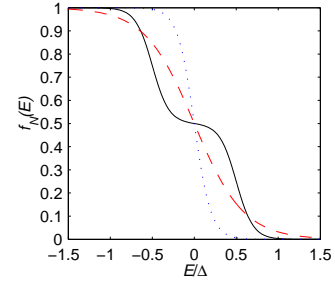}
\caption{(Color in online edition): Quasiparticle energy
distribution function in the center of a normal-metal wire placed
between two normal-metal reservoirs and biased with a voltage
$eV=10 k_B T$. The three lines correspond to three extreme limits:
$L \ll \ell_{e-e}, \ell_{e-ph}$ (nonequilibrium limit, black solid
line), $\ell_{e-e} \ll L \ll \ell_{e-ph}$ (quasiequilibrium limit,
red dashed line), and $\ell_{e-e}, \ell_{e-ph} \ll L$ (equilibrium
limit, blue dotted line).} \label{fig:distfunclimits}
\end{figure}

In the quasiequilibrium limit, we get $f(E,x)=f_{\rm
eq}(E;\mu(x),T_e(x))$, where
\begin{subequations}
\begin{align}
\mu(x)&=\mu_L+eV \frac{x}{L}\\
T_e(x)^2&=T_L^2+(T_R^2-T_L^2)\frac{x}{L}+\frac{V^2}{\Lor}\frac{x}{L}\left(1-\frac{x}{L}\right).
\end{align}
\end{subequations}
In both cases, the current is simply given by $\cur=V/R_D$ and the
heat current in the limit $V=0$ by $\jQ=\Lor (T_R^2-T_L^2)/R_D$.
For $V \neq 0$, the resistor dissipates power and $\jQ$ is not
conserved. The electron distribution function in the center of the
wire, $x=L/2$ is plotted in Fig.~\ref{fig:distfunclimits} for the
nonequilibrium, quasiequilibrium and local equilibrium (strong
electron--phonon scattering) limits.

To obtain estimates for the thermoelectric effects due to the
particle-hole symmetry breaking, let us lift the assumption of
energy independent diffusion constant and density of states,
expanding them as $D(E) \approx D_0 + c_D \frac{E-E_F}{E_F}$ and
$\nu(E) \approx \nu_F + c_N \frac{E-E_F}{E_F}$. For linear
response, the resulting expressions for the charge and heat
currents are \cite{cutler:1336}
\begin{subequations}
\begin{align}
\cur&=-G L\nabla \mu/e + G L\alpha \nabla T\label{eq:ncur}\\
\jQ&=-\Pi G L\nabla \mu/e+G_{\rm th} L\nabla T.\label{eq:nheatcur}
\end{align}
\label{eq:ncurs}
\end{subequations}
Here $G=e^2 \nu_F D_0 A/L$ is the Drude conductance, $G_{\rm
th}=\Lor G T=\kappa A/L$ ({\it Wiedemann-Franz law}) is the heat
conductance, $\alpha=e\Lor G' T/G$ ({\it Mott law}) is the Seebeck
coefficient describing the thermoelectric power, $\Pi=\alpha T$
({\it Onsager relation}) is the Peltier coefficient, and
$G'=e^2(c_D \nu_F+D_0 c_N)A/(L E_F)$ describes effects due to the
particle-hole symmetry breaking. We see that the thermoelectric
effects are in general of the order $k_B T/E_F$; in good metals at
temperatures of the order of 1 K they can hence be typically
ignored. Therefore, the Peltier refrigerators discussed in
Subs.~\ref{subs:peltier} rely on materials with a low $E_F$.

\begin{figure}[tb]
\begin{center}
\includegraphics[width=\columnwidth,clip]{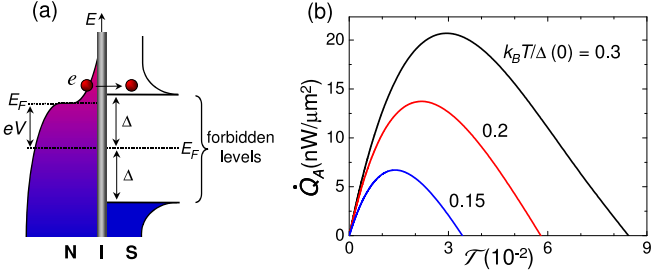}
\end{center}
\caption{Electron cooling at NIS junctions. (a) Sketch of the energy
band diagram of a voltage biased NIS junction. Upon biasing the
structure, the most energetic electrons ($e$) can most easily tunnel
into the superconductor. As a result the electron gas in the N
electrode is \textit{cooled}. (b) Maximum cooling power surface
density $\dot{Q}_A$ vs interface transmissivity $\tprob$ at
different temperatures calculated for a NS contact.}
\label{fig:coolprinciple}
\end{figure}

\subsubsection{Superconducting tunnel structures}
\label{subs:suptuns}

Consider a NIS tunnel structure, coupling a large superconducting
reservoir with temperature $T_{e,S}$ to a large normal-metal
reservoir with temperature $T_{e,N}$ via a tunnel junction with
resistance $R_T$. Let us then assume a voltage $V$ applied over
the system. In this case, the heat current (cooling power) from
the normal metal is given by Eq.~\eqref{eq:tunnelheatcurrent} with
$N_L(\energy) = 1$, $N_R(\energy) = N_S(\energy)$, $f_L(\energy)
\equiv
f_{\rm eq}(E-eV,T_{e,N})$ and $f_R(\energy) \equiv f_{\rm eq}(E,T_{e,S})$. %
For small pair breaking inside the superconductor, i.e., $\Gamma
\ll \Delta$, $\jQ_{\rm NIS}$ is positive for $eV<\Delta$, i.e., it
{\it cools} the normal metal. It is straightforward to show that
$\jQ_{\rm NIS}(V)=\jQ_{\rm NIS}(-V)$. This is in contrast with
Peltier cooling, where the sign of the current determines the
direction of the heat current. For $eV > \Delta$, the current
through the junction increases strongly, resulting in Joule
heating and making $\jQ_{\rm NIS}$ negative. The cooling power is
maximal near $eV \approx \Delta$.

\begin{figure}[t!b!]
\centering
\includegraphics[height=0.42\columnwidth]{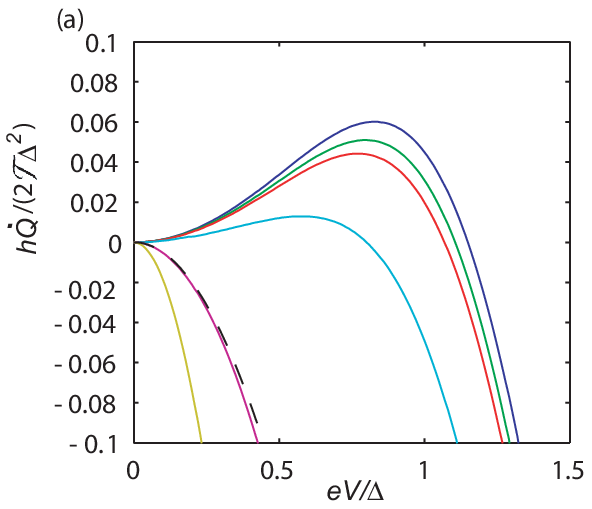}
\includegraphics[height=0.42\columnwidth]{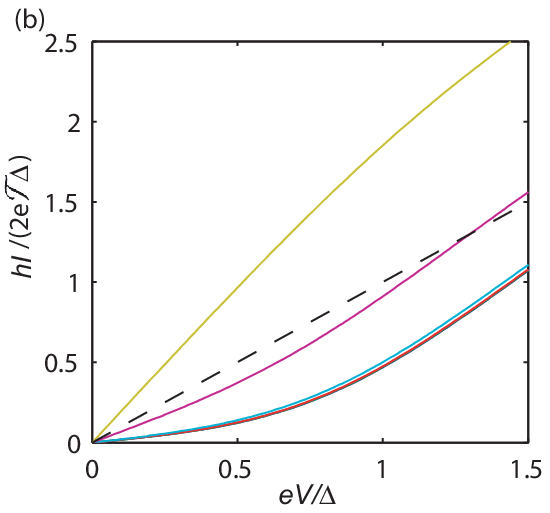}
\caption{NS point contact characteristics. (a) Heat current from the
N side as a function of voltage for different transparencies
$\tprob$ (from top to bottom, tunneling limit $\tprob \rightarrow
0$, $\tprob=0.005$, $\tprob=0.01$, $\tprob=0.05$, $\tprob=0.5$ and
$\tprob=1$) at $k_B T=0.3\Delta$. For $\tprob \lesssim 0.05$, the
heat current is positive, corresponding to cooling. (b)
Current-voltage characteristics for the same values of $T$ and
$\tprob$ as in (a) (now $\tprob$ increases from bottom to top). The
first four curves lie essentially on top of each other. The
corresponding NIN curves are shown with the dashed lines.}
\label{fig:niscurrentsvstau}
\end{figure}

In order to understand the basic mechanism for cooling in such
systems, let us consider the simplified energy band diagram of a
NIS tunnel junction biased at voltage $V$, as depicted in Fig.
\ref{fig:coolprinciple}(a). The physical mechanism underlying
quasiparticle cooling  is rather simple: owing to the presence of
the superconductor, in the tunneling process quasiparticles with
energy $E<\Delta$ cannot tunnel inside the forbidden energy gap,
but the more energetic electrons (with $E>\Delta$) are removed
from the N electrode. As a consequence of this "selective"
tunneling of hot particles, the electron distribution function in
the N region becomes sharper: the NIS junction thus behaves as an
electron cooler.

The role of barrier transmissivity in governing heat flux across
the NIS structure was analyzed by \textcite{bardas:12873}. %
They pointed out the interplay between single-particle tunneling
and Andreev reflection \cite{andreev:1823} on the heat current. In
the following it is useful to summarize their main results.

The cooling regime requires a tunnel contact. The effect of
transmissivity is illustrated in Fig.~\ref{fig:coolprinciple}(b),
which shows the maximum of the heat current density (i.e., the
specific cooling power) $\dot{Q}_{A}$ versus interface
transmissivity $\tprob$ at different temperatures. This can be
calculated for a generic NS junction using
Eqs.~\eqref{eq:incohbcns}, \eqref{eq:cohfactors} and
\eqref{eq:pecurs}. The quantity $\dot{Q}_{A}$ is a non-monotonic
function of interface transmissivity, vanishing both at low and
high values of $\tprob$. In the low transparency regime,
$\dot{Q}_{A}$ turns out to be linear in $\tprob$, showing that
electron transport is dominated by single particle tunneling. Upon
increasing barrier transmissivity, Andreev reflection begins to
dominate quasiparticle transport, thus suppressing heat current
extraction from the N portion of the structure. The heat current
$\dot{Q}_{A}$ is maximized between these two regimes at an optimal
barrier transmissivity, which is temperature dependent.
Furthermore, by decreasing the latter leads to a reduction of both
the optimal $\tprob$ and of the transmissivity window where the
cooling takes place. In real NIS contacts used for cooling
applications, the average $\tprob$ is typically in the range
$10^{-6}...10^{-4}$ \cite{nahum:3123,leivo:1996} corresponding to
junction specific resistances $R_c$ (i.e., the product of the
junction normal state resistance and the contact area) from tens
to several thousands $\Omega\,\mu$m$^2$. This limits  the
achievable $\dot{Q}_{A}$ to some pW$/\mu$m$^{2}$. From the above
discussion it appears that exploiting low-$R_c$ tunnel contacts is
an important requirement in order to achieve large cooling power
through NIS junctions. However, in real low-$R_c$ barriers,
pinholes with a large $\tprob$ appear. They contribute with a
large Joule heating (see Fig.~\ref{fig:niscurrentsvstau}, which
shows the heat and charge currents through the NIS interface as
functions of voltage for different $\tprob$), and therefore tend
to degrade the cooling performance at the lowest temperatures, and
to overheat the superconductor at the junction region due to
strong power injection. Barrier optimization in terms of both
materials and technology seems to be still nowadays a challenging
task (see also Sec.~\ref{subs:oxidebarriers}).

In a SINIS system in the quasiequilibrium limit, the temperatures
$T_{e,N}$ and $T_{e,S}$ of the N and S parts may differ. If the
superconductors are good reservoirs, $T_{e,S}$ equals the lattice
temperature. In the experimentally relevant case when the
resistance $R_D$ of the normal-metal island is much lower than the
resistances $R_L$ and $R_R$ of the tunnel junctions, the
normal-metal temperature $T_{e,N}$ is determined from the heat
balance equation
\begin{equation}
\jQ_{\rm SINIS}(V;T_{e,S},T_{e,N})=P_{\rm coll},
\end{equation}
where $\jQ_{\rm SINIS}=2\jQ_{NIS}$ and $P_{\rm coll}$ describes
inelastic scattering due to phonons (Eq.~\eqref{eq:phononpower})
and/or due to the electromagnetic environment
(Eq.~\eqref{eq:empower}).

For further details about NIS/SINIS cooling in the
quasiequilibrium limit, we refer to Sec.~\ref{sec:SINIS} and
\cite{anghel:197}.

\begin{figure}[tb]
\centering
\includegraphics[width=0.49\columnwidth]{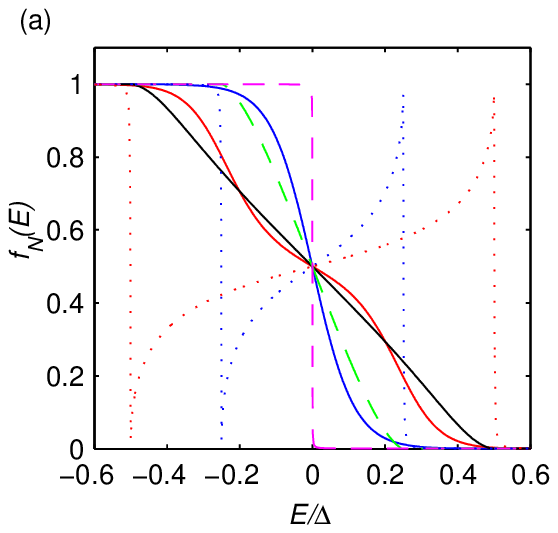}
\includegraphics[width=0.49\columnwidth]{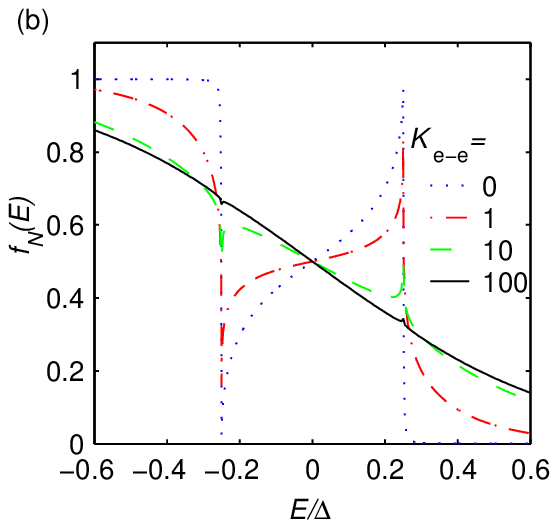}
\caption{Nonequilibrium distribution inside the SINIS island,
calculated from Eq. (42): (a) No inelastic scattering, ${\mathcal
I}_{\rm coll}=0$, $T=0.1 T_c$, and using a depairing strength
$\Gamma=10^{-4}\Delta$ inside the superconductors. The voltage runs
from zero to $V=3\Delta/e$ in steps of $\Delta/(2e)$ and spans three
regimes: (i) anomalous heating regime ($eV=0 \dots \Delta$, solid
lines, widening with an increasing $V$) discussed in
\cite{pekola:056804}, (ii) cooling regime ($eV=1.5 \Delta$ and
$eV=2\Delta$, dashed lines, narrowing with an increasing $V$) and
the strong nonequilibrium heating regime ($eV=2.5 \Delta$ and
$eV=3\Delta$, dotted lines, widening with an increasing $V$). (b)
Distribution at $eV=2.5\Delta$ for different strengths of the
electron--electron interaction, parametrized by the parameter
$K_{e-e}$ from Eq.~\eqref{eq:colleepref} with $R_D$ replaced by
$R_T$ and $\energy^*$ by $\Delta$, as in this case these describe
the collision integral.} \label{fig:sinisdist}
\end{figure}

Let us now consider the limit of full nonequilibrium, neglecting
the proximity effect from the superconductors on the normal-metal
island.\footnote{This is justified in the limit $R_L, R_R \gg
R_D$.} Then the distribution function inside the normal metal may
be obtained by solving the kinetic equation
Eq.~\eqref{eq:diffusiveboltzmann} in the static case along with
the boundary conditions given by Eq.~\eqref{eq:incohbc}. For
simplicity, let us assume $R_L=R_R \equiv R_T \gg R_D$. In this
limit, the distribution function $f_N(\energy)$ in the
normal-metal island is almost independent of the spatial
coordinate. Then, in Eq.~\eqref{eq:diffusiveboltzmann} we can use
$\partial_x^2 f \approx (\partial_x f(L)-\partial_x f(0))/L$,
where $L$ is the length of the N wire. From
Eq.~\eqref{eq:tunnellimitbc} we then get
\begin{equation}
\frac{R_D}{R_T}\left(N_R (f_N-f_R) - N_L (f_L-f_N)\right) = \tau_D
{\mathcal I}_{\rm coll},
\end{equation}
where $\tau_D = L_N^2/D$ is the diffusion time through the island.
As a result \cite{heslinga:5157,giazotto:137001}, the distribution
function in the central wire can be expressed as
\begin{equation}
f_N=\frac{f_R N_R+f_L N_L+R_I \Vol e^2 \nu_F {\mathcal I}_{\rm
coll}}{N_R+N_L}. \label{eq:sinisdist}
\end{equation}
Here $\Vol=LA$ is the volume of the island. In the presence of
inelastic scattering, this is an implicit equation, as ${\mathcal
I}_{\rm coll}$ is a functional of $f_N$. Examples of the effect of
inelastic scattering have been considered in \cite{heslinga:5157}
in the relaxation time approach, and in \cite{giazotto:137001}
including the full electron--electron scattering collision
integral. The distribution function $f_N$ in
Eq.~\eqref{eq:sinisdist} is plotted in Fig.~\ref{fig:sinisdist}
for a few values of the voltage for ${\mathcal I}_{\rm coll}=0$
and for a few strengths of electron--electron scattering at
$eV=2.5\Delta$.

\subsubsection{Superconductor-normal-metal structures with
transparent contacts}

If a superconductor is placed in a good electric contact with a
normal metal, a finite pairing amplitude is induced in the normal
metal within a thermal coherence length $\xi_T = \sqrt{\hbar
D/(2\pi k_B T)}$ near the interface. This superconducting {\em
proximity effect} modifies the properties of the normal metal
\cite{lambert:901,belzig:1251} and makes it possible to transport
supercurrent through it. It also changes the local distribution
functions by modifying the kinetic coefficients in
Eq.~\eqref{eq:kinetic}. Superconducting proximity effect is
generated by Andreev reflection \cite{andreev:1228} at the
normal-superconducting interface, where an electron scatters from
the interface as a hole and vice versa. Andreev reflection forbids
the sub-gap energy transport into the superconductor, and thus
modifies the boundary conditions for the distribution functions.
In certain cases, the proximity effect is not relevant for the
physical observables whereas the Andreev reflection can still
contribute. This incoherent regime is realized if one is
interested in length scales much longer than the extent of the
proximity effect. Below, we first study such "pure"
Andreev-reflection effects, and then go on to explain how they are
modified in the presence of the proximity effect.

In the incoherent regime, for $\energy < \Delta$, Andreev
reflection can be described through the boundary conditions
\cite{pierre:1078}
\begin{subequations}\label{eq:incohsupbc}
\begin{align}
&f(\mu_S+\energy)=1-f(\mu_S-\energy)\\
&\mathbf{\hat{n}}\cdot \nabla
[f(\mu_S-\energy)-f(\mu_S+\energy)]=0,
\end{align}
\end{subequations}
evaluated at the normal-superconducting interface. Here
$\mathbf{\hat{n}}$ is the unit vector normal to the interface and
$\mu_S$ is the chemical potential of the superconductor. The
former equation guarantees the absence of charge imbalance in the
superconductor, while the latter describes the vanishing energy
current into it. For $E>\Delta$, the usual normal-metal boundary
conditions are used. Note that these boundary conditions are valid
as long as the resistance of the wire by far exceeds that of the
interface; in the general case, one should apply
Eqs.~\eqref{eq:incohbcns}.

Assume now a system where a normal-metal wire is placed between
normal-metal (at $x=L$, potential $\mu_N$ and temperature
$T=T_{e,N}$) and superconducting reservoirs (at $x=0$, $\mu_S=0$,
$T=T_{e,S}$). Solving the Boltzmann equation
\eqref{eq:diffusiveboltzmann} then yields \cite{nagaev:081301}
\begin{equation}
f(E)=\begin{cases}
\frac{1}{2}\left(1+\frac{x}{L}\right)f_N(E)+\frac{1}{2}\left(1-\frac{x}{L}\right)\bar{f}_N(E),
\quad E < \Delta\\
\left(1-\frac{x}{L}\right)f_{\rm
eq}(E;\mu=0,T_{e,S})+\frac{x}{L}f_N(E), \quad E > \Delta.
\end{cases}
\end{equation}
Here $f_N(E)=f_{\rm eq}(E;\mu_N,T_{e,N})$ and $\bar{f}_N(E)=f_{\rm
eq}(E;-\mu_N,T_{e,N})$. This function is plotted in
Fig.~\ref{fig:noneqdistfuncs} (b). In the quasiequilibrium regime,
the problem can be analytically solved in the case $eV, k_B T \ll
\Delta$, i.e., when Eqs.~\eqref{eq:incohsupbc} apply for all
relevant energies. Then the boundary conditions are $\mu=\mu_S=0$
and $\mathbf{\hat{n}}\cdot \nabla T_e=0$ at the NS interface.
Thus, the quasiequilibrium distribution function is given by
$f(E,x)=f(E;\mu(x),T_e(x))$ with $\mu(x)=\mu_N (1-x/L)$ and
\begin{equation}
T_e(x)=\sqrt{T_{e,N}^2+\frac{V^2}{\Lor}\frac{x}{L}\left(2-\frac{x}{L}\right)}.
\end{equation}
Note that this result is independent of the temperature $T_{e,S}$
of the superconducting terminal.

In this incoherent regime, the electrical conductance is
unmodified compared to its value when the superconductor is
replaced by a normal-metal electrode: Andreev reflection
effectively doubles both the length of the normal conductor and
the conductance for a single transmission channel
\cite{beenakker:12841}, and thus the total conductance is
unmodified. Wiedemann-Franz law is violated by the Andreev
reflection: there is no heat current into the superconductor at
subgap energies. However, the sub-gap current induces Joule
heating into the normal metal and this by far overcompensates any
cooling effect from the states above $\Delta$ (see
Fig.~\ref{fig:niscurrentsvstau}).

The SNS system in the incoherent regime has also been analyzed by
\textcite{bezuglyi:14439} and \textcite{pierre:1078}. Using the
boundary conditions in Eqs.~\eqref{eq:incohsupbc} at both NS
interfaces with different potentials of the two superconductors
leads to a set of recursion equations that determine the
distribution functions for each energy. The recursion is
terminated for energies above the gap, where the distribution
functions are connected simply to those of the superconductors.
This process is called the multiple Andreev reflection: in a
single coherent process, a quasiparticle with energy $E<-\Delta$
entering the normal-metal region from the left superconductor
undergoes multiple Andreev reflections, and its energy is
increased by the applied voltage during its traversal between the
superconductors. Finally, when it has Andreev reflected $\sim
(2\Delta/eV)$ times, its energy is increased enough to overcome
the energy gap in the second superconductor. The resulting energy
distribution function is a staircase pattern, and it is described
in detail in \cite{pierre:1078}. The width of this distribution is
approximately $2\Delta$, and it thus corresponds to extremely
strong heating even at low applied voltages.

The superconducting proximity effect gives rise to two types of
important contributions to the electrical and thermal properties
of the metals in contact to the superconductors: it modifies the
charge and energy diffusion constants in Eqs.~\eqref{eq:kinetic},
and allows for finite supercurrent to flow in the normal-metal
wires.

The simplest modification due to the proximity effect is a
correction to the conductance in NN'S systems, where N' is a
phase-coherent wire of length $L$, connected to a normal and a
superconducting reservoir via transparent contacts. In this case,
$\jS=\T=0$ and the kinetic equation for the charge current reduces
to the conservation of $\jT=\DT \partial_x f^T$. This can be
straightforwardly integrated, yielding the current
\begin{equation}
I=G_N/e \int d\energy f^T_{\rm eq}(\energy;V) D_T(\energy),
\label{eq:pecorrectedg}
\end{equation}
where $f^T_{\rm eq}(\energy;V)=\{\tanh[(\energy+eV)/(2 k_B
T)]-\tanh[(\energy-eV)/(2 k_B T)]\}/2$, $T$ is the temperature in
the normal-metal reservoir, and we assumed the normal metal in
potential $-eV$. The proximity effect can be seen in the term
$D_T=(\int_0^L \frac{dx}{\DT(x;\energy)}/L)^{-1}$. For $T=0$, the
differential conductance is $dI/dV=D_T(eV)G_N$. A detailed
investigation of $D_T(\energy)$ requires typically a numerical
solution of the retarded/advanced part of the Usadel equation
\cite{golubov:1123}. In general, it depends on two energy scales,
the Thouless energy $\energy_T=\hbar D/L^2$ of the N' wire and the
superconducting energy gap $\Delta$. The behavior of the
differential conductance as a function of the voltage and of the
linear conductance as a function of the temperature are
qualitatively similar, exhibiting the reentrance effect
\cite{golubov:1123,charlat:4950,hartog:13738}: for $eV, k_B T \ll
\energy_T$ and for $\max(eV,k_B T) \gg \energy_T$, they tend to
the normal-state value $G_N$ whereas for intermediate
voltages/temperatures, the conductance is larger than $G_N$,
showing a maximum for $\max(eV,k_B T)$ of the order of
$\energy_T$.

\begin{figure}[tb]
\centering
\includegraphics[width=0.7\columnwidth]{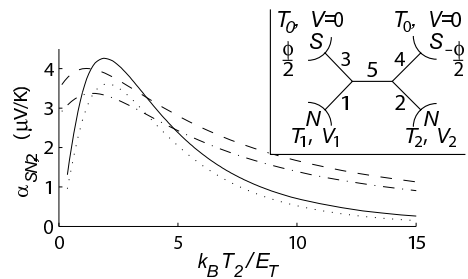}
\caption{Supercurrent-induced N-S thermopower $\alpha_{NS}$ for
the system depicted in the inset. Solid line: $\alpha_{NS}$ at
$T_1 \approx T_2$. Dotted line: approximation
\eqref{eq:thermovoltage1}. The correction
\eqref{eq:thermovoltage2} accounts for most of the difference.
Dashed line: $\alpha_{NS}$ at $k_B T_1=3.6 E_T$ with varying
$T_2$. Dash-dotted line: the corresponding approximation. Inset:
Andreev interferometer where thermoelectric effects are studied.
Two superconducting reservoirs with phase difference $\phi$ are
connected to two normal-metal reservoirs through diffusive
normal-metal wires. Adapted from \cite{virtanen:177004}.}
\label{fig:nsthermopower}
\end{figure}

The proximity-effect modification to the conductance can be tuned
in an Andreev interferometer, where a normal-metal wire is
connected to two normal-metal reservoirs and two superconductors
\cite{golubov:1123,nazarov:823,pothier:2488}. This system is
schematized in the inset of Fig.~\ref{fig:nsthermopower}. Due to
the proximity effect, the conductance of the normal-metal wire is
approximatively of the form $G(\phi)=G_N+\delta G
(1+\cos(\phi))/2$, where $\phi$ is the phase difference between
the two superconducting contacts and $\delta G$ is a positive
temperature and voltage-dependent correction to the conductance.
Its magnitude for typical geometries is at maximum some $0.1 G_N$.
The proximity-induced conductance correction is widely studied in
the literature and we refer to \cite{lambert:901,belzig:1251} for
a more detailed list of references on this topic.

The thermal conductance of Andreev interferometers has been
studied very recently. The formulation of the problem is very
similar as for the conductance correction. For $\energy \ll
\Delta$, there is no energy current into the superconductors.
Therefore, it is enough to solve for the energy current $\jL=\DL
\partial_x f^L$ in the wires 1, 2 and 5. This yields
\begin{equation*}
\jQ=G_N/e^2 \int d\energy \energy D_L(\energy) (f^L_{\rm
eq}(\energy;T_2)-f^L_{\rm eq}(\energy;T_1)),
\end{equation*}
where $f^L_{\rm eq}(\energy;T)=\tanh[\energy/(2 k_B T)]$ and the
thermal conductance correction is obtained from
$D_L(\energy)=(\int_0^L \frac{dx}{\DL(x;\energy)}/L)^{-1}$. Here
the spatial integral runs along the wire between the two
normal-metal reservoirs. In the general case, $D_L$ has to be
calculated numerically. The thermal conductance correction has
been analyzed by \textcite{bezuglyi:137002} and
\textcite{jiang:147002}. They found that it can be strongly
modulated with the phase $\phi$: in the short-junction limit where
$E_T \gg \Delta$, for $\phi=0$, $D_L$ almost vanishes, whereas for
$\phi=\pi$, $D_L$ approaches unity and thus the thermal
conductance approaches its normal-state value. For a long junction
with $E_T \lesssim \Delta$, the effect becomes smaller, but still
clearly observable. The first measurements \cite{jiang:0501478} of
the proximity-induced correction to the thermal conductance show
the predicted tendency of the phase-dependent decrease of $\kappa$
compared to the normal-state (Wiedemann-Franz) value.

Prior to the experiments on heat conductance in proximity
structures, the thermoelectric power $\alpha$ was experimentally
studied in Andreev interferometers
\cite{eom:437,dikin:012511,dikin:564,parsons:140502,parsons:316,jiangup04}.
The observed thermopower was surprisingly large, of the order of
100 neV/K
--- at least one to two orders of magnitude larger than the
thermopower in normal-metal samples. Also contrary to the Mott
relation (c.f., below Eqs.~\eqref{eq:ncurs}), this value depends
nonmonotonically on the temperature \cite{eom:437} at the
temperatures of the order of a few hundred mK, and even a sign
change could be found \cite{parsons:140502}. Moreover, the
thermopower was found to oscillate as a function of the phase
$\phi$. The symmetry of the oscillations has in most cases been
found to be antisymmetric with respect to $\phi=0$, i.e., $\alpha$
vanishes for $\phi=0$. However, in some measurements, the
Chandrasekhar's group \cite{eom:437,jiangup04} have found
symmetric oscillations, i.e., $\alpha$ had the same phase as the
conductance.

The observed behavior of the thermopower is not completely
understood, but the major features can be explained. The first
theoretical predictions were given by \textcite{claughton:6605},
who constructed a scattering theory to describe the effect of
Andreev reflection on the thermoelectric properties of proximity
systems. Based on their work, it was shown \cite{heikkila:1862}
that the presence of Andreev reflection can lead to a violation of
the Mott relation. This means that a finite thermopower can arise
even in the presence of electron-hole symmetry.
\textcite{seviour:6116}, \textcite{kogan:875}, and
\textcite{virtanen:177004,virtanen:401} showed that in Andreev
interferometers carrying a supercurrent, a large voltage can be
induced by the temperature gradient both between the normal-metal
reservoirs and between the normal metals and the superconducting
contacts. \textcite{virtanen:177004} showed that in long
junctions, the induced voltages between the two normal-metal
reservoirs and the superconductors can be related to the
temperature dependent equilibrium supercurrent $I_S(T)$ flowing
between the two superconductors via
\begin{equation}
V_{1/2}^0=\frac{1}{2}\frac{R_5 (2
R_{4/3}+R_5)R_{3/4}\left[I_S(T_1)-I_S(T_2)\right]}{R_{\rm NNN}
R_{\rm SNS}}. \label{eq:thermovoltage1}
\end{equation}
Here $R_{\rm NNN}=R_1+R_2+R_5$, $R_{\rm SNS}=R_3+R_4+R_5$ and
$R_k$ are the resistances of the five wires defined in the inset
of Fig.~\ref{fig:nsthermopower}. This is in most situations the
dominant term and it can be also phenomenologically argued based
on the temperature dependence of the supercurrent and the
conservation of total current (supercurrent plus quasiparticle
current) in the circuit \cite{virtanen:177004}. A similar result
can also be obtained in the quasiequilibrium limit for the
linear-response thermopower \cite{virtanen:401}. In addition to
this term, the main correction in the long-junction limit comes
from the anomalous coefficient $\T$,
\begin{equation}
eV_{1/2}^1=\frac{\mp R_{1/2}}{R_{\rm NNN}}\langle \T[1/2]\rangle
\mp \frac{R_{3/4} R_5}{R_{\rm NNN} R_{\rm SNS}} \langle \T[5]
\rangle. \label{eq:thermovoltage2}
\end{equation}
Here $\langle \T[k] \rangle \equiv \frac{1}{L_k} \int_0^{L_k} dx
\int_0^\infty d\energy \T[k](\energy) [f_{\rm
eq}^L(\energy,T_1)-f_{\rm eq}^L(\energy,T_2)] $ and $L_k$ is the
length of wire $k$. One finds that for a "cross" system without
the central wire (i.e., $R_5=0$), this term dominates $V_{1/2}^0$.
Further corrections to the result \eqref{eq:thermovoltage1} are
discussed by \textcite{kogan:875} and \textcite{virtanen:401}.

The above theoretical results explain the observed magnitude and
temperature dependence of the thermopower and also predict an
induced voltage oscillating with the phase $\phi$. However, the
thermopower calculated by \textcite{seviour:6116},
\textcite{kogan:875} and \textcite{virtanen:177004,virtanen:401}
is always an antisymmetric function of $\phi$, and it vanishes for
a vanishing supercurrent in the junction (including all the
correction terms). Therefore, the symmetric oscillations of
$\alpha$ cannot be explained with this theory.

The presence of the supercurrent breaks the time-reversal symmetry
and hence the Onsager relation $\Pi=T\alpha$ (see
Eqs.~\eqref{eq:ncurs} and below) need not to be valid for $\phi
\neq 0$. Heikkil\"a, {\it et al.} predicted a nonequilibrium
Peltier-type effect \cite{heikkila:100502} where the supercurrent
controls the local effective temperature in an out-of-equilibrium
normal-metal wire. However, it seems that the induced changes are
always smaller than those due to Joule heating and thus no real
cooling can be realized with this setup.

Recently, the thermoelectric effects in coherent SNS Josephson
point contacts have been analyzed in
\cite{zhao:077003,zhao:134503}. In such point contacts, the heat
transport is strongly influenced by the Andreev bound states
forming between the two superconductors.

\subsection{Heat transport by phonons}
\label{subs:phonons}

When the electrons are thermalized by the phonons, they may also
heat or cool the phonon system in the film. Therefore, it is
important to know how these phonons further thermalize with the
substrate, and ultimately with the heat bath on the sample holder
that is typically cooled via external means (typically by either a
dilution or a magnetic refrigerator). Albeit slow electron-phonon
relaxation often poses the dominating thermal resistance in
mesoscopic structures at low temperatures, the poor phonon thermal
conduction itself can also prevent full thermal equilibration
throughout the whole lattice. This is particularly the case when
insulating geometric constrictions and thin films separate the
electronic structure from the bulky phonon reservoir (see
Fig.~\ref{fig:setup}). In the present section we concentrate on
the thermal transport in the part of the chain of
Fig.~\ref{fig:setup} beyond the sub-systems determined by
electronic properties of the structure.

The bulky three-dimensional bodies cease to conduct heat at low temperatures according to the well appreciated $\kappa \propto T^3$ 
law in crystalline solids arising from Debye heat capacity via
\begin{equation} \label{kappa}
\kappa = Cv_S\ell_{\rm el,ph}/3,
\end{equation}
where $\kappa$ is the thermal conductivity and $C$ is the heat capacity per unit volume, $v_S$ is the speed of sound and $\ell_{\rm el,ph}$ is the 
mean free path of phonons in the solid \cite{ashcroftmermin}.
Thermal conductivity of glasses follows the universal $\propto T^2$
law, as was discovered by \textcite{zeller:2029}, which dependence
is approximately followed by non-crystalline materials in general
\cite{pobell:96}. These laws are to be contrasted to $\propto T$
thermal conductivities of pure normal metals (Wiedemann-Franz law,
c.f., below Eq.~\eqref{eq:ncurs}).
Despite the rapid weakening of thermal conductivity toward low temperatures, the dielectric materials in crystalline bulk are 
relatively good thermal conductors. One important observation here is that the absolute value of the bulk thermal conductivity in 
clean crystalline insulators at low temperatures does not provide the full basis of thermal analysis without a proper knowledge of the 
geometry of the structure, because the mean free paths often exceed the dimensions of the structures. For example, in pure silicon 
crystals, measured at sub-kelvin temperatures \cite{klitsner:6551}, thermal conductivity $\kappa \simeq 10$ Wm$^{-1}$K$^{-4}T^3$, heat 
capacity $C \simeq 0.6$ JK$^{-4}$m$^{-3}T^3$ and velocity $v_S\simeq 5700$ m/s imply by Eq. (\ref{kappa}) a mean free path of $\simeq 10$ 
mm, which is more than an order of magnitude longer than the thickness of a typical silicon wafer. Therefore phonons tend to propagate 
ballistically in silicon substrates.

What makes things even more interesting, but at the same time more complex, e.g., in terms of practical thermal design, is that at 
sub-kelvin temperatures the dominant thermal wavelength of the phonons, $\lambda_{ph} \simeq hv_S/k_{\rm B}T$,  is of the order of 0.1 
$\mu$m, and it can exceed 1 $\mu$m at the low temperature end of a
typical experiment (see discussion in Subs.~\ref{subs:eph}).
A direct consequence of this fact is that the 
phonon systems in mesoscopic samples cannot typically be treated as three-dimensional, but the sub-wavelength dimensions determine the 
actual dimensionality of the phonon gas. Metallic thin films and narrow thin film wires, but also thin dielectric films and wires are 
to be treated with constraints due to the confinement of phonons
in reduced dimensions.

The issue of how thermal conductivity and heat capacity of thin membranes and wires get modified due to geometrical constraints on the 
scale of the thermal wavelength of phonons has been addressed by several authors experimentally 
\cite{leivo:1305,holmes:2250,woodcraft:1968} and theoretically
(see, e.g., \cite{anghel:2958,kuhn:125425}).
The main conclusion is that structures with one or two dimensions $d < \lambda_{ph}$ restrict the propagation of (ballistic) phonons into 
the remaining "large" dimensions, and thereby reduce the magnitude of the corresponding quantities $\kappa$ and $C$ at typical 
sub-kelvin temperatures, but at the same time the temperature
dependences get weaker.

\begin{figure}[tb]
\begin{center}
\includegraphics[width=\columnwidth]{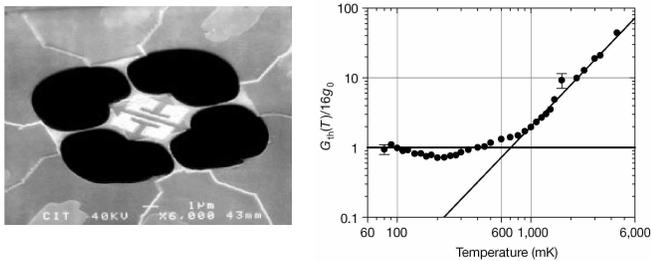}
\caption{The suspended silicon nitride bridge structure on the left, which was used by \textcite{schwab:974} to measure the quantum of 
thermal conductance shown in the graph on the right: thermal conductance levels off at the value $16g_0$ at temperatures well below 1 
K. Adapted from \cite{schwab:974}.}\label{fig:schwab}
\end{center}
\end{figure}

In the limit of narrow short wires at low temperatures the phonon thermal conductance gets quantized, as was experimentally 
demonstrated by \textcite{schwab:974}. This limit had been
theoretically addressed by
\textcite{angelescu:673}, \textcite{rego:232} and \textcite{blencowe:4992}, somewhat in analogy to the well-known Landauer result on electrical conduction through 
quasi-one-dimensional constrictions \cite{landauer:223}. The quantum of thermal conductance is $g_0\equiv\pi^2k_{\rm B}^2T/(3h)$, and 
there are four phonon modes at low temperature due to four mechanical degrees of freedom each adding $g_0$ to the thermal conductance 
of the quantum wire. In an experiment, see Fig. \ref{fig:schwab} four such wires in parallel thus carried heat with conductance 
$g=16g_0$. There are remarkable differences, however, in this result as compared to the electrical quantized conductance. In the 
thermal case, only one quantized level of conductance, $4g_0$ per wire, could be observed, and since the quantity transported is 
energy, the "quantum" of thermal conductance carries $\propto T$
in its expression besides the constants of nature.

The thermal boundary resistance (Kapitza resistance, after the
Russian physicist P. Kapitza) between two bulk materials is
$\propto T^{-3}$ due to acoustic mismatch \cite{lounasmaa:74}. A
remaining open issue 
is the question of thermal boundary resistance in a structure
where at least one of the phonon baths facing each other is
restricted, such that thermal phonons perpendicular to the
interface do not exist in this particular subsystem. Classically
the penetrating phonons need, however, to be perpendicular enough
in order to avoid total reflection at the surface
\cite{pobell:96}. In practice though, the phonon systems in the
two subsystems cannot be considered as independent. We are not
aware of direct experimental investigations on this problem.

On the device level reduced dimensions can be beneficial, e.g., in isolating thermally those parts of the devices to be refrigerated 
from those of the surrounding heat bath. This has been the method by which NIS based phonon coolers, i.e., refrigerators of the 
lattice have been realized experimentally utilizing thin silicon nitride films and narrow silicon nitride bridges 
\cite{manninen:1885,fisher:2705,luukanen:281,clark:173508}. These
devices are discussed in detail in Section V.C.

Detectors utilizing phonon engineering are discussed in
Sec.~\ref{sec:thermaldetectors}.

\subsection{Heat transport in a metallic reservoir}
\label{subs:reservoir}

Energy dissipated per unit time in a biased mesoscopic structure is given quite generally as $\dot{Q}=IV$, where $I$ is the current 
through and $V$ is the voltage across the device. This power is often so large that its influence on the 
thermal budget has to be carefully considered when designing a
circuit on a chip. For instance, the NIS cooler of Sec.~\ref{sec:SINIS} has a coefficient of performance given by Eq.~(\ref{eta}), with a typical value in the range 0.01$\dots$0.1. This simply 
means that the total power dissipated is 10 to 100 times higher than the net power one evacuates from the system of normal electrons. 
Yet this tiny fraction of the dissipated power is enough to cool the electron system far below the lattice temperature due to the 
weakness of electron--phonon coupling. This observation implies that 10 to 100 times higher dissipated power outside the normal island 
tends to overheat the connecting electrode significantly, again because of the weakness of the electron--phonon coupling. Therefore it 
is vitally important to make an effort to thermalize the connecting reservoirs to the surrounding thermal bath efficiently. In the 
case of normal-metal reservoirs, heat can be conveniently conducted along the electron gas to an electrode with a large volume in 
which electrons can then cool via electron--phonon relaxation. In the case of a superconducting reservoir, e.g., in a NIS refrigerator, 
the situation is more problematic because of the very weak thermal
conductivity at temperatures well below the transition temperature $T_c$. In this case the superconducting reservoirs should either be especially thick, or they should be attached to 
normal metal conductors ("quasiparticle traps") as near as
possible to the source of dissipation (see discussion in
Sec.~\ref{sec:SINIS}). The latter approach is, however, not always
welcome, because, especially in the case of a good metallic
contact between the two conductors, the operation of the device
itself can be harmfully affected by the superconducting proximity
effect.

\begin{figure}[tb]
\begin{center}
\includegraphics[width=\columnwidth]{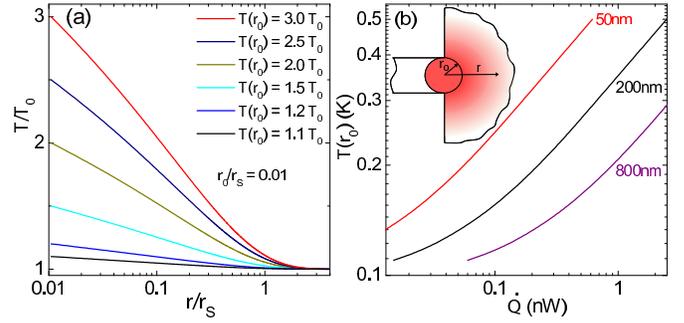} \caption{(Color in online edition): Temperature rise according
to the presented reservoir heating model. For details see
text.}\label{tradial}
\end{center}
\end{figure}

Let us consider heat transport in a normal metal reservoir. In the first example we approximate the reservoir geometry by a 
semi-circle, connected to a biased sample with a hot spot of radius $r_0$ at its origin (see the inset of Fig.~\ref{tradial} (b)). This hot spot can 
approximate, for example, a tunnel junction of area $\pi r_0^2$.
The results depend only logarithmically on $r_0$, and therefore its exact value is irrelevant when making estimates. We first assume that the electrons carry the heat away with negligible 
coupling to the lattice up to a distance $r$. According to
Eq.~\eqref{eq:quasieqtemp}, we can then write the radial flux of
heat in the quasiequilibrium limit in the form
\begin{equation} \label{radial}
\dot{Q}(r)=-\kappa S dT/dr,
\end{equation}
where $\kappa$ is the electronic thermal conductivity, $S=\pi rt$ is the conduction area at distance $r$ in a film of thickness $t$, 
and $T$ is the temperature at radius $r$. According to the
Wiedemann-Franz law and the temperature independent residual
electrical resistivity in metals, one has $\kappa=\Lor \sigma T
\equiv \kappa_0 T$ (see below Eq.~\eqref{eq:ncurs}).
With these assumptions, using Eq.~(\ref{radial}), one finds a radial 
distribution of temperature
\begin{equation} \label{radial2}
T(r_1)=\sqrt{T(r_2)^2 + \frac{\dot {Q}R_{\Box}}{\pi \Lor} \ln
(r_2/r_1)},
\end{equation}
where $r_1$ and $r_2$ are two distances from the hot spot, and we defined the square resistance as $R_{\Box} = \rho / t$. Thus 
making the reservoir thicker helps to thermalize it. The model above is strictly appropriate in the case where a thin film in form of 
a semi-circle is connected to a perfect thermal sink at its perimeter (at $r=r_2$).
A more adequate model in a typical experimental case is 
obtained by assuming a uniform semi-infinite film connected at its side to a hot spot as above,
but now assuming that the film 
thermalises via electron--phonon coupling. Using Eq.~(\ref{radial})
and energy conservation one then obtains (see
Sec.~\ref{subs:limits})
\begin{equation}
\frac{d\dot{Q}}{dr} + \Sigma \pi rt(T^p - T_0^p) = 0,
\end{equation}
where $T_0$ is the lattice temperature, and $p$ is the exponent of electron phonon relaxation, typically $p=5$. We can write this 
equation into a dimensionless form
\begin{equation} \label{radial3}
\frac{d^2u}{d\rho^2} + \frac{1}{\rho}\frac{du}{d\rho} =u^{p/2}-1.
\end{equation}
Here we have defined $u\equiv (T(r)/T_0)^2$ and $\rho \equiv r/r_{\rm S}$,
where $r_{\rm S} \equiv \sqrt{\kappa 
_0/(2\Sigma)}/T_0^{p/2-1}$ is the length scale of temperature over
which it relaxes towards $T_0$.
Figure \ref{tradial} (a) shows the solutions of Eq.~(\ref{radial3}) for different values (1.1, 1.2, 1.5, 2.0, 2.5, and 3.0) of 
relative temperature rise $T(r_0)/T_0$. We see, indeed, that
$r_{\rm S}$ determines the relaxation length.
To have a concrete example let us consider a copper film with thickness $t=200$ nm. For copper, using $p=5$, we have $\kappa _0 \simeq 1$ 
WK$^{-2}$m$^{-1}$ and $\Sigma \simeq 2\cdot 10^9$ WK$^{-5}$m$^{-3}$,
which leads to a healing length of $r_{\rm S} \sim$ 500 $\mu$m at 
the bath temperature of $T_0 =$ 100 mK. The temperature rise 
versus input power $\dot{Q}$ has been plotted in Fig.
\ref{tradial} (b) for copper films with different thicknesses. In
particular for $t=200$ nm, we obtain a linear response of
$(T(r_0,\dot{Q})/T_0-1)/\dot{Q} \simeq 7\cdot 10^{-4}$/pW.

A superconducting reservoir, which is a necessity in some devices, poses a much more serious overheating problem. Heat is transported 
only by the unpaired electrons whose number is decreasing exponentially as $\propto \exp(-\Delta/k_{\rm B}T)$ towards low 
temperatures. Therefore the electronic thermal transport is reduced by approximately the same factor, as compared to the corresponding 
normal metal reservoir. Theoretically then $\kappa$ is suppressed by nine orders of magnitude from the normal state value in aluminium 
at $T=100$ mK. It is obvious that in this case other thermal
conduction channels, like electron--phonon relaxation, become
relevant, but this is a serious problem in any case. At higher
temperatures, say at $T/T_{\rm C} \ge 0.3$, a significantly
thicker superconducting reservoir can help \cite{clark:625}.

\section{Thermometry on mesoscopic scale}
\label{sec:thermometry}
Any quantity that changes with temperature can in principle be used as a thermometer. Yet the usefulness of a particular thermometric 
quantity in each application is determined by how well it satisfies a number of other criteria. These include, with a weighting factor 
that depends on the particular application: wide operation range
with simple and monotonic dependence on temperature, low 
self-heating, fast response and measurement time,
ease of operation, immunity to external parameters,
in particular to magnetic field, 
small size and small thermal mass. One further important issue in thermometry in general 
terms is the classification of thermometers into {\sl primary} thermometers, i.e., those that provide the absolute temperature without 
calibration, and into {\sl secondary} thermometers, which need a calibration at least at one known temperature. Primary thermometers 
are rare, they are typically difficult to operate, but nevertheless they are very valuable, e.g., in calibrating the secondary 
thermometers. The latter ones are often easier to operate and
thereby more common in research laboratories and in industry.

In this review we discuss a few mesoscopic thermometers that can be used at cryogenic temperatures.
Excellent and thorough reviews of 
general purpose cryogenic thermometers, other than mesoscopic ones,
can be found
in many text books and review articles, see, e.g., 
\cite{lounasmaa:74,pobell:96,quinn:83} and many references therein.

Modern micro- and nanolithography allows for new thermometer
concepts
and realizations where sensors can be very small, thermal 
relaxation times are typically short, but which generally do not
allow except very tiny amounts of self-heating. The heat
flux between electrons and phonons gets extremely weak at low 
temperatures whereby electrons decouple thermally from the lattice
typically at sub-100 mK temperatures, unless special care is taken 
to avoid this. Therefore, especially at these low temperatures the
lattice temperature and the electron temperature measured by such
thermometers often deviate from each other. An important example of
this is the electron temperature in NIS electron coolers to be
discussed in Section V.

A typical electron thermometer relies on a fairly easily and
accurately measurable quantity $M$ that is related to the electron
energy distribution function $f(\varepsilon)$ via
\begin{equation}
M=\int_{-\infty}^\infty d\varepsilon k(\varepsilon)
g[f(\varepsilon)]. \label{eq:genericthermometer}
\end{equation}
Here the kernel $k(\varepsilon)$ describes the process which is
used to measure $f(\varepsilon)$ and $g[f]$ is a functional of
$f(\varepsilon)$. The quantity $M$ typically refers to an average
current or voltage, in which case $g[f]=f-f_0$ is a linear
function of $f(\varepsilon)$ with some constant function $f_0$; or
it can refer to the noise power, in which case $g[f]$ is quadratic
in $f$. For the thermometer to be easy to calibrate,
$k(\varepsilon)$ should be a simple function dependent only on a
few parameters that need to be calibrated. Moreover, if
$k(\varepsilon)$ has sufficiently sharp features, it can be used
to measure the shape of $f(\varepsilon)$ also in the
nonequilibrium limit.

\subsection{Hybrid junctions}
\label{subs:tjthermometry}

Tunnelling characteristics through a barrier separating two conductors with non-equal densities of states (DOSs) are usually 
temperature dependent. The barrier B may be a solid insulating layer (I), a Schottky barrier formed between a semiconductor and a 
metal (Sc), a vacuum gap (I), or a normal metal weak link (N). We are going to discuss thermometers based on tunnelling in a CBC' 
structure. C and C' stand for a normal metal (N), a superconductor (S), or a semiconductor (Sm). As it turns out, the current-voltage 
($I-V$) characteristics of the simplest combination, i.e., of a NIN tunnel junction,
exhibit no temperature dependence in the limit of 
a very high tunnel barrier. Yet NIN junctions form elements of presently actively investigated thermometers (Coulomb blockade 
thermometer and shot noise thermometer) to be discussed separately.
The NIN junction based thermometers are suitable for general 
purpose thermometry, because their characteristics are typically not sensitive to external magnetic fields. Superconductor based 
junctions are, on the contrary, normally extremely sensitive to magnetic fields, and therefore they are suitable only in experiments 
where external fields can be avoided or at least accurately controlled.

In SBS' junctions one has to distinguish between two tunnelling mechanisms, Cooper pair tunnelling (Josephson effect) and 
quasiparticle tunnelling. The former occurs at low bias voltage and temperature, whereas the latter is enhanced by increased 
temperature and bias voltage. In the beginning of this section we
discuss quasiparticle tunnelling only.

Let us consider tunnelling between two normal-metal conductors
through an insulating barrier. I-V characteristics of such a
junction were given by Eq.~\eqref{eq:tunnelcurrent}. We assume
quasi-equilibrium with temperatures $T_i$, $i=L,R$ on the two
sides of the barrier. Since $N_i(E) = 1$ to high precision
at all relevant energies ($k_{\rm B}T_i,eV \ll 
E_{\rm F}=$ Fermi energy), Eq.~\eqref{eq:tunnelcurrent} can be integrated easily to yield $I=V/R_{\rm T}$. Therefore the $I-V$ characteristics are 
ohmic, and they do not depend on temperature, and a NIN junction
appears to be unsuitable for thermometry.
There is, however, a weak correction to this result, due to the finite 
height of the tunnel barrier \cite{simmons:1793}, which will be
discussed in subsection III.D.

\subsubsection{NIS thermometer}
\label{subs:nisthermometer}

As a first example of an on-chip thermometer, let us discuss a tunnel junction between a normal metal and a superconductor (NIS 
junction) \cite{rowell:2456}.
I-V characteristics of a NIS junction have the very important property that they depend on the temperature of the N 
electrode only, which is easily verified, e.g., by writing $I(V)$ of
Eq.~\eqref{eq:tunnelcurrent} with $N_L(E)={\rm const.}$ and
$N_R(E)=N_S(E)$ in a symmetric form
\begin{equation} \label{Isymm}
I(V)= \frac{1}{2eR_{\rm T}}\int _{-\infty}^{\infty} N_S(E)[f_{\rm
N}(E-eV)-f_{\rm N}(E+eV)]dE.
\end{equation}
This insensitivity to the temperature of the superconductor holds
naturally only up to temperatures where the superconducting energy
gap can be assumed to have its zero-temperature value. This is
true in practice up to $T/T_{c}\lesssim 0.4$.

\begin{figure}[tb]
\begin{center}
\includegraphics[width=0.4
\textwidth]{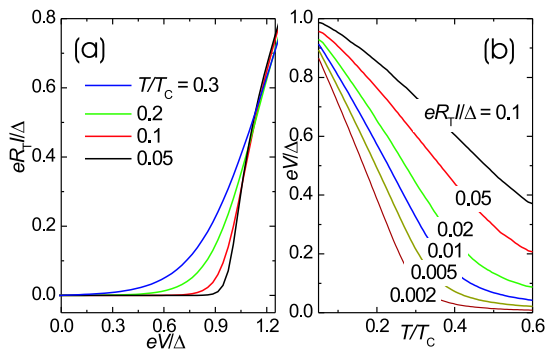}
\caption{NIS thermometer characteristics. Calculated I-V characteristics of a normal metal - superconductor tunnel junction at various relative temperatures 
$T/T_{c}$ in (a). The corresponding voltage over the junction as a
function of temperature
when the junction is biased at a 
constant current is shown in (b). Characteristics at a few values of
current are shown.}\label{fig:nis}
\end{center}
\end{figure}

Employing Eq.~(\ref{Isymm}) one finds that a measurement of voltage $V$ at a constant current $I=I_0$ yields a direct measure of 
$f_{\rm N}(E)$, and in quasi-equilibrium, where the distribution follows thermal Fermi-Dirac distribution, it also yields temperature in 
principle without fit parameters. Figure \ref{fig:nis} (a) shows the calculated I-V characteristics of
a NIS junction at 
a few temperatures $T/T_{c}$. Figure \ref{fig:nis} (b) gives the corresponding thermometer calibration: junction voltage has 
been plotted against temperature at a few values of (constant) measuring current.

A NIS thermometer has a number of features which make it attractive in applications. The sensing element can be made very small and 
thereby NIS junctions can probe temperature locally and detect temperature gradients. Junctions made by electron beam lithography can 
be much smaller than 1 $\mu$m in linear dimension \cite{nahum:3075}. Using a scanning tunnelling microscope with a superconducting tip 
as a NIS junction one can most likely probe the temperature of the surface locally on nanometer
scales in an instrument like those of 
\textcite{moussy:128,vinet:165420}. Self-heating can be made very small by operating in the sub-gap voltage range $e|V| < \Delta$ (see Fig. 
\ref{fig:nis}), where current is very small. The superconducting probe is thermally decoupled from the normal region whose 
temperature is monitored by it. The drawbacks of this technique include high sensitivity to external magnetic field, high impedance of 
the sensor especially at low temperatures, and sample-to-sample deviations from the ideal theoretical behaviour. Due to this last 
reason, a NIS junction can hardly be considered as a primary thermometer: deviations arise especially at low temperatures, one 
prominent problem is saturation due to subgap leakage due to
non-zero DOS within the gap and Andreev reflection.

A fast version of a NIS thermometer was implemented by \textcite{schmidt:1002}. They achieved sub-$\mu$s readout times 
(bandwidth about 100 MHz) by imbedding the NIS junction in an $LC$ resonant circuit. This rf-NIS read-out is possibly very helpful in 
studying thermal relaxation rates in metals, and in fast far-infrared bolometry.

NIS junction thermometry has been applied in x-ray detectors \cite{nahum:733}, far-infrared bolometers \cite{mees:2329,chouvaev:985}, 
in probing the energy distribution of electrons in a metal \cite{pothier:178}, and as a thermometer in electronic coolers at 
sub-kelvin temperatures \cite{nahum:3123,leivo:1996}. It has also been suggested to be used as a far-infrared photon counter 
\cite{anghel:295}. In many of the application fields of a NIS thermometer it is not the only choice:
for example, a superconducting 
transition edge sensor (TES) can be used very conveniently in the
bolometry and calorimetry applications (see Sec.~IV).

Recently Schottky contacts between silicon and superconducting metal have been shown to exhibit similar characteristics as fully 
metallic NIS junctions \cite{savin:1471,buonomo:7784}. These
structures are discussed in subsection V.C.3.

\subsubsection{SIS thermometer}

\begin{figure}[tb]
\begin{center}
\includegraphics[width=0.5
\textwidth]{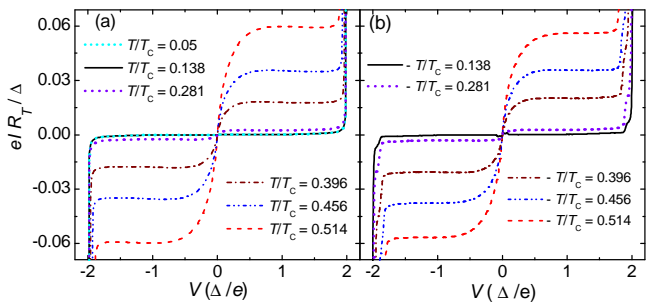} \caption{SIS junction as a thermometer.
(a) Calculated and (b) measured $I-V$ curves of a SIS tunnel
junction at a few temperatures. Supercurrent has been suppressed.
Experimental data from \cite{savin:2005}.}\label{fig:sis}
\end{center}
\end{figure}

A tunnel junction between two superconductors supports supercurrent, whose critical value $I_{\rm C}$ has a magnitude which depends on 
temperature according to \cite{ambegaokar:486}: $I_{\rm C} = \frac{\pi \Delta} 
{2eR_{\rm T}}\tanh(\frac{\Delta} {2k_{\rm B}T})$. This can naturally be used to indicate temperature because of the temperature dependence of 
the energy gap and the explicit hyperbolic dependence on $T$. Yet,
these dependencies are exponentially weak at low temperatures. 
Another possibility is to suppress $I_{\rm C}$ by magnetic field, e.g., in a SQUID configuration,
and to work at non-zero bias voltage and measure the quasiparticle 
current, which can be estimated by Eq.~\eqref{eq:tunnelcurrent} again with both DOSs given by Eq.~\eqref{eq:sdos} now. The resulting current depends 
approximately exponentially on temperature, but the favourable fact is that the absolute magnitude of 
the current is increased because it is proportional to the product of the two almost infinite DOSs matching at low bias voltages, and, 
in practical terms, the method as a probe of a quasiparticle distribution is more robust against magnetic noise. Figure \ref{fig:sis} 
shows the calculated and measured dependences of $I(V)$ at a few
values of temperature $T/T_{c}$. Although the curves corresponding
to the lowest temperatures almost overlap here, it is
straightforward to verify that a standard measurement of current can
resolve temperatures down to below $0.1 T_{c}$ using an ordinary
tunnel junction.

SIS junctions have not found much use as thermometers in the traditional sense, but they are extensively investigated and used as 
photon and particle detectors because of their high energy
resolution (see, e.g., \cite{booth:493} and Section IV). In this
context SIS detectors are called STJ detectors, i.e.,
superconducting tunnel junction detectors. Another application of
SIS structures is their use as mixers \cite{tinkham:96}. While
superficially similar to STJ detectors, their theory and operation
differ significantly from each other. SIS junctions are also
suitable for studies of quasiparticle dynamics and fluctuations in
general (see, e.g., \cite{wilson:067004}).

\subsubsection{Proximity effect thermometry}

For applications requiring low-impedance ($\lesssim 1$ $\Omega$)
thermometers at sub-micron scale, the use of clean NS contacts may
be more preferable than thermometers applying tunnel contacts. In
an SNS system, one may again employ either the supercurrent or
quasiparticle current as the thermometer. For a given phase $\phi$
between the two superconductors, the former can be expressed
through \cite{belzig:1251}
\begin{equation}
I_S(\phi)=\frac{1}{e R_N} \int_{0}^\infty d\energy
j_S(\energy;\phi) (1-2f(\energy)), \label{eq:snssupercur}
\end{equation}
where $R_N$ is the normal-state resistance of the weak link,
$j_S(\energy;\phi)$ is the spectral supercurrent
\cite{heikkila:184513}, and the energies are measured from the
chemical potential of the superconductors. Hence, the supercurrent
has the form of Eq.~\eqref{eq:genericthermometer}. In practice, one
does not necessarily measure $I_S(\phi)$, but the critical current
$I_C=\max_\phi I_S(\phi)$. In diffusive junctions, this is obtained
typically for $\phi$ near $\pi/2$, although the maximum point
depends slightly on temperature.

The problem in SNS thermometry is in the fact that the
supercurrent spectrum $j_S(\varepsilon)$ depends on the quality of
the interface, on the specific geometry of the system, and most
importantly, on the distance $L$ between the two superconductors
compared to the superconducting coherence length
$\xi_0=\sqrt{\hbar D/(2\Delta)}$ \cite{heikkila:184513}.
Therefore, $I_C(T)$ dependence is not universal. However, the size
of the junction can be tuned to meet the specific temperature
range of interest. In the limit of short junctions, $L \lesssim
\xi_0$ \cite{kulik:142}, the temperature scale for the critical
current is given by the superconducting energy gap $\Delta$ and
for $k_B T \ll \Delta$, the supercurrent depends very weakly on
the temperature.  In a typical case $\xi_0 \sim 100 \dots 200$ nm,
and thus already a weak link with $L$ of the order of 1 $\mu$m,
easily realisable by standard lithography techniques, lies in the
"long" limit. There, the critical current is $I_C = c (k_B
T)^{3/2} \exp(-\sqrt{2\pi k_B T/E_T})/(e R_N\sqrt{E_T})$
\cite{zaikin:184}. This equation is valid for $k_B T \gtrsim 5
E_T$. Here $E_T=\hbar D/L^2$ and the prefactor $c$ depends on the
geometry \cite{heikkila:184513}, for example for a two-probe
configuration $c=64\sqrt{2\pi^3}/(3+2\sqrt{2})$. The exponential
temperature dependence and the crossover between the long- and
short-junction limits were experimentally investigated by
\textcite{dubos:064502} and the above theoretical predictions were
confirmed.

Using a four-probe configuration with SNS junctions of different
lengths, \textcite{jiang:2190} exploited the strong temperature
dependence of the supercurrent to measure the local temperature.
The device worked in the regime where part of the junctions were
in the supercurrent-carrying state and part of them in the
dissipative state. In the limit where supercurrent is completely
suppressed, there is still a weaker temperature dependence of the
conductance, due to the proximity-effect correction (c.f.,
Eq.~\eqref{eq:pecorrectedg}) \cite{charlat:4950}. This can also be
used for thermometry \cite{aumentado:3554}, but due to the much
smaller effect of temperature on the conductance, it is less
sensitive.

Besides being just a thermometer of the electron gas, critical
current of a SNS Josephson junction has been shown to probe the
electron energy distribution
\cite{baselmans:43,baselmans:094504,huang:020507,giazotto:137001}
as indicated by Eq.~\eqref{eq:snssupercur}.

\subsection{Coulomb blockade thermometer, CBT}

\begin{figure}[tb]
\begin{center}
\includegraphics[width=0.4
\textwidth]{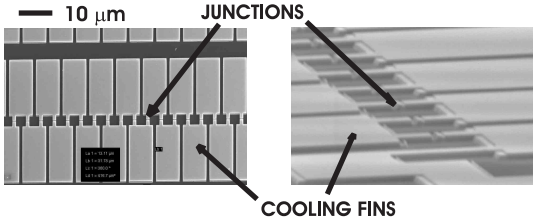}
\caption{A typical CBT sensor for the temperature range 20 mK - 1 K. The structure has been fabricated by electron beam lithography, 
combined with aluminium and copper vacuum evaporation. Both top view and a view at an oblique angle are shown; the scale indicated 
refers to the top view. Figure adapted from
\cite{meschke:1119}.}\label{fig:cbtsem}
\end{center}
\end{figure}

Single electron tunnelling (SET) effects were foreseen in micro-lithographic structures
in the middle of 1980's \cite{averin:345}.
Such effects in granular structures had been known to exist already much earlier \cite{neugebauer:74,giaever:1504,lambe:1371}. The 
first lithographic SET device was demonstrated by \textcite{fulton:109}. Since that time SET effects have formed 
a strong subfield in mesoscopic physics. Typically SET devices operate in the full Coulomb blockade regime, where temperature is so 
low that its influence on electrical transport characteristics can be neglected; at low bias voltages current is blocked due to the 
charging energy of single electrons. Coulomb blockade thermometer (CBT) operates in a different regime, where single electron effects 
still play a role but temperature predominantly influences the electrical transport characteristics 
\cite{pekola:2903,farhangfar:191,delsing:1264,meschke:1119}.
CBT is a primary thermometer, whose operation is based on competition between thermal energy $k_{\rm B}T$ at temperature $T$, 
electrostatic energy $eV$ at bias voltage $V$, and charging energy due to extra or missing individual electrons with unit of charging 
energy $E_{\rm C} = e^2/(2C^*)$, where $C^*$ is the effective capacitance of the system. Figure \ref{fig:cbtsem} shows an SEM 
micrograph of a part of a typical CBT sensor suitable for the temperature range 0.02$\dots$1 K. This sensor consists of four parallel 
one-dimensional arrays with 40 junctions in each.

\begin{figure}[tb]
\begin{center}
\includegraphics[width=0.5
\textwidth]{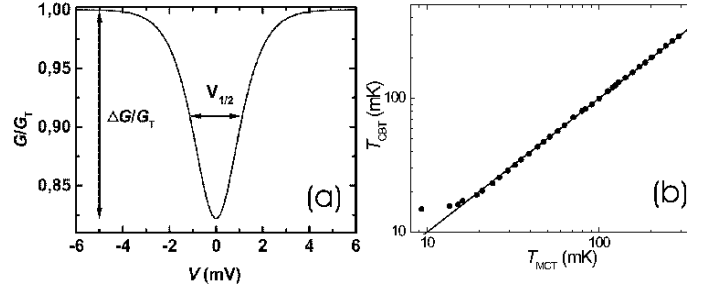} \caption{Performance of a CBT
thermometer. (a) Typical measurement: $G(V)/G_{\rm T}$ is the
differential conductance scaled by its
asymptotic value at large positive and negative voltages, plotted as a function of bias voltage $V$. $V_{1/2}$ indicates the full 
width at half
minimum of the characteristics, which is the main thermometric
parameter. The full depth of the line, $\Delta G/G_{\rm T}$, is another parameter to determine temperature. In (b) temperature deduced 
by CBT has been compared to that obtained by a $^3$He melting curve thermometer. Saturation of CBT below 20 mK indicates typical 
thermal decoupling between electrons and phonons. Figure from Ref.
\cite{meschke:1119}.}\label{fig:cbtg}
\end{center}
\end{figure}

In the partial Coulomb blockade regime the $I-V$ characteristics of
a CBT array with $N$ junctions in series do not display a sharp
Coulomb blockade gap, but, instead, they are smeared over a bias
range $eV \sim Nk_{\rm B}T$. The asymptotes of the $I-V$ at large
positive and negative voltages have, however, the same offsets as at
low $T$, determined by the Coulomb gap. In the partial Coulomb
blockade regime it is convenient to measure not the $I-V$ directly,
but the differential conductance, i.e., the slope of the $I-V$
curve, $G\equiv dI/dV$, vs. $V$. The result is a nearly bell shaped
dip in conductance around zero bias. Figure \ref{fig:cbtg}
illustrates a measured conductance curve, scaled by asymptotic
conductance $G_{\rm T}$ at large positive and negative voltages. The
important property of this characteristic is that the full width of
this dip at half minimum, $V_{1/2}$, approximately proportional to
$T$, determines the temperature without any fit, or any material or
geometry dependent
parameters, i.e., it is a primary measure of temperature. In the lowest order in $\frac{E_{\rm C}}{k_{\rm B}T}$, one finds for the 
symmetric linear array of $N$ junctions of capacitance $C$ in series \cite{pekola:2903,farhangfar:191}
\begin{equation} \label{cb3}
G(V)/G_{\rm T} = 1 - \frac{E _{\rm C}}{k_{\rm B}T}\hspace{1mm}
g(\frac{eV}{Nk_{\rm B }T}).
\end{equation}
For such an array $C^*=NC/[2(N-1)]$ and $g(x) = [x \sinh (x) - 4
\sinh ^2 (x/2)]/[8 \sinh ^4 (x/2)]$. The full width at half minimum
of the conductance dip has the value
\begin{equation} \label{cb5}
V_{1/2} \simeq 5.439Nk_{\rm B}T/e,
\end{equation}
which allows one to determine $T$ without calibration.

It is
often convenient to make use of the secondary mode of CBT, in
which one measures the depth of the dip, $\Delta G/G_{\rm T}$,
which is proportional to the inverse temperature, again in the lowest order in $\frac{E_{\rm C}}{k_{\rm B}T}$:
\begin{equation} \label{cb7}
\Delta G / G_{\rm T} = E _{\rm C}/(6k_{\rm B}T).
\end{equation}

Measuring the inflection point of the conductance curve provides an alternative means to determine $T$ without calibration. In 
\cite{delsing:1264} this technique was used for fast thermometry on two dimensional junction arrays. A simple and fast alternative 
measurement of CBT can be achieved by detecting the third harmonic current in a pure AC voltage biased configuration 
\cite{meschke:05}.

There are corrections to results (\ref{cb3}),
(\ref{cb5}) and (\ref{cb7}) due to both next order terms in
$E_{\rm C}/k_{\rm B}T$ and due to non-uniformities in the array \cite{farhangfar:191}. Higher order corrections do sustain the primary 
nature of CBT, which implies wider temperature range of operation.
The primary nature of the conductance curve exhibited by Eq.
(\ref{cb3}) through the bias dependence $g(\frac{eV}{Nk_{\rm B }T})$
is preserved also irrespective of the capacitances of the array,
even if they would not be equal. Only a non-uniform distribution of
tunnel resistances leads to a deviation from the basic result of
bias dependence of Eq. (\ref{cb3}). Fortunately the influence of the
inhomogeneity on temperature reading is quite weak, and it leads
typically to less than 1 \% systematic error in $T$.

The useful temperature range of a CBT array is limited at high
temperature by the vanishing signal ($\Delta G/G_{\rm T} \propto
T^{-1}$). In practice the dip must be deeper than $\sim 0.1$ \% to
be resolvable from the background. The low temperature end of the
useful temperature range is set by the appearance of charge
sensitivity of the device, i.e., the background charges start to
influence the characteristics of the thermometer. This happens
when $\Delta G/G_{\rm T} \simeq 0.5$ \cite{farhangfar:191}. With these
conditions it is obvious that the dynamic range of one CBT sensor
spans about two decades in temperature.

The absolute temperature range of CBT techniques is presently
mainly limited by materials issues. At high temperatures, the
measuring bias range gets wider, $V_{1/2} \propto T$, and the
conductance becomes bias dependent also due to the finite (energy)
height of the tunnel barrier \cite{simmons:1793,gloos:2915}.
Because of this, the present high temperature limit of CBTs is
several tens of K. The (absolute) low temperature end of the CBT
technique is determined by self-heating due to biasing the device.
One can measure temperatures reliably down to about 20 mK at
present with better than 1 \% absolute accuracy in the range
0.05$\dots$4 K, and about 3 \% down to 20 mK \cite{meschke:1119}.

Immunity to magnetic field is a desired but rare property among thermometers. For instance,
resistance thermometers have usually strong magnetoresistance. 
Therefore one could expect CBT to be also "magnetoresistive". On the contrary, CBT has proven to be immune to even the strongest 
magnetic fields ($> 20$ T) \cite{pekola:5582,pekola:263,linden:273}. This happens because CBT operation is based on electrostatic properties: 
tunnelling rates are determined by charging energies. Hence, CBT should be perfectly immune to magnetic field %
as long as energies of electrons 
with up and down spins do not split appreciably, which is always
the case in experiments.

Coulomb blockade is also a suitable probe of non-equilibrium
energy distributions as demonstrated by \textcite{anthore:076806}.

\subsection{Shot noise thermometer, SNT}

Noise current or voltage of a resistor (resistance $R$) has been known to yield absolute temperature since 1920's \cite{johnson:97, 
nyquist:110,blanter:1}. Till very recently only equilibrium noise, i.e., noise across an unbiased resistor, was employed to measure 
temperature. In this case Johnson noise voltage $v_{\rm n}$ and current $i_{\rm n}$ squared have expectation values 
$\langle v_{\rm n}^2 \rangle = R^2\langle i_{\rm n}^2 \rangle = 4k_{\rm B}T R \Delta f$, where $\Delta f$ is the frequency band of the 
measurement. In other words, the current spectral density is given by $S_{\rm I}=4k_{\rm B}T/R$, and voltage spectral density by 
$S_{\rm V}=4k_{\rm B}TR$. There are several critical issues in measuring temperature through Johnson noise. First, the bandwidth has 
to be known exactly. Secondly, the small signal has to be amplified, and the gain must be known to high accuracy. Thirdly, the noise 
signal becomes extremely small at cryogenic temperatures, which in turn means that the measurement has to be able to detect a very 
small current or voltage, and at the same time no other noise sources should cause comparable voltages or currents.

Success in noise thermometry depends critically on the performance of the amplifiers used to detect the tiny noise current (or 
voltage). Constant progress in improving SQUIDs (Superconducting Quantum Interference Device) and in optimizing their 
operation has pushed the lowest temperature measurable by a Johnson noise thermometer down to below 1 mK \cite{lusher:1}.

It is not only the equilibrium noise that can be useful in thermometry. In the opposite limit, at $eV \gg k_{\rm B}T$, the dominating noise, 
e.g., in tunnel junctions, is shot noise
\cite{schottky:541,blanter:1}, whose current spectral density is
given asymptotically by $S_I = F2e|I|$. Fano factor $F$ equals $1$
for a tunnel junction. The way Johnson noise transforms into shot
noise upon increasing the bias obeys the relation
(see Eq. (\ref{eq:nnnoise}) with $\tprob_n \rightarrow 0$, and, e.g., 
 \cite{ziel:86})
\begin{equation} \label{gen}
S_I(T)=2eI\coth(\frac{eV}{2k_{\rm B}T}).
\end{equation}
At $|V| \gg k_{\rm B}T/e$, Eq. (\ref{gen}) obtains the Poisson expression above, whereas at $|V| \ll k_{\rm B}T/e$ it assumes the thermal 
Johnson form $S_I \simeq 4k_{\rm B}T/R_T$.
\begin{figure}[tb]
\begin{center}
\includegraphics[width=0.4
\textwidth]{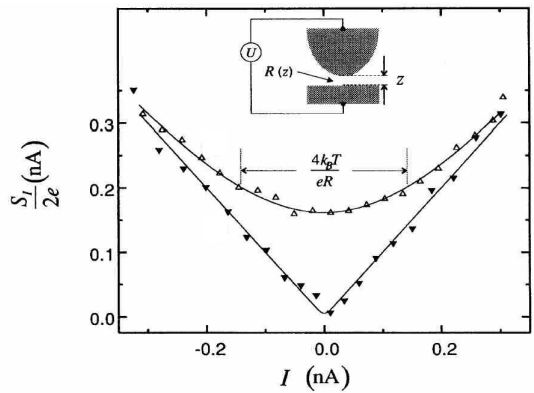} \caption{Cross-over from Johnson noise
to shot noise when increasing the bias voltage of a tunnel junction.
The data shown by open symbols
has been 
measured at 300 K with $R_{\rm T}=0.32$ G$\Omega$, and data shown by
solid symbols at 77 K with $R_{\rm T}=2.7$ G$\Omega$. Adapted from
\cite{birk:1610}.}\label{fig:snb}
\end{center}
\end{figure}

\begin{figure}[tb]
\begin{center}
\includegraphics[width=0.3
\textwidth]{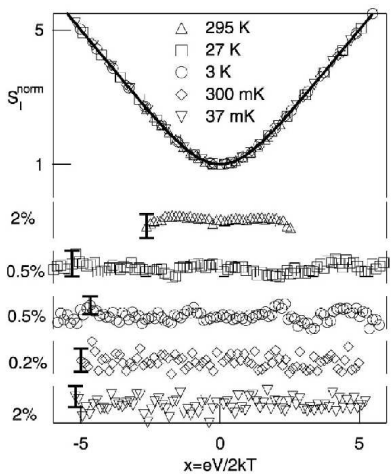}
\caption{Normalised junction noise versus normalised voltage at various temperatures. The residuals from the expected $x\coth(x)$ 
law are shown in the bottom half of the figure. From
\cite{spietz:1929}.}\label{fig:spietz}
\end{center}
\end{figure}

Temperature dependent cross-over characteristics from Johnson to shot noise according to Eq. (\ref{gen}) have been demonstrated 
experimentally, e.g., in scanning tunnelling microscope experiments \cite{birk:1610} as shown in Fig. \ref{fig:snb}.
Recently, \textcite{spietz:1929} employed this cross-over in a lithographic tunnel junction between metallic films for 
thermometry (shot noise thermometer, SNT) in the temperature range from few tens of mK up to room temperature. At both ends of this 
range there are systematic errors in the reading, which can possibly be corrected by a more careful design of the sensor. In a rather 
wide range around 1 K (about 0.1$\dots$10 K), the absolute accuracy is better than 1 \%. The crossover voltage depends only on $k_{\rm 
B}T/e$, which means that the thermometer is indeed primary; ideally the temperature reading does not depend on the materials or 
geometry of the sensor. Figure \ref{fig:spietz} shows the experimental data of \textcite{spietz:1929} at several temperatures: normalised (by 
Johnson noise) current noise has been plotted against normalised voltage $x \equiv eV/(2k_{\rm B}T)$, whereby all curves at different 
temperatures should lie at $S_{\rm I}^{\rm norm}=x\coth(x)$. As seen in the bottom half of the figure, the residuals are smallest 
around 1 K.

As discussed above, noise measurements are difficult to perform especially at low temperature where signal gets very small. At any 
temperature it is critically important to know the frequency window of the measurement and the gains in the amplifiers in Johnson 
noise thermometry. Yet the SNT avoids some of these problems. It is based on the cross-over between two noise mechanisms, both of 
which represent white noise, whereby the frequency window is ideally not a concern, since the same readout system is used in all the 
bias regimes. Moreover, the gains of the amplifiers are not that critical either because of the same argument. One can also use 
relatively high bandwidth which increases the absolute noise signal to be measured, and thereby makes the measurement faster.

The SNT technique has a few further attractive features. It is likely that its operation can be easily extended up to higher 
temperatures despite the deviations observed in the first
experiments.
The sensor consists of just one, relatively large size tunnel junction, which means that it is easy to fabricate with high precision. 
Also it is likely, although not yet demonstrated, that the SNT is not 
sensitive to magnetic field, since its operation is based on
tunnelling characteristics in a NIN tunnel junction as in CBT.

Finally, noise measurements can in principle be used to measure
the distribution function in non-equilibrium as well, as proposed
by ~\textcite{pistolesi:214518}.

\subsection{Thermometry based on the temperature dependent conductance of planar tunnel junctions}

The effect of temperature on the current across a tunnel barrier
with finite
height is a suitable basis for thermometry in a wide temperature range \cite{gloos:2915}. \textcite{simmons:1793} showed 
that the tunnelling conductance at zero bias across a thin insulating barrier depends on temperature as
\begin{equation} \label{simmons}
G(T)= G_0[1+(T/T_0)^2],
\end{equation}
where $G_0$ is the temperature independent part of conductance and the scaling temperature $T_0$ depends on the barrier height $\phi 
_0$. For a rectangular barrier of width $s$ one has $T_0^2 = \frac{3\hbar^2\phi _0}{\pi^2k_{\rm B}^2ms^2}$, where $m$ is the effective 
mass of the electrons within the insulating barrier. Experiments over a temperature range from 50 K up to 400 K on Al-AlOx-Al tunnel 
junctions have demonstrated that Eq. (\ref{simmons}) is obeyed remarkably well \cite{gloos:2915,suoknuuti:01}. Moreover, in these 
measurements the scaling temperature was found to be $T_0 \simeq 720$ K in all samples, without a clear dependence on the specific 
(zero temperature) conductance of the barrier, which varied over three orders of magnitude from $3$ $\mu$S/$\mu$m$^2$ up 
to 3000 $\mu$S/$\mu$m$^2$. This property makes the method attractive in wide range thermometry, and $T_0$ can indeed be considered as 
a material specific, but geometry and thickness independent
parameter up to a certain accuracy.

\subsection{Anderson-insulator thin film thermometry}

\begin{figure}[tb]
\begin{center}
\includegraphics[width=0.3
\textwidth]{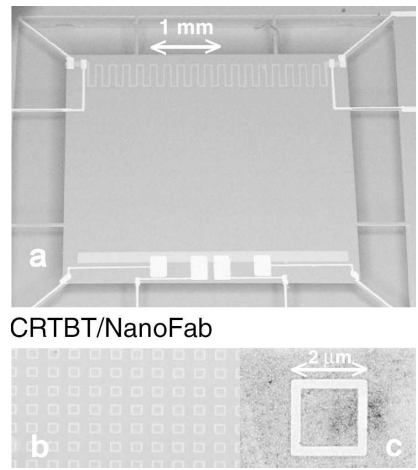} \caption{The suspended thermal
sensor employed in \cite{bourgeois:057007}. The
Nb$_{x'}$N$_{1-x'}$ thermometer can be seen in the lower part of
the rectangular silicon membrane. The 450 000 Al superconducting
rings are located in the middle part of the membrane; examples of
them are shown in (b) and (c). Figure from
\cite{bourgeois:057007}.}\label{fig:bourgeois}
\end{center}
\end{figure}

As regards to temperature read-out of micro-calorimetric devices,
resistive thin film thermometers near the metal-insulator transition
(MIT) are relatively popular. Electrical resistivity properties on
both sides of the MIT are rather well understood \cite{belitz:261},
and in general resistance of such thin films shows strong
temperature dependence, suitable for thermometry and in particular
for calorimetry. On the insulator side resistivity $\rho$ is
determined by hopping, and it has typically $\rho \propto
e^{(T_0/T)^n}$ temperature dependence, with $T_0$ and $n$ constants.
On the metallic side, weaker dependence can be found. In practice,
both Nb$_x$Si$_{1-x}$
\cite{denlinger:946,marnieros:862,marnieros:2469} and
Nb$_{x'}$N$_{1-x'}$ \cite{fominaya:4191,bourgeois:057007} thin film
based thermometers have been successfully employed. The suitable
conduction regime can be tailored by adjusting $x$ ($x'$) in
electron beam co-evaporation \cite{denlinger:946} or in dc magnetron
sputtering of Nb in a nitrogen atmosphere \cite{fominaya:4191}.

Bolometric and calorimetric radiation detectors are discussed in
detail in Sec.~\ref{sec:thermaldetectors}. Here we briefly mention
the application of a Nb$_{x'}$N$_{1-x'}$ thermometer in a
measurement of the heat capacity of 450 000 superconducting thin
film loops on a silicon membrane \cite{bourgeois:057007}, see
Fig.~\ref{fig:bourgeois}. The heat capacity of the loops is
proportional to their total mass, which was about 80 ng in this
case. Vortices entering simultaneously into the 450 000 loops under
application of magnetic field could be observed. A similar
measurement \cite{lindell:1884}, employing a NIS thermometer could
resolve the specific heat jump at $T_{\rm c}$ of 14 thin film
titanium disks with total mass of 1 ng on a silicon nitride
membrane.

\section{Thermal detectors and their characteristics}
\label{sec:thermaldetectors}

The absorption of electromagnetic radiation by matter almost
always ends in the situation where the incident optical power has,
possibly via cascades of different physical processes, transformed
into aggravated random motion of lattice ions, i.e. into heat.
Here lies the fundamental principle behind the thermal detection
of radiation: the transformation of the input electromagnetic
energy to heat. In many cases, this state of maximum entropy has
lost all the coherent properties that the incident radiation might
have possessed, but yet there is information in this messy final
state of the system: the rise in the system temperature. In this
section, we give a short overview of thermal detectors, their
theory and operation, and discuss some examples of thermal
detectors and their applications.

There is a very small difference between thermometry and
bolometry, the thermal radiation detection. Yet thermometry often implies
measurement of temperature changes over a large fractional
temperature range, whereas in the typical bolometric application,
the observed temperature variations are extremely small.
This allows us in the following to
concentrate on the limit of small temperature changes around some
mean well-defined value, i.e., we assume the operation in the
quasiequilibrium regime (see Sec.~II). In essence, bolometry is
really high-precision thermometry, nothing more, with the added
ingredient of finding efficient ways of coupling incident
radiation to the device.

Although thermal detectors have developed enormous diversity since
their introduction in 1880 by Samuel P. Langley, and major
advances in the way the temperature rise is measured and the
detectors are constructed have been made, the basic principle
remains the same. The operating principle of a thermal detector
can be traced back to the generalized thermal model, shown in Fig.
\ref{fig:setup}.  The four main parts of a thermal detector are a
thermally isolated element, a thermal link with a thermal
conductance $G_\mathrm{th}$ (here $G_{e-ph}$
or $G_{ph-sub}$), a thermal sensing element (i.e., a thermometer),
and a coupling structure (e.g., an impedance matching structure
for electromagnetic radiation) that serves to maximize the
absorbtion of the incident radiation, be it in the forms of alpha
particles or microwaves.

\subsection{Effect of operating temperature on the performance of thermal detectors}
\label{subsec:cooleffect}

Regardless of the exact architecture of the thermal detector,
lowering the operating temperature will improve the performance
significantly. This fact introduces the cooling techniques to the
use of thermal detectors. Because of the diversity of the
technological approaches it is hard to summarize the effect of the
operating temperature in a fully universal fashion. However, some
general trends can be evaluated. The figure of merit for thermal
detectors is the noise equivalent power (NEP), which relates to
the signal to noise ratio by $\mathrm{SNR}=\dot
Q_\mathrm{opt}/(\mathrm{NEP}\sqrt{2 \tau_\mathrm{int}})$, where
$\dot Q_\mathrm{opt}$ is the incident optical power and
$\tau_\mathrm{int}$ is the post-detection integration time. The
limiting NEP for an optimized thermal detector is given by the
thermal fluctuation noise (TFN) arising from random fluctuations
of energy across the thermal link with a thermal conductance
$G_\mathrm{th}\equiv d\dot Q/dT$ which result in variations in the
temperature of the device (see also discussion in
Subs.~\ref{subsubs:noise}). The TFN limited noise equivalent power
(NEP) in the linear order is given by the fluctuation-dissipation
result \cite{mather1}
\begin{align}
\mathrm{NEP_{TFN}} \approx  \sqrt{4 k_\mathrm{B}    T^2G_\mathrm{th}} \left\{ \begin{array}{ll}
\propto T^3 & \textrm{e-p decoupling}\\
\propto T^{5/2} & \textrm{lattice isolation}\\
\end{array} \right.
\end{align}
with the two cases relating the temperature dependence to the
location of the thermal bottleneck. The expression for the lattice isolated case is generally valid
at temperatures $T \lesssim \theta_\mathrm{D}/10$ with
$\theta_\mathrm{D}$ the Debye temperature of the insulating
material. The temperature $T$ is the highest of the temperatures
present in the system, i.e., the temperature of the thermally
isolated element or that of the heat sink.

Refrigeration, combined with the advances in nanolithography
techniques have recently opened a whole new realm for the
application of electron-phonon decoupling to improve the thermal
isolation of bolometric detectors. The operation of these
so-called hot electron devices is usually limited to low ($<$ 1 K)
temperatures and very small thermally active volumes $ \Vol$ as
the TFN between the electron gas and the lattice is given by
$\sqrt{5 k_\mathrm{B}\Sigma_\mathrm{e-p}\Vol(T_e^6+T_{ph}^6)}$
\cite{golwala:64} where $\Sigma$ varies between 1-4
nW$/\mu$m$^3/$K$^5$ in metals (See Table \ref{table:Sigma}). The
use of the lattice for the thermal isolation lends itself to
operation over much broader temperature range as the geometry of
the thermal link can be used to increase the thermal resistance.

Typically, the bath temperatures of cryogenic thermal detectors
are centered around four temperature ranges: 4.2 K, the boiling
temperature of liquid He at 1 atm, around 300 mK a temperature
attainable with $^3$He sorption refrigerators, around 100 mK,
easily attainable using a compact adiabatic demagnetization
refrigerator (ADR), or below $\sim$ 50 mK, when a dilution
refrigerator is used. In the recent years, the use of electronic
refrigeration has become appealing with the development of the
SINIS coolers. These coolers could enable the operation of thermal
detectors at a much lower temperature than is 'apparent' to the
user. For instance, the $^3$He sorption coolers are quite
attracting due to their compactness, low cost and simple
operation. Used together with a SINIS cooler would allow for an
affordable cryogenic detector system without having to sacrifice
in performance.

Since the thermal fluctuation noise in a thermal detector is
proportional to $T_e^n+T_{\rm ph}^n$, maximum performance
improvement is achieved by the reduction of the bath temperature
\cite{anghel:556}. Modest performance increase is possible with
direct electronic cooling, such as is the case in many of the SINIS
bolometer experiments.

In addition to improved noise performance, direct coupling of a
SINIS cooler to a thermal detector could allow for increasing the
dynamic range of for example bolometers based on transition edge
sensors (TES): A SINIS cooler can be used to draw a constant power
from a TES that would otherwise saturate due to an optical load.

\subsection{Bolometers: Continuous excitation}

Thermal detectors that are used to detect variations in the
incident flux of photons or particles are called bolometers (from
the Greek word {\em bol$\overline{e}$} - beam). This condition is
generally met when the mean time between incident quanta of energy
that arrive at the detector is much shorter than the recovery time
of the bolometer  \footnote{The opposite (calorimetric) limit will
be discussed in section \ref{subsec:calorimeters}}. The bolometric
operating principle is very simple: Change $\Delta \dot
Q_\mathrm{opt}$ in the incident optical power creates a change in
the temperature of a thermally isolated element by $\Delta
T=\Delta \dot Q_\mathrm{opt}/G_\mathrm{th}$. A sensitive
thermometer is used to measure this temperature change. The
recovery time $\tau_0$ is determined by the heat capacity $C$ of
the bolometer, and the thermal conductance $G_\mathrm{th}$ with
$\tau_0=C/G_\mathrm{th}$, analogously to an electrical $RC$
circuit.

Bolometric detectors remain popular today, 125 years after their
first introduction. Probably the most important advantage of
thermal detectors is their versatility: Bolometers can detect
radiation from $\alpha$ -particles to radio waves, their dynamic
range can be easily adapted for a variety of signal or background
levels. As an extreme example,  bolometers have been used to
detect infrared radiation from nuclear fireballs \cite{stubbs},
and the cosmic microwave
 background. \cite{lamarre:730}.

In the early days, bolometers typically utilized hand-crafted
construction (dental floss, cigarette paper and balsa wood are
examples of typical materials used in the construction)
\cite{davis}. If detecting electromagnetic waves, typically the
absorber consisted of a metal with a suitable thickness yielding a
square resistance of 377 $\Omega$, i.e. matching the impedance of
the vacuum. Further improvements on matching were achieved by
placing the bolometer in a resonant cavity. One major setback with
bolometers in their early days was unavoidably slow speed, caused
by the large heat capacity resulting from the macroscopic size of
the components used. The dawn of modern microfabrication
techniques has all but eliminated this shortcoming, with
bolometers of high sensitivity achieving time constants as short
as a few hundred nanoseconds.

Today, the most common type of cryogenic resistive bolometers
utilize transition edge sensors for the thermometry. In a TES
bolometer, a superconducting film with a critical temperature
$T_\mathrm{c}$ is biased within its superconductor - normal metal
transition where small changes in the film temperature result to
changes in the current through the device (or the voltage across
the film). In most cases, the TES consists of two or more
sandwiched superconductor-normal metal layers. The relative
thicknesses of the S and N layers are used to tune the transition
temperature to a desirable value by the proximity effect.

Transition-edge sensors are by no means a novel type of a thermal
detector, as first suggestions for their use came out already in
the late thirties \cite{andrews:132,goetz:1270}, and first
experimental results by 1941 \cite{andrews:1045}. Two principal
problems prohibited the wide use of this type of thermal detectors
for some fifty years: Typically the normal-state resistance of the
superconducting films was too low in order to obtain adequate
noise matching with field effect
transistor (FET) preamplifiers, and the lack of good
transimpedance amplifiers usually required the films to be current
biased with a voltage readout. This introduced a requirement to
tune the heat bath temperature very accurately within the narrow
range of temperatures in the superconducting transition. This also
made the devices exceedingly sensitive to small variations in the
bath temperature, introducing stringent requirements for the heat
bath stability.

These limitations can be overcome by the use of an external
negative feedback circuit that maintains the film within its
transition temperature and above the bath temperature. Such an
approach was adopted by \textcite{clarke:664}, who were able to
demonstrate NEP=$1.7\cdot10^{-15}$ W$/\sqrt{\mathrm{Hz}}$ at an
operating temperature of 1.27 K, using a transformer-coupled FET
as the voltage readout. Introducing a negative feedback has
similar advantages as in the case of operational amplifiers:
linearity is improved, sensitivity to internal parameters of the
amplifier (or bolometer) is reduced, and the speed is increased.
Interestingly, the use of an external negative feedback in
conjunction with superconducting transition edge sensors never
found widespread use, possibly due to the (slightly) more
complicated read-out architecture, and the need for a matching
transformer.

The breakthrough of TESs came in 1995, when superconducting
quantum interference device (SQUID) ammeters, which are inherently
well suited in matching to low load impedances, were introduced as
the readout devices for TESs \cite{irwin:2690,irwin:1998}. This
allowed for the use of voltage biasing, which introduces strong
negative electrothermal feedback (ETF) that causes the thermally
isolated film to self-regulate within its superconducting
transition. The local nature of the ETF makes the operation of
these detectors very simple as no external regulation is
necessary. As with an external negative feedback, an important
advantage of the voltage biased TES is the fact that once the bath
temperature is below $\sim T_c/2$, the need for bath temperature
regulation is significantly relaxed. Finally, the increased speed
due to the strong negative ETF increases the bandwidth of the
detector, allowing for either detecting faster signal changes in a
bolometer, or higher count rates in a calorimeter. A comprehensive
review of the theory and operation of voltage biased TESs has been
recently published \cite{enss:0}.

The behaviour of thermal detectors is generally well understood.
In the following discussion we shall summarize the main results
for the theory of thermal detectors. The results are quite
generally applicable to any resistive bolometers, but the
treatment is geared towards voltage biased TES, firstly as they
are currently the most popular type of thermal detectors, and
secondly because the electrothermal effects are very prominent in
these devices, having a major impact on the device speed and
linearity.

We start by writing the equation governing the thermal circuit
(here we limit the discussion to a simple case of one thermal
resistance and heat capacity). Generally, the power flow to the
heat sink is given by $\dot Q_{\rm out}=K(T^n-T_0^n)$, where $K$
is a constant which depends on materials parameters and the
geometry of the link. The time dependence of the bolometer
temperature can be solved from the heat equation for a bolometer
with a bias point resistance of $R=V/I$ absorbing a time-varying
optical signal $\dot Q_{\rm opt}(t)=\dot Q_{\rm o}e^{i \omega t}$
(c.f., Eq.~\eqref{eq:quasieqtemp}):
\begin{align}
C \frac{d(\delta T e^{i \omega
t})}{dt}+K(T^n-T_0^n)+G_\mathrm{th}\delta T \nonumber \\=\dot
Q_{\rm bias}+\dot Q_{\rm o}e^{i \omega t}+\frac{d\dot Q_{\rm
bias}}{dT}\delta T, \label{eq:heatbal2}
\end{align}
where $\delta T$ is used to denote the temperature change due to
the signal power and $\dot Q_{\rm bias}$ describes the incoming
heat flow due to the detector bias. Equating the steady state
components of the equation yields $\dot Q_{\rm
bias}=K(T^n-T_0^n)$, from which one can obtain the result for the
average operating temperature of the bolometer, given by $T=\dot
Q_{\rm bias}/\overline G_\mathrm{th} +T_0$ where an average
thermal conductance $\overline G_\mathrm{th}$ is defined by
$\overline G_\mathrm{th}=K(T^n-T_0^n)/(T-T_0)=\dot Q_{\rm bias}/[(
\dot Q_{\rm bias}/K+T_0^n )^{1/n}-T_0]$.

The electrothermal feedback manifests itself through the fact that
the change in input signal power modifies the bias dissipation, an
effect described by the last term in Eq.~(\ref{eq:heatbal2}).
Taking a closer look at the temperature change we obtain
\begin{equation}
\delta T=\frac{\dot Q_o}{G_\mathrm{th}+i\omega C - d \dot Q_{\rm
bias}/dT} \label{eq:dynterms}
\end{equation}
where $G_\mathrm{th}=d\dot Q_{\rm out}/dT \approx nKT^{n-1}$ is
the dynamic thermal conductance. Now, considering the
electrothermal term
\begin{equation}
\frac{d \dot Q_{\rm bias}}{dT}
=-\frac{\dot Q_{\rm bias}\alpha}{T} \beta(\omega),
\label{eq:dPdT}
\end{equation}
where $\alpha=d\log{R}/d\log{T}$ describes the sensitivity of the
detector resistance to the temperature changes and
$\beta(\omega)\equiv [R-Z_{\rm S}(\omega)]/[R+Z_{\rm S}(\omega)]$
is the effect of the bias circuit (with an embedding impedance of
$Z_{\rm S}$) on the ETF. Taking into account the thermal cut-off
of the bolometer, the frequency-dependent loop gain is defined as
\begin{equation}
{\cal L}(\omega)\equiv {\cal L}_0 \frac{\beta(\omega)}{\sqrt{1+
\omega^2 \tau_0^2}}, \label{eq:loopgain1}
\end{equation}
where ${\cal L}_0 \equiv \dot Q_{\rm bias}\alpha/(G_{\rm th}T)$
and $\tau_0=C/G_\mathrm{th}$ is the intrinsic thermal time
constant of the bolometer. The electro-thermal loop gain describes
the effect of varying incident optical power to the bias power
dissipated in the detector. For positive bolometers with $\alpha
>0$ the loop gain is positive for current bias (since Re$[\beta(\omega)]>0$) and negative for voltage bias (as Re$[\beta(\omega)<0]$). For bolometers with a negative temperature coefficient of resistance the situation is reversed. For metallic bolometers operated at room temperature $\alpha \sim 1$ and the loop
gain is typically small (${\cal L}_0\lesssim$1) so that the role
of ETF is negligible. On the contrary, superconducting detectors
with $\alpha \sim 100$ and $G_\mathrm{th}$ some three orders of
magnitude smaller than for room temperature devices can have large
loop gain (${\cal L}_0 \gtrsim 50$), so that ETF plays a
significant role in the detector characteristics. A major impact
of negative ETF is that the bolometer time constant is reduced
from $\tau_0$ to $\tau_{\rm eff}=\tau_0/[1+\beta(0){\cal L}_0]$.
This reduction in the time constant is one of the major benefits
of strong negative ETF.

Voltage biasing conditions are typically reached by driving a
constant current $I_0$ through a parallel combination of the
bolometer and a load resistor $Z_{\rm S}$. As long as $Z_{\rm
S}\ll R$, the bolometer is effectively voltage biased. The
responsivity of a voltage biased bolometer can be derived as
follows: The current responsivity is defined as ${\cal R}_{\rm I}
\equiv dI/d\dot Q_{\rm o}$. Using Eqs. (\ref{eq:dynterms}),
(\ref{eq:dPdT}) and (\ref{eq:loopgain1}), $V=I_0 Z_{\rm
S}R/(Z_{\rm S}+R)$, and $I=I_0 Z_{\rm S}/(Z_{\rm S}+R)$, the
result for the current responsivity becomes
\begin{eqnarray}
{\cal R}_{\rm I}(\omega)&=& -\frac{1}{V} \frac{(1+\beta){\cal L}_0}{2(1+\beta {\cal L}_0)}\frac{1}{\sqrt{1+\omega^2 \tau_{\rm eff}^2}}.
\label{eq:currentresp}
\end{eqnarray}
In the limiting case at $\omega=0$ with $\beta=1$ (perfect voltage bias) and ${\cal L}_0 \gg 1$,
\begin{equation}
{\cal R}_{\rm I}(0)=-\frac{1}{V}.
\label{eq:silimit}
\end{equation}

In order to evaluate the NEP for a bolometer, a discussion on the
noise sources in bolometers is merited. In general, the NEP, noise
spectral density $S$, and responsivity are related by
NEP$=\sqrt{S_{V,I}}/|{\cal R}_{V,I}|$, where the subscripts refer
to voltage or current noise and responsivity, respectively. The
noise in bolometers is due to several uncorrelated sources. The
most important of them is the thermal fluctuation noise, mentioned
already above. Generally, this contribution is given by
\begin{equation}
\mathrm{NEP}_{\rm TFN}=\sqrt{4 \gamma k_{\rm B}T_{\rm c}^2 G_\mathrm{th}}.
\label{eq:TFN}
\end{equation}
Here $\gamma$ describes the effect of the temperature gradient across the thermal link between the sensor 
and the heat sink \cite{mather1} in the diffusive limit (i.e., no
ballistic heat transport present) as
\begin{equation}
\gamma={\frac {\left (b+1\right )\left ({{ T_{\rm c}}}^{b+3}-{{ T_{\rm c}}}^{-b}{{ T_0}}^{3+2\,b}\right )}{\left (3+2\,b\right ){{ 
T_{\rm c}}}^{2}\left ({{ T_{\rm c}}}^{b+1}-{{ T_0}}^{b+1}\right
)}} \overset{T_{\rm c} \gg T_0}{\rightarrow} \frac{b+1}{2b+3}.
\label{eq:gamma}
\end{equation}
Here we assumed that the thermal conductivities obey $\kappa
\propto T^b$. For resistive bolometers, another important
contribution to the NEP is due to the Johnson noise, given by
\begin{equation}
\mathrm{NEP}_{\rm J}=\sqrt{\frac{4 k_{\rm B}T_{\rm c}}{R}} \frac{V}{{\cal L}_0} \sqrt{1+ \omega^2 \tau_0^2}.
\label{eq:JN}
\end{equation}
In addition to these noise sources, the current noise of the
amplifier, $S_{i_\mathrm{n,amp}}$, adds a contribution
$\mathrm{NEP_{amp}}=\sqrt{S_{i_\mathrm{n,amp}}}/{\cal
R}_\mathrm{I}$. The total NEP of a bolometer is then
\begin{align}
\mathrm{NEP}^2_\mathrm{tot} = &
\mathrm{NEP}^2_\mathrm{TFN}+\mathrm{NEP}^2_\mathrm{J} \nonumber \\
& +\mathrm{NEP}^2_\mathrm{amp}+\mathrm{NEP}^2_\mathrm{excess},
\label{eq:totnep}
\end{align}
where the slightly ambiguous contribution $\mathrm{NEP_{excess}}$
encompasses various contributions from additional external and
internal noise sources. Typical excess external noise
contributions arise from heat bath temperature fluctuations and
pickup in the cabling to name a few. In addition to the external
excessive noise, it has in recent years become clear that there
are also internal noise sources that are not fully accounted for.
For instance, TESs with significant internal thermal resistance
can no longer be treated using the simple lumped element model,
and they suffer from internal thermal fluctuation noise (ITFN),
adding a contribution \cite{hoevers:4422}
\begin{equation}
\mathrm{NEP_{ITFN}}=\sqrt{\frac{4 k_\mathrm{B} T R}{\Lor}}
G_\mathrm{th} \sqrt{1+\omega^2\tau_0^2}, \label{eq:itfn}
\end{equation}
where $\Lor$ is the Lorenz number.

For some devices, these noise sources are sufficient to explain
all the observed noise. However, many groups are developing X-ray
microcalorimeters (see below) which are often operated at a small
fraction of the normal state resistance exhibit noise that
increases rapidly as the bias point resistance is decreased, and
has a significant influence on the performance of the detectors.
Several possible explanations have been put forth, e.g., noise
arising from the fluctuations in magnetic domains or phase-slip
lines \cite{knoedler:2773,wollman:196}. A systematic study of the
excess noise in different TES geometries has been published
recently \cite{ullom:4206}, showing that there exists a clear
correlation between $\alpha$ and the observed excess noise, with
the magnitude of the excess noise scaling roughly as
$0.2\sqrt{\alpha}$. A quantitative agreement with the measured
excess noise spectrum has so far been achieved in one experiment
\cite{luukanen:238306}, where the TES consisted of an annular (so
called Corbino) geometry with a superconducting center contact,
and a concentric current return at the outer perimeter of the
annular TES. This geometry results in strictly radial current
flow, enabling a simple analytical expression for the current
density in the TES. The $1/r$ dependence in the current density
together with a small radial temperature gradient causes the TES
to separate to two annular superconducting and normal state
regions.

For any system undergoing a second order phase transition, order
parameter fluctuations will take place. The excess noise arises
from the thermally driven fluctuation of the phase boundary which
manifest as resistance fluctuations. The volume associated with
the order parameter fluctuation can be obtained by noting that the
Ginzburg-Landau free energy $\delta F$ associated with the
fluctuation is $\delta F \sim k_\mathrm{B}T_\mathrm{c}$. As the TES is
biased towards smaller $R$, the relative contribution of the order
parameter fluctuations becomes larger until it fully dominates
over the other noise contributions. For the Corbino-geometry TES,
the contribution due to the fluctuation superconductivity noise
(FSN) to the NEP is
\begin{equation}
\mathrm{NEP_{FSN}}=\frac{0.24 \Lor T_\mathrm{c}^2G}{V^2 \alpha}
\Gamma \sqrt{1+\omega^2 \tau_0^2}, \label{eq:NEPfsn}
\end{equation}
where $\Gamma \approx 10^{-8}$ K$/\sqrt{\mathrm{Hz}}$ is a constant dependent on the TES parameters.

So far, this noise model has not successfully been applied to TESs
in the more conventional square geometry, mainly due to the fact
that the current distribution varies with bias point and is not
easily calculable. A solution could be obtained by solving the
full 2D Ginzburg-Landau equations for a square geometry.

\subsubsection{Hot electron bolometers}

In principle, modern bolometers can be artificially divided into
two major sub-classes depending on where the dominant thermal
bottleneck lies. The so-called hot-electron bolometers (HEBs)
utilize the decoupling of the electron gas in a metal or a
semiconductor from the phonon heat bath. The earliest HEBs were
based on InSb, where the weak coupling of the electrons to the
lattice at temperatures around 4 K allows the electrons to be
heated to a temperature significantly above that of the lattice
even in a bulk sample. Mobility in InSb is limited by ionized impurity
scattering which results in decreasing resistivity with an
increasing electric field. This is the basis of the detection
mechanism \cite{rollin:1102,brown:213}. Often HEBs have sufficient speed for mixing, and in fact HEB mixers are a current topic of considerable interest. Most HEBs mixers operate at 4 K, and in order to maintain our focus on phenomena and devices at temperatures below 1 K, we unfortunately will not discuss them within this review.

HEB direct detectors \cite{richards:1,ali:184,karasik:188} have been a subject of considerable interest, but in general a full
optical demonstration remains to be carried out. The attractive
features of such devices include very short time constant (well
below 1 $\mu$s), potentially very good NEP performance (below
$10^{-19}$ W/$\sqrt{\mathrm{Hz}}$ when operated at or below 0.3
K), and simple construction that does not require surface
micromachining steps. The architecture is essentially very similar
to that of the HEB mixers: a small superconducting TES film coupled to the feed of a lithographic antenna. The application of the SQUID readout scheme
utilized in typical hot phonon microbolometers and
microcalorimeters might not be as straightforward as one could
expect as the introduction of the SQUID input coil inductance to
the voltage biasing circuit can make the system unstable due to
the interaction of the poles of the electrical circuit and the
thermal circuit. The approximate criterion for the stability of a
voltage biased TES is that the effective time constant of the TES
should be about one order of magnitude longer than the electrical
time constant of the bias circuit \cite{irwin:3978}.

A hot-electron bolometer that has been demonstrated is based on
electron thermometry with NIS junctions
\cite{nahum:3075,nahum:3203}. Here the incident optical power
elevates the electron temperature in a small normal metal island
weakly thermally coupled to the lattice phonons, and the change in
the electron temperature can be sensed as a change in the
tunneling current (see Subs.~\ref{subs:nisthermometer}). Noise
equivalent powers below $10^{-19}$ W$/\sqrt{\mathrm{Hz}}$ have
been predicted \cite{kuzmin:349} but remain to be experimentally
verified. An additional attractive feature of the SINIS bolometer
is that a DC bias on the device can be used to refrigerate the
electrons to a temperature below that of the bath temperature (see
Subs.~\ref{sec:SINIS}). Another benefit over TESs is the fact that
the SINIS bolometer saturates much more gently compared to the
TESs, which basically have no response at all once the device is
overheated above $T_c$. The self-cooling property of the SINIS
bolometer can also be used to compensate for excessive background
loading, thus effectively giving it a larger dynamic range. The
main obstacle towards constructing large arrays of (SI)NIS based
HEBs is that their impedance (typically 1 k$\Omega$ - 100
k$\Omega$) is hard to match to the existing cryogenic SQUID
multiplexers
\cite{chervenak:4043,dekorte:3807,yoon:371,reintsema:4500,lanting:112511}.
In principle, one could apply superconducting transformers to
match the SQUID noise, but transformers with sufficient impedance
transformation range are quite large, which makes this approach
unpractical. A novel readout method that lends itself for array
readouts is a microwave reflectometric measurement, in which the
SINIS bolometer is connected in series with a tuning inductor
\cite{schmidt:1002,schmidt:140301}. The $LC$ resonance frequency
of the inductor and the stray capacitance of the junctions is
tuned to fall within the bandwidth of the cryogenic microwave
amplifer (400-600 MHz), facilitating good impedance match. The
dynamic resistance of the device is highly sensitive to the
electron temperature, and thus temperature changes cause
modulation of $Q$ of the resonance circuit. This modulation is
sensed by sending a small RF signal to the resonant circuit, and
measuring the reflected power. The electrical NEP inferred from
noise measurements was in these experiments $1.6 \times 10^{-17}$
W/$\sqrt{\mathrm{Hz}}$.

\subsubsection{Hot phonon bolometers}

\begin{figure}[h]
\begin{center}
\includegraphics[width=60mm]{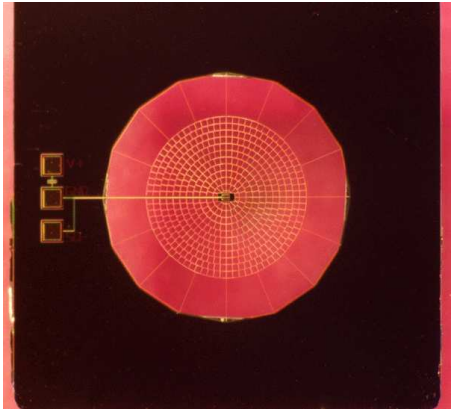}
\caption{(Color in online edition): A micrograph of a "spider-web"
bolometer. A NTD Ge thermistor is located at the centre of the
web. Image courtesy of NASA/JPL-Caltech.} \label{fig:spiderweb}
\end{center}
\end{figure}

The second major class of bolometers are hot phonon bolometers
(HPBs). They rely on a geometrical design of the heat link
$G_\mathrm{geom}$ so that the thermal bottleneck lies between two
phonon populations. This approach is the most common, and allows
for operation at temperatures up to and beyond room temperature.
The earliest and most widely used of the contemporary HPBs are the
so-called spider-web bolometers \cite{mauskopf:765}, operated at
300 mK and below, where a free-standing Si$_3$N$_4$ mesh is used
to support a thermal sensing element. Narrow Si$_3$N$_4$ legs
provide the thermal isolation for the mesh. Before the
introduction of TESs and SQUIDs the thermometer of choice was a
small crystal of neutron-transmutated (NTD) Ge due to its
relatively high temperature coefficient of resistance ($d\log R/d
\log T|_{T=0.3 \mathrm{K}}\approx -6$) and large resistance ($\sim
25$ M$\Omega$) that allowed for good noise matching with FET
preamplifiers. On the other hand this made the devices very
microphonic. Efficient optical coupling was possible since
radiation at wavelengths smaller than the mesh period are absorbed
to a resistive film deposited on the Si$_3$N$_4$ mesh. An example
of a spider-web bolometer using a NTD Ge is shown in Fig.
\ref{fig:spiderweb}. Later versions of the spider-web bolometers
have adopted the use of TES as thermometers, coupled to a SQUID
readout \cite{gildemeister:868}.

Alternatively to the spider-web absorber, a $\lambda/4$ resonant
cavity can be used to maximize the optical efficiency over a
limited bandwidth. The SCUBA-2 instrument
\cite{duncan:19,holland:1} on the James Clerk Maxwell telescope is
an ambitious overtaking with a pixel count of over 12 000. The
bolometers consist of thermally isolated, (1 mm$^2$)$\times$ 60 nm
silicon 'bricks' with the front surface of the Si degenerately
doped with phosphorous to 377 $\Omega$ per square. The resonant
cavity is formed by the Si in between the doped layer, and a Mo-Cu
TES deposited to the back side of the pixel.

Even though the superconducting TES thermometers are becoming
increasingly popular, bolometers utilizing lithographic
semiconducting thermistors still yield impressive performance and
are more forgiving in terms of the saturation power. In the
so-called pop-up bolometers, a lithographic, doped Si is used  as
the thermistor, while an ingenious torsional bending method of the
Si$_3$N$_4$ legs is used to bend the legs and wiring layers
perpendicular to the absorbers. This architecture enables the
construction of arrays with a very high filling factor
\cite{Voellmer:63}. Another promising lithographic resistive
thermometer technology is based on thin films of NbSi
\cite{camus:419}. The high resistivity of these films allows for
impedance levels that provide a good noise matching to room
temperature JFET amplifiers.

\begin{figure}[h]
\begin{center}
\includegraphics[width=75mm]{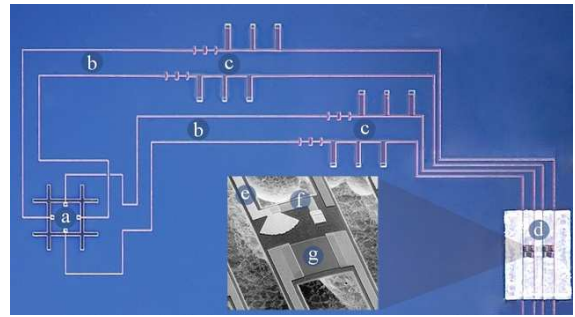}
\caption{(Color in online edition): Micrograph of an
antenna-coupled TES bolometer. A dual-polarized double-slot
antenna (a) is coupled via microstrip transmission lines (b) with
band-pass and low-pass filters (c) to two Al/Ti bilayer TES
thermometers located on suspended Si$_3$N$_4$ membranes (d). The
inset shows in detail the termination of the microstrip line (e)
to a resistor (f), and the Al/Ti TES (g). Figure courtesy A. T.
Lee, UC Berkeley.} \label{fig:berkeley_acmb}
\end{center}
\end{figure}

Instead of the surface absorbing approach, another method is based
on the use of a lithographic antenna \cite{hwang:773,neikirk:245},
and terminating the induced currents to a thermally isolated
bolometer, an approach often used in the HEB mixers discussed
above. The attractive feature of this method is that the thermally
sensing volume can be made much smaller compared to the case of
the surface-absorbing bolometers, making these devices much
faster. The low heat capacity often allows for lower NEP, as in
many cases the NEP$_{\rm TFN}$ is limited by the maximum time
constant $\tau_{\rm max}=C/G_{\rm th, min}$ allowed by the
application. In this case NEP$_{\rm TFN}=\sqrt{4 \gamma k_{\rm
B}T^2 C/ \tau_{\rm max}}$. Moreover, the small size of the
thermally sensitive volume makes it far less sensitive to out of
band stray light, relaxing filtering and baffling requirements of
the incoming radiation. Antenna coupling also lends itself to the
construction of integrated, on-chip filters
 for defining the bands for an array of bolometers \cite{mees:2329,hunt:318,myers:114103}. Arrays of antenna-coupled bolometers can also utilize the  inherent polarization selectivity of the antennas. An example of an antenna-coupled TES bolometer that incorporates on-chip transmission line impedance transformers and band-pass filters is shown in Fig. \ref{fig:berkeley_acmb}. No NEPs have yet been measured on this device, but the expected NEP is $\sim 10^{-16}$ W$/\sqrt{\mathrm{Hz}}$  for a device with $T_\mathrm{c}=450$
 mK.

\begin{figure}[h]
\begin{center}
\includegraphics[width=75mm]{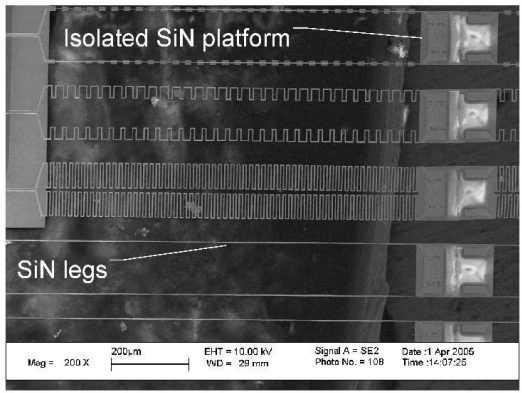}
\caption{SEM micrograph of an isolated Si$_3$N$_4$ platforms with
Mo/Au TESs. As an extreme example of thermal isolation, the
meandering Si$_3$N$_4$ legs have an aspect ratio of $\sim$ 2800:1,
yielding a thermal conductance $G_{\rm th}\approx 100$ fW/K at a
bath temperature of 210 mK. Figure courtesy M.E. Kenyon, JPL.}
\label{fig:kenyonbolo}
\end{center}
\end{figure}

While not published yet, the best performance of an
antenna-coupled HPB utilizing a similar design has been obtained
by the JPL-Caltech bolometer group, with an electrical NEP=$5\cdot
10^{-19}$ W$/\sqrt{\mathrm{Hz}}$ at 230 mK \cite{kenyon:perscomm}.
The architecture is shown in Fig. \ref{fig:kenyonbolo}.

In the simplest case, the bolometer is simply a strip of metal
placed to the feed of the lithographic antenna. Although this
approach is by far the simplest, it introduces limitations for the
performance of a HPB: Unlike in the case of HEB mixers, maximizing
the thermal isolation of the bolometer is always desirable.
However, the bolometer resistance should be matched to the
impedance of the antenna (typically $\sim$ 100 $\Omega$ for
broadband lithographic antennas on Si substrates), and thus the thermal
conductance to the bath through the antenna,$G_{\rm ant}$ , is
fixed by the Wiedemann-Franz law, $G_\mathrm{ant}=\Lor
T/\mathrm{Re}({Z_\mathrm{a}})$, where $Z_\mathrm{a}$ is the
antenna impedance. A parallel heat loss path is to the substrate
below the bolometer film, with conductance $G_\mathrm{sub}\propto
LW$ where $L$ and $W$ are the length and width of the bolometer
film, respectively. For substrate mounted antenna-coupled HPBs it
is thus beneficial to minimize the size of the bolometer. This
requirement can be relaxed if the bolometer strip is released from
the underlying substrate to form an air-bridge \cite{neikirk:153}.
For air-bridged devices at temperatures higher than 4 K,
NEP$_\mathrm{TFN} \propto T^{3/2}$. Recently, an air-bridge
bolometer operating at 4.2 K was demonstrated, showing potential
for background-limited performance when observing 300 K
blackbodies \cite{luukanen:3970,luukanen:0}. The potential
applications for these devices include passive detection of
concealed weapons under clothing, remote trace detection, and
terrestrial submillimetre-wave imaging.

\subsection{Calorimeters: Pulsed excitation}
\label{subsec:calorimeters}

In the limit opposite to the bolometric detection, i.e. when the
mean time between the quanta of energy  arriving at the detector
exceeds the device relaxation time, thermal detectors are known as
calorimeters. While the topic of calorimetry also encompasses heat
capacity measurements especially in mesoscopic samples, the
following discussion concentrates on the detection of radiation
only in order to keep the scope of our review in reasonable
limits. For those interested in microcalorimetry in the sense of
heat capacity measurements, we direct the reader to references
\cite{denlinger:946,fominaya:4191,lindell:1884,bourgeois:057007}
and \cite{marnieros:862}.

Cryogenic calorimeters are used today in a large
variety of applications, from the detection of weakly interacting
massive particles (WIMPs) in dark matter search
\cite{akerib:082002,bravin:107}, X-ray
\cite{dekorte:167,kelley:114,moseley:1257} and $\gamma$-ray
\cite{vandenberg:436} astrophysics to secure optical
communications \cite{miller:791,nam:523}. As is the case with
bolometers, calorimeters are usually operated at temperatures
below 1 K. The theory and operation of calorimetric thermal
detectors is very similar to bolometers, and generally the
theoretical treatment above is valid. However, the optimization of
calorimetric detectors can be quite different. The quantum of
energy $E$ deposited by either a photon, charged particle, WIMP
etc. can be determined from the temperature rise $\Delta T=E/C$,
where $C$ is the heat capacity of the calorimeter. This
temperature rise then decays exponentially with a time constant
$\tau_\mathrm{eff}$ to its equilibrium value. Thermometry in the
calorimeters is most often done either using semiconducting or TES
thermometers, as is the case with bolometers. In addition,
thermometry based on the change of magnetization of a paramagnetic
sensor is appealing due to its non-dissipative nature, and has
yielded some promising results
\cite{schonefeld:211,fleischmann:3947}.

The figure of merit for a calorimeter is the energy resolution,
$\Delta E$, of full-width at half-maximum (FWHM),  related to NEP
through \cite{moseley:1257}
\begin{align}
\Delta E = & 2 \sqrt{2 \ln{2}} \left( \int_0^\infty
\frac{4}{\mathrm{NEP}^2(f)_{\rm tot}}  df \right) ^{-1/2}
\nonumber\\
& \approx 2 \sqrt{2 \ln{2}} \mathrm{NEP}(0)_{\rm tot}
\sqrt{\tau_\mathrm{eff}}, \label{eq:energyres}
\end{align}
For a 'classic' calorimeter with a white noise spectrum the energy
resolution is in terms of the operating temperature and the heat
capacity given by
\begin{equation}
\Delta E=2 \sqrt{2 \ln{2}} \sqrt{k_\mathrm{B}T^2C},
\label{eq:energyres2}
\end{equation}
indicating that $\Delta E$ scales $\propto T^{5/2}$ or $\propto
T^{3/2}$ depending whether the heat capacity of the sensor is
dominated by the lattice or the electronic system, respectively.

\begin{figure}[h]
\begin{center}
\includegraphics[width=60mm]{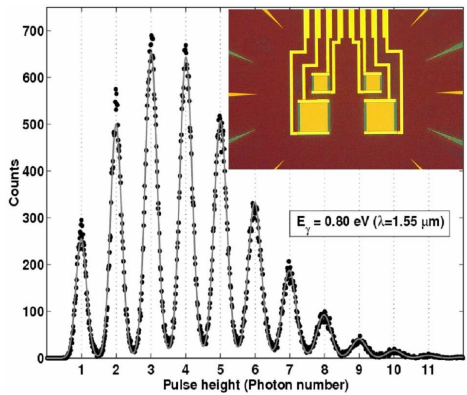}
\caption{(Color in online edition): The energy spectrum of a
pulsed 1550 nm laser, measured with an optical TES
microcalorimeter. The peaks in the plot correspond to the
photon-number state of the incoming pulses. The inset shows a
micrograph of the devices. Figure courtesy of A.J. Miller, NIST.}
\label{fig:opticalTES}
\end{center}
\end{figure}

In most of the devices under development today the lattice
temperature is being measured, as sufficient cross section to the
incoming energy often requires a rather large volume of a device.
An exception to the norm are TES calorimeters optimized for
optical single photon detection for applications in secure quantum
key distribution \cite{nam:523}, shown in Fig.
\ref{fig:opticalTES}. In these devices, the thermal isolation is
via electron-phonon decoupling. A trade-off is made between energy
resolution and the speed of the detectors. Energy resolution of
the detectors is sufficient to determine the photon-number state
of the incoming photons, while maintaining a speed that is
adequate for fast information transfer.

\begin{figure}[h]
\begin{center}
\includegraphics[width=55mm]{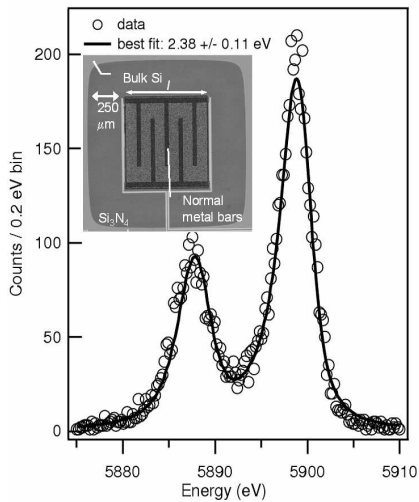}
\caption{An energy spectrum of the $^{55}$Mn $K_{\alpha}$ complex,
obtained with a Mo/Cu TES microcalorimeter. The width of the
measured characteristic x-ray lines is a convolution of the
intrinsic x-ray linewidth and the gaussian detector contribution
with a FWHM of 2.38 $\pm 0.11$ eV. The TES is fabricated on a
free-standing Si$_3$N$_4$ membrane. The inset shows a SEM
micrograph of the TES. The normal metal bars extending partially
across the TES have been experimentally verified to improve the
energy resolution of the detector \cite{ullom:4206}. Figure
courtesy of J. Ullom, NIST.} \label{fig:NISTcalorimeter}
\end{center}
\end{figure}

\begin{figure}[h]
\begin{center}
\includegraphics[width=75mm]{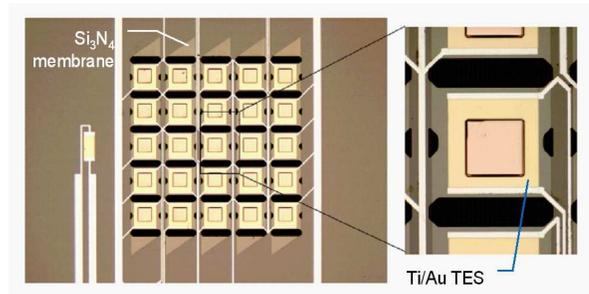}
\caption{(Color in online edition): A prototype Ti/Au
microcalorimeter array. The dark regions within the Si$_3$N$_4$
membrane are holes in the membrane. Figure courtesy of
SRON-MESA.}\label{fig:SRONarray}
\end{center}
\end{figure}

As with bolometers, the TESs currently outperform the competition
in terms of the sensitivity. The best reported energy resolution
for any energy dispersive detector was recently obtained with a
Mo-Cu calorimeter (see Fig. \ref{fig:NISTcalorimeter}), yielding
an energy resolution of 2.38$\pm$ 0.11 eV at a photon energy of
5.89 keV \cite{enss:0}.
The driving application in the development of X-ray
microcalorimeters are two major X-ray astrophysics missions
planned by the European Space Agency (ESA) and the U.S. National
Astronautics and Space Administration (NASA). Both missions, {\em
X}-ray {\em E}volving {\em U}niverse {\em S}pectroscopy ({\em
XEUS}) \cite{xeus:website} mission  and the {\em Constellation-X}
mission \cite{conx:website} will employ X-ray microcalorimeters as
their primary instrument. The energy resolution of the state of
the art microcalorimeters have already reached the requirements
stated in the science goals for these missions. Thus, the current
primary focus of the technical research is on the development of
large arrays of microcalorimeters, and more importantly, SQUID
multiplexing read-outs for the detector arrays. A prototype $5
\times 5$ array of Ti/Au TES microcalorimeters is shown in Fig.
\ref{fig:SRONarray}.

\subsection{Future directions}

While cryogenic thermal detectors are the most sensitive radiation
detectors around, further significant performance increases are
envisioned in the future. The push is from single pixels to large
staring focal plane arrays of detectors for numerous astrophysics
applications. For this reason, novel thermal detector concepts
must be able to be integrated into large arrays.

The performance of hot electron bolometer mixers has been
improving rapidly over the past years. Noise temperatures
approaching $10 \times h \nu/(2 k_B)$ are being reported, and the
emphasis is in pushing the operating frequency deeper into the THz
region. In principle, frequencies in the infrared range are not
out of the question. Lithographic antennas have demonstrated good
performance up to 30 THz \cite{grossman:3225}, while it is clear
that the fabrication becomes increasingly challenging as the
required feature size is reduced.

In the single pixel direct detector development, the quest for
ever better sensitivity is still ongoing. For bolometers, the improvement
in NEP directly translates to the capability of observing over
less pre-detection bandwidth without sacrificing signal to noise
ratio. This would enable the construction of arrays capable of
yielding spectroscopic information. It is here where the functions
of HEB mixers, bolometers and calorimeters conjoin - in the
capability of performing single photon spectrometry at
far-infrared wavelengths.

\section{Electronic refrigeration}
\label{sec:ecool}

Thermoelectric effects and, in particular, thermoelectric cooling
have been discovered more than 170 years ago \cite{peltier:371}.
During the last 40 years considerable progress has been made in
developing practical thermoelectric refrigerators for industrial
and scientific applications \cite{rowe:157,nolas:}. The
temperature range of interest has been, however, far above
cryogenic temperatures. Yet, during the last decade, solid state
refrigerators for low temperature applications and, in particular,
operating in the sub-kelvin temperature range have been
intensively investigated. The motivations of this activity stem
from the successful development and implementation of
ultrasensitive radiation sensors and quantum circuits which
require on-chip cooling \cite {pekola:41} for proper operation at
cryogenic temperatures. Solid state refrigerators have typically
lower efficiency as compared to more traditional systems (e.g.,
Joule-Thomson or Stirling gas-based refrigerators). By contrast,
they are more reliable, cheaper and, what is more relevant, they
can be easily scaled down to mesoscopic scale. All this gives a
unique opportunity to combine on-chip refrigeration with different
micro- and nanodevices.

The aim of this part of the review is to describe the existing solid
state electron refrigerators operating at cryogenic temperatures
(in particular below liquid helium temperature), and to give an
overview of some novel ideas and refrigeration schemes.

\subsection{General principles}

The physical principle at the basis of thermoelectric cooling is
that energy transfer is associated with quasiparticle electric
current, as shown in Sec. II. Under suitable conditions thermal
currents can be exploited for heat pumping, and if heat is
transferred from a cold region to a hot region of the system, the
device acts as a refrigerator. The term \emph{refrigeration} is
associated throughout this Review to a process of lowering the
temperature of a system with respect to the ambient temperature,
while \emph{cooling}, in general, means just heat removal from the
system. It is noteworthy to mention that the maximum cooling power
of a refrigerator is achieved at a vanishing temperature gradient,
while the maximum temperature difference is achieved at zero
cooling power. The efficiency of a refrigerator is normally
characterized by its coefficient of performance ($\eta$), i.e.,
the ratio between the refrigerator cooling power
($\dot{Q}_{cooler}$) and the total input power ($P_{total}$):
\begin{equation}
\mathcal{\eta}=\frac{\dot{Q}_{cooler}}{P_{total}}.
\label{eta}
\end{equation}
Irreversible processes (e.g., thermal
conductivity and Joule heating) degrade the efficiency of
 refrigerators, and are  essential elements that need to be carefully evaluated for the optimization
of any device.

The basic principles  of the design are in general similar for
different types of refrigerators. The increase of temperature
gradient can be achieved by realizing a multistage refrigerator.
In this case, the stage operating at a higher temperature should be
designed for larger cooling power in order to efficiently extract the heat released
from the lower-temperature stage ("pyramid design").
The enhancement of the refrigerator cooling power can be
achieved by connecting several refrigerators in parallel.
The parallel design is more effective both
for an efficient heat evacuation from the hotter regions of the device and
for the application of higher electric currents to the refrigerator.
It also allows more freedom to properly bias
 all cooling elements.

The temperature dependence of
the electric and thermal properties of  the active parts in the refrigerator may limit
their exploitation at low temperatures. The reduction of thermal conductivity by lowering the temperature has both positive (better thermal insulation between cold and hot
regions) and negative (difficulty in removing heat from the
system) effects.

At cryogenic temperatures different types of superconductors can be efficiently exploited.
They can be used both as
\emph{passive} and \emph{active} elements:
in the former case, owing to their low thermal conductivity and zero
electric resistance (e.g., as one of the two arms in Peltier refrigerators),
in the latter as materials with an energy gap in the density of states
for energy-dependent electron tunneling (e.g., in NIS coolers).

Currently, the development of refrigerating techniques follows two main
directions: search of new materials with improved characteristics suitable for existing refrigeration schemes, and development of new
refrigeration methods and principles.

\subsection{Peltier refrigerators}
\label{subs:peltier}

\begin{figure}[tb]
\begin{center}
\includegraphics[width=4.5cm,clip]{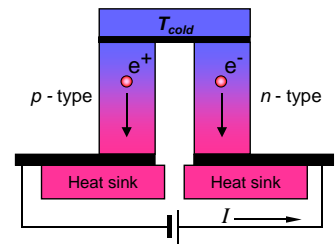}
\end{center}
\caption{Basic Peltier thermoelement.} \label{PeltierElement}
\end{figure}

Thermoelectric (Peltier) refrigeration is widely used for cooling
different electronic devices \cite{rowe:157, nolas:, phelan:356}.
Nowadays Peltier refrigerators providing temperature reduction
down to $100...200$ K and cooling power up to 100 W are available.
Peltier cooling (or heating) occurs when an electric current is
driven through the junction of two different materials. The heat
released or absorbed, $\dot{Q}_{Peltier}$, depending on the
direction of the electric current at the junction, is proportional
to the electric current ($I$) driven through the circuit,
$\dot{Q}_{Peltier}=\Pi_{AB} I$, where $\Pi_{AB}=\alpha_{AB}T$, and
$\Pi_{AB}$ and $\alpha_{AB}$ are the Peltier and Seebeck
coefficients of the contact, respectively (see also Eq.
(\ref{eq:ncurs})). In order to obtain enhanced values of the
Peltier coefficient, conventional Peltier refrigerators generally
consist of \emph{p}- and \emph{n}-type semiconductors with
opposite sign of $\Pi$ coefficients (see Fig.
\ref{PeltierElement}). The efficiency of a Peltier refrigerator is
not  determined only by the coefficient
$\Pi_{AB}=\Pi_{A}-\Pi_{B}$, but also by thermal conductivities
($\kappa$) of both materials across the contact. Furthermore their
electric resistances ($\rho$) are responsible of Joule heating
affecting the coefficient of performance. The maximum temperature
difference ($\Delta T_{max}$) achievable with a Peltier
refrigerator is given by \cite{nolas:} $\Delta
T_{max}=ZT^{2}_{cold}/2,$ where
$Z=\frac{\alpha^{2}_{AB}}{\rho\kappa}$ is a figure of merit of the
refrigerator, and $T_{cold}$ is the temperature of the cold
junction (see Fig. \ref{PeltierElement}). More often, however, the
refrigerator efficiency is characterized by the dimensionless
figure of merit $ZT$. We recall that $ZT\propto
(\frac{k_BT}{E_F})^2$. Most of the materials used in
thermoelectric applications have $ZT \sim 1 $ at room temperature
\cite{min:860, disalvo:703}. In general, at low temperatures only
one single thermoelectric material is needed, because a
superconductor can be used as one of the two arms of the
refrigerator \cite{nolas:, goldsmid:344}.

\begin{figure}[t!b!]
\begin{center}
\includegraphics[width=\columnwidth,clip]{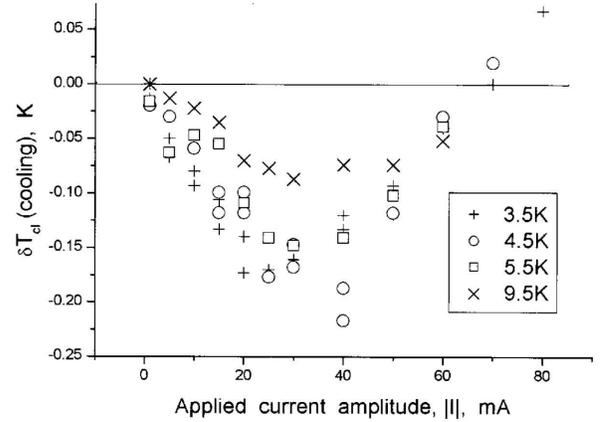}
\end{center}
\caption{Thermoelectric refrigeration at cryogenic temperatures
using cerium hexaboride. Adapted from \cite{harutyunyan:2142}.}
\label{Harutyunyan214}
\end{figure}

In spite of the drastic reduction of cooling efficiency at low
temperature, there is, however, some development of new materials
and devices suitable for operation at cryogenic temperatures.
Recently, Peltier cooling with $\Delta T_{max} \approx 0.2$ K was
demonstrated below $10$ K \cite{harutyunyan:2142} (see Fig.
\ref{Harutyunyan214}). Crystals of CeB$_{6}$ were used to exploit
the strong thermoelectric coefficients arising from the Kondo
effect. The dimensionless figure of merit of this material is
$0.1...0.25$ in the temperature range between $4$ K and $10$ K.
The authors claim that a proper optimization of the refrigerator
design would allow more than 10\% temperature reduction below 4K
with a single-stage refrigerator.

At millikelvin temperatures lattice specific heat and thermal
conductivity decrease drastically, and thermoelectric
refrigeration might become feasible. Following this idea and
taking into account the Wiedemann-Franz relation, values of $ZT$
as high as 20 at temperatures below 10 mK and the possibility to
achieve thermoelectric refrigeration were predicted
\cite{goldsmid:289,nolas:}.

\begin{figure}[tb]
\begin{center}
\includegraphics[width=\columnwidth,clip]{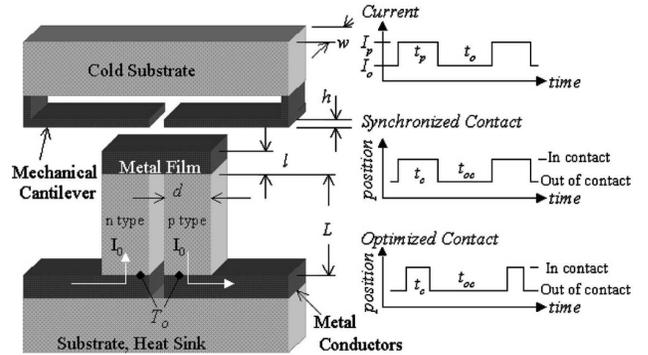}
\end{center}
\caption{Schematic diagram of the thermoelectromechanical cooler,
time sequences of the pulsed current applied to the device, and
the two modes of cantilever contact: synchronized and optimized
\cite{miner:1176}.} \label{Miner1176}
\end{figure}

\textcite{Kapitulnik:180} proposed to exploit a metal close to its
metal-insulator transition for the implementation of a
thermoelectric refrigerator operating below liquid-He
temperatures. The basic concept behind this proposal is that near
the metal-insulator transition, the transport coefficients acquire
anomalous power laws such that their relevant combination
describing the figure of merit for efficient cooling also becomes
large.

Further improvement of the efficiency of conventional
thermoelectric coolers could be, in principle, achieved through a thermoelectromechanical refrigerator \cite{miner:1176,thonhauser:3247}.
In such a device, a periodic variation of the electric current through a Peltier element
combined with
a synchronized mechanical thermal switch would allow
to improve the overall cooling performance (see Fig. \ref{Miner1176}).
The enhancement of refrigeration is
due to the spatial separation of the Peltier cooling and Joule heating.

The analysis of thermoelectric devices is usually based on the
parameters typical of bulk materials. Significant progress in low
temperature Peltier refrigeration might be achieved by using
exotic materials \cite{rontani:3033,goltsev:2272} and
low-dimensional structures \cite{disalvo:703, hicks:3230,
balandin:415}, such as composite thin films, modulation-doped
heterostructures, quantum wires, nanotubes, quantum dots, etc.
These systems offer, in general, more degrees of freedom to
optimize those quantities that affect the efficiency of
thermoelectric refrigerators.

\subsection{Superconducting electron refrigerators}
\label{sec:SEC}

\subsubsection{(SI)NIS structures}
\label{sec:SINIS}

\begin{figure}[b!]
\begin{center}
\includegraphics[width=\columnwidth,clip]{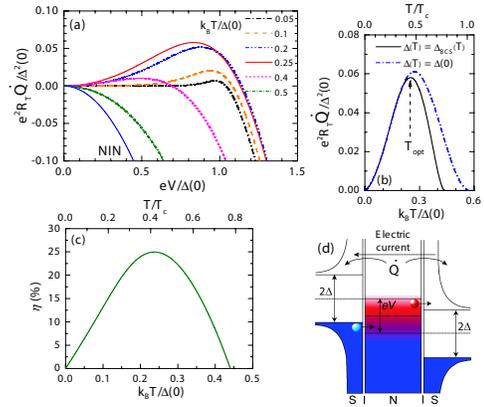}
\end{center}
\caption{Characteristics of (SI)NIS refrigerators. (a) Calculated
cooling power $\dot{Q}$ of a NIS junction vs bias voltage $V$ for
several temperatures $T=T_{e,N}=T_{e,S}$. Also shown is the behavior
of a NIN junction. (b) $\dot{Q}$ calculated at the optimal bias
voltage as a function of temperature, assuming both a
temperature-independent energy gap (dash-dotted blue line) and the
real BCS dependence (black line). $T_{opt}$ indicates the
temperature value that maximizes $\dot{Q}$. (c) Coefficient of
performance $\eta$ calculated at the optimal bias voltage versus
temperature. (d) Scheme of the energy band diagram of a voltage
biased SINIS junction. The electric current flows into the normal
region through one tunnel junction and out through the other, while
the heat current $\dot{Q}$ flows out of the N electrode through both
tunnel junctions.} \label{fig:niscoolp}
\end{figure}

Although heat transport in superconducting microstructures
originally dates back more than 40 years ago in SIS junctions
\cite{parmenter:274,melton:1858,gray:633,chi:4465}, it is
instructive to start our description of this topic from NIS tunnel
junction structures. Figure \ref{fig:niscoolp}(a) shows the
calculated $\dot{Q}$ for a NIS tunnel junction (see Sec. II.G.2)
versus bias voltage at different temperatures
($T=T_{e,N}=T_{e,S}$). When $\dot{Q}$ is positive, it implies heat
removal from the N electrode, i.e., \textit{hot} excitations are
transferred to the superconductor. For each temperature there is
an optimal voltage that maximizes $\dot{Q}$ and, by decreasing the
temperature, the heat current results to be peaked around $V\simeq
\Delta/e$. Figure \ref{fig:niscoolp}(b) displays the heat current
versus temperature calculated at each optimal bias voltage. The
quantity $\dot{Q}(T)$ is maximized at $T\approx
0.25~\Delta/k_B=T_{opt}$ (as indicated by the arrow in the figure)
where it reaches $\dot{Q}\simeq 6\times 10^{-2}
\Delta^{2}/e^2R_T$, decreasing both at lower and higher
temperatures. Also shown in the figure is $\dot{Q}(T)$ obtained
assuming a temperature-independent energy gap. Such a  comparison
shows that this latter assumption is fully justified for $T\leq
0.2~ \Delta /k_B$. In the low temperature limit ($T_{e,N}\leq
T_{e,S}\ll \Delta/k_B$) it is possible to give an approximate
expression \cite{anghel:197} for the optimal bias voltage
($V_{opt}$), $V_{opt}\approx (\Delta -0.66k_B T_{e,N})/e$, as well
as for the maximum cooling power at $V_{opt}$,
\begin{equation}
\label{eq:jQapprox}
\begin{array}{c}
\dot{Q}_{opt}\approx\frac{\Delta^2}{e^2R_T}[0.59(\frac{k_BT_{e,N}}{\Delta})^{3/2}\\
-\sqrt{\frac{2\pi k_BT_{e,S}}{\Delta}}\textrm{exp}(-\frac{\Delta}{k_BT_{e,S}})].
\end{array}
\end{equation}
Equation (\ref{eq:jQapprox}) is useful for getting quantitative
estimates on the performance of realistic coolers. In the same
temperature limit and for $V=V_{opt}$ the current through the
NIS junction can be approximated as $I\approx 0.48 \frac{\Delta}{eR_T}\sqrt{\frac{k_BT_{e,N}}{\Delta}}$.
The NIS junction coefficient of performance $\eta$ is given by
$\eta(V)=\dot{Q}(V)/[I(V)V]$. For $V\approx \Delta/e$ and in the low temperature limit, $\eta$ thus
obtains the approximate value
\begin{equation}
\eta_{opt}\approx 0.7\,\frac{T_{e,N}}{T_c},
\label{eq:etasin}
\end{equation}
where we assumed $\Delta=1.764\,k_BT_c$, and $T_c$ is the critical
temperature of the superconductor. Equation (\ref{eq:etasin})
shows that the efficiency of  a NIS junction is around or below
$20\%$ at the typical operation temperatures.
The full behavior of $\eta$ versus temperature calculated
at each optimal bias voltage is displayed in Fig. \ref{fig:niscoolp}(c).
The simple results
presented above point out how the optimized operation of a
superconducting tunnel junction as a building block of
microrefrigerators stems from a delicate balance among several
factors such as the contact resistance, the operation temperature,
the superconducting gap $\Delta$ as well as the bias voltage
across the junction.

The first observation of heat extraction from a normal metal dates
back to 1994 \cite{nahum:3123}, where cooling of conduction
electrons  in Cu below the lattice temperature was demonstrated
using an Al/Al$_2$O$_3$/Cu tunnel junction. A significant
improvement was made two years later, still in the
Al/Al$_2$O$_3$/Cu system, by \textcite{leivo:1996}, which
recognized that using two NIS junctions in series and arranged in
a symmetric configuration (i.e., in a SINIS fashion) leads to a
much stronger cooling effect. This fact can be understood keeping
in mind that $\dot{Q}$ is a symmetric function of $V$ so that, at
fixed voltage across the structure, quasi-electrons above $\Delta$
are extracted from the N region through one junction, while at the
same time quasi-holes are filled  in N below $-\Delta$  from the
other junction (see Fig. \ref{fig:niscoolp}(d)). In this
configuration, a reduction of the electron temperature from $300$
mK to about $100$ mK was obtained. Later on, several other
experimental evidences of electron cooling in SINIS metallic
structures were reported
\cite{leoni:3877,fisher:2705,clark:625,arutyunov:326,leoni:3572,tarasov:714,pekola:2782,pekola:485,vystavkin:598,fisher:561,leivo:227,pekola:056804,luukanen:281}.
In these experiments NIS junctions are used to alter the electron
temperature in the N region as well as to measure it. In order to
measure the temperature, the N region is normally connected to
additional NIS contacts (i.e., "probe" junctions) acting as
thermometers (previously calibrated by varying the bath
temperature of the cryostat), and operating along the lines
described in Sec. III.A.1. Moreover, the differential conductance
of the probe junctions gives also detailed information about the
actual quasiparticle distribution function in the N region
\cite{pothier:3490,pekola:056804}.

\begin{figure}[tb!]
\begin{center}
\includegraphics[width=\columnwidth,clip]{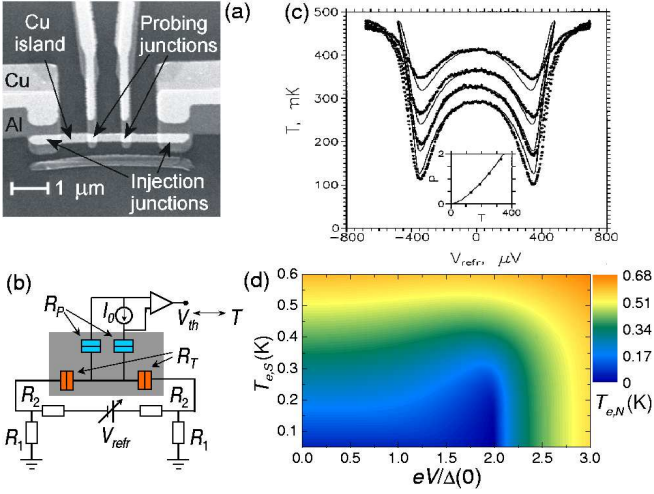}
\end{center}
\caption{SINIS refrigeration. (a) Scanning electron micrograph of a
typical SINIS microrefrigerator. The structure  was fabricated by
standard electron beam lithography combined with Al and Cu UHV
e-beam evaporation. The probe junction resistance usually satisfies
the condition $R_p\gg R_T$ in order to prevent self-cooling. (b)
Sketch of a typical measurement setup for electron cooling. (c)
Electron temperature in the N region versus $V_{refr}$ measured at
different bath temperatures. Dots represent experimental data, while
solid lines are theoretical fits. The inset shows the cooling power
(pW) against temperature obtained from the fits (dots). (d) Contour
plot of the theoretical electron temperature $T_{e,N}$ vs bias
voltage $V$ and $T_{e,S}$ for a SINIS cooler, assuming thermal load
due to the electron-phonon interaction (see text). (a) is adapted
from \cite{pekola:056804}; (c) from \cite{leivo:1996}.}
\label{fig:realsinis}
\end{figure}

Figure \ref{fig:realsinis}(a) shows the SEM micrograph of a
typical Al/Al$_2$O$_3$/Cu SINIS refrigerator including the
superconducting probe junctions. The schematic of a commonly used
experimental setup for electron refrigeration and temperature
measurement is shown in Fig. \ref{fig:realsinis}(b). The voltage
bias $V_{refr}$ across the SINIS structure allows to change the
electron temperature in the N region; at the same time, a measure
of the voltage drop  across the two probe junctions ($V_{th}$) at
a constant bias current ($I_0$)  yields the electron temperature
$T_{e,N}$ in the normal electrode \cite{rowell:2456}. Figure
\ref{fig:realsinis}(c) illustrates the experimental data of
\textcite{leivo:1996} of the measured electron temperature
$T\equiv T_{e,N}$ versus  $V_{refr}$, taken at different
bath temperatures (i.e., those at $V_{refr}=0$). As can be readily
seen, the electron temperature rapidly decreases by increasing the
voltage bias across the SINIS structure, reaching the lowest value
around $V_{refr}\approx 2\Delta \simeq 360~\mu$eV (note that now
two junctions in series are involved in the process). Then, by
further increasing the bias, the electron temperature rises due to
the heat flux induced into the metal. Furthermore, the minimum
electron temperature strongly depends on the starting bath
temperature. The inset shows the extracted maximum cooling power
(dots) that obtains values as high as $1.5$ pW at $T=300$ mK,
corresponding to about 2 pW$/\mu$m$^2$ (in the present experiment,
submicron  NIS junctions were exploited, with barrier resistances
$R_T\simeq 1~$k$\Omega$).

The general operation of a SINIS microrefrigerator like that shown
in Fig.  \ref{fig:realsinis}(a) can be understood by recalling
that the final electron temperature  in the N region stems from
the balance among several factors that tend to drive power into
the electron  system (i.e., power losses), as discussed in detail
in Sec. II. Most of metallic SINIS refrigerators fabricated so far
operate in a regime  where strong inelastic electron-electron
interaction  tends to drive the system into quasiequilibrium,
where the electron system can be described by a Fermi function at
a  temperature $T_{e,N}$ that differs, in general, from that of
the lattice ($T_{ph}$). At the cryogenic temperatures of interest
(i.e., typically below 1 K), the dominant contribution comes from
electron-phonon scattering that exchanges energy between electrons
and the lattice phonons. The refrigerator cooling power
($\dot{Q}_{refr}$) can be defined as the maximum power load the
(SI)NIS device can sustain while keeping the N region at
temperature $T_{e,N}$, and can be generally expressed as
\begin{equation}
\label{eq:ephbaleq}
\dot{Q}_{refr}=2\dot{Q}(V/2;T_{e,N},T_{e,S})-\dot{Q}_{e-ph}(T_{e,N},T_{ph}),
\end{equation}
where the factor 2 takes into account the presence of two
identical NIS junctions, and it is often assumed that in the
superconductors $T_{e,S}=T_{ph}$. The minimum temperature
$T_{e,N}$ reached by the N metal thus fulfills the condition
$\dot{Q}_{refr}=0$. An example of $T_{e,N}$ versus $V$ behavior
along the lines of Eq. (\ref{eq:ephbaleq}) is represented by the
solid curves plotted in Fig. \ref{fig:realsinis}(c). Furthermore,
any additional  term that adds thermal load into the electron
system can be properly taken into account by including its
specific contribution to the right-hand side of Eq.
(\ref{eq:ephbaleq}) \cite{fisher:2705,ullom:2036}. Figure
\ref{fig:realsinis} (d) shows an example of the calculated
electron temperature $T_{e,N}$ vs voltage and $T_{e,S}$ as
obtained from the solution of Eq. (\ref{eq:ephbaleq}),
$\dot{Q}_{refr}=0$. Here we considered a typical Al/Al$_2$O$_3$/Cu
SINIS cooler with volume $\mathcal{V}=0.15\,\mu$m$^3$, $R_T=1$
k$\Omega$, $\Sigma=2$ nW$/$$\mu$m$^{3}$K$^{5}$ (Cu), and
$T_{e,S}=T_{ph}$. As it can be readily seen, the $T_{e,N}$  value
strongly depends on the bias voltage as well as on the lattice
temperature. Depending on the thermal load due to the
electron-phonon interaction and the operating $V$, the SINIS
device can thus behave as a cooler or as a refrigerator.

Let us now highlight some of the basic requirements of SINIS
structures for \textit{electron} cooling operation (their use for
\textit{lattice} cooling is addressed in Sec. \ref{sec:extcool}).
In particular, the working temperature range of these structures
depends mainly on the type of the superconductor (through its
energy gap $\Delta$ and hence its critical temperature $T_c$), on
the strength of the electron-phonon interaction $\Sigma$, on the
junction resistance $R_T$, and on the N region volume
$\mathcal{V}$. First of all, a reduction of the active volume to
be cooled is the most straightforward method to increase the
efficiency of the refrigeration process, this being more relevant
at high lattice temperatures (according to the assumed
$T_{ph}^5$-dependence of electron-phonon interaction). In
addition, cooling power maximization in the \textit{
high-temperature}  regime (let us say $1... 4.2$ K) would require
both to lower the electron-phonon coupling constant and to
increase $\Delta$ through a proper choice of the superconductor.
The first issue can be solved up to some extent with a variety of
materials with different $\Sigma$ (see Table \ref{table:Sigma}).
As far as the second issue is concerned, there is a vast choice of
superconducting metals with $T_c$ covering the temperature range
up to about $20$ K \cite{kittel:0}.

\begin{figure}[tb!]
\begin{center}
\includegraphics[width=\columnwidth,clip]{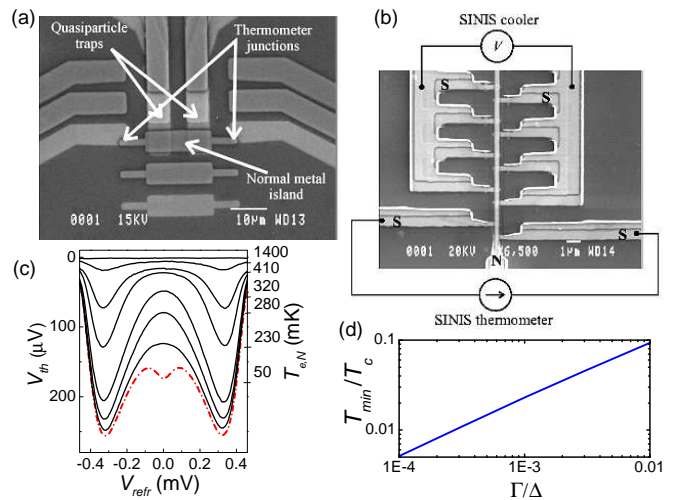}
\end{center}
\caption{Optimization of SINIS refrigerators. (a) SEM micrograph of
an Al/Al$_2$O$_3$/Cu SINIS microrefrigerator exploiting large-area
junctions ($\sim 10~\mu$m$^2$) with quasiparticle traps. (b) SEM
image of  a part of comb-like SINIS structure with $10+10$ junctions
for cooling and a SINIS thermometer. (c) Electron temperature
$T_{e,N}$ in the N region of an Al/Al$_2$O$_3$Cu SINIS refrigerator
versus $V_{refr}$ measured at different bath temperatures. The
lowest curve (red dash-dotted line) shows the anomalous heating
effect observable at the lowest temperatures and attributed to the
presence of quasiparticle states within the superconducting gap. (d)
Theoretical ultimate minimum electron temperature of a SINIS cooler
$T_{min}/T_c$ at $V\simeq 2~\Delta/e$ versus $\Gamma/\Delta$ in
quasiequilibrium. (a) is adapted from \cite{pekola:2782};  (b) from
\cite{luukanen:281}; (c) and (d) from \cite{pekola:056804}.}
\label{fig:traps}
\end{figure}
On the other hand, the question of junction resistance requires a
careful discussion. In general, from Eq. (\ref{eq:jQapprox}),
$\dot{Q}$ enhancement is expected from a reduction of $R_T$. This
latter issue can  be accomplished both by making thinner barriers
and by increasing the junction area. The first issue is not so
straightforward, as already discussed in Subs.~\ref{subs:suptuns},
due to the intrinsic difficulty in fabricating high-quality
low-$R_c$ barriers, although optimized barriers are currently
under investigation (see Sec.~\ref{subs:oxidebarriers}). Latter
option was experimentally addressed by Fisher \textit{et al.}
(1999) in Al/Al$_2$O$_3$/Ag refrigerators, where large cooling
powers of a few tens of pW were obtained with junctions of
$20\times 20~\mu$m$^2$ surface area. The reduction in electron
temperature was, however, much inferior to that achievable with
sub-micron sized junction. The problem intrinsic to junctions with
large overlap (especially at the lowest temperatures) stems from
the larger density of quasiparticles present in the
superconductor, due to the fact that quasiparticles require a
larger time to exit the junction region and escape from the
superconductor. Therefore, this excess of quasiparticles alters in
general the refrigerator performance by returning energy to the
normal electrode, mainly due to back-tunneling from the
superconductor as well as due to recombination processes, where
phonons can enter and heat up the N region
\cite{ullom:2036,jochum:3217,kaplan:4854}. In addition, inelastic
scattering with phonons and dynamic impurities can also lead to an
excess of quasiparticles. These contributions can easily
overcompensate the junction cooling power, so that it is crucial
to remove this excess of quasiparticles from the superconductor.
Toward this end a number of techniques exist among which we
mention the exploitation of defect-free and thick S electrodes
(that allow quasiparticles to escape \textit{ballistically} from
the junction area as well as to decrease their density near the
barrier) and the exploitation of quasiparticle "traps"
\cite{parmenter:274}, i.e., normal metal films connected to the
superconductor in the junction region through a tunnel barrier or
in direct metallic contact \cite{irwin:5322,pekola:2782} (see Fig.
\ref{fig:traps}(a)). Such traps act as sinks of quasiparticles absorbing almost
all the excess quasiparticle energy present in the superconductor,
and have proved to efficiently help heat removal from the
superconductor leading to a significant improvement of SINIS
device cooling performance
\cite{pekola:2782,pekola:056804,luukanen:281}. All these are
commonly exploited tricks for the thermalization of the
superconducting electrodes, and in these conditions, it is a
reasonable approximation to set $T_{e,S}=T_{ph}$. Furthermore, the
experiments \cite{pekola:2782} demonstrated that the trap
performance is in general superior when it is in direct metallic
contact at short distance from the cooling junction (typically
below $1~\mu$m, although the optimal distance mainly depends on
the superconductor coherence length), nevertheless in small-area
junctions even a contact through an insulating barrier seems
sufficient for the purpose. The effectiveness of trapping in SINIS
structures was theoretically addressed in detail  by
\textcite{voutilainen:054505} and \textcite{golubev:165}. Another
possibility to achieve high cooling by maximizing the ratio of the
junction area to the size of the region to be cooled is to use
several small-area junctions (with size in the sub-micron range)
connected in parallel to the N electrode (to limit the drawbacks
typical of large junctions)
\cite{luukanen:281,leoni:3877,arutyunov:326,manninen:3020}, as
shown in Fig. \ref{fig:traps}(b).

The issue of tunnel junction \emph{asymmetry} in SINIS
refrigerators was addressed by \textcite{pekola:485}. This effect
is fortunately weak: these authors theoretically showed that the
maximum cooling power is reduced by $7\%$ in the case the junction
resistances differ by a factor of two, as compared to a symmetric
structure with the same total junction area. Furthermore, the
reduction is only about $25\%$ even when the resistance ratio is
four. This effect stems from the "self aligning" character of the
double junction structure: the voltage drop across each junction
is simultaneously close to $\Delta/e$, thus corresponding to the
maximum cooling power, when the voltage across the whole SINIS
system is close to $2\Delta/e$. This fact is due to the high
non-linearity of the current-voltage characteristics of the two
junctions which carry the same current. The experiments have
confirmed such a weak dependence of the cooling power on the
structure asymmetry \cite{pekola:485}.

In the \textit{low-temperature} regime the situation is  rather
different. While power load from electron-phonon interaction
becomes less and less dominant by decreasing the temperature,  and
typically below $0.1$ K the cooler behavior can be described as if
the lattice would not exist at all, other factors can limit the
achievable cooling power as well as the lowest attainable electron
temperature. First of all, at the lowest temperatures the maximum
achievable cooling power of a NIS junction is intrinsically
limited, given by $\dot{Q}\propto (e^2R_T)^{-1}\Delta
^{1/2}(k_BT_{e,N})^{3/2}$. Then non-idealities in the tunnel
barriers, where the possible presence of Andreev-like transport
channels (e.g., pin-holes) may strongly degrade the cooling power
(in the high temperature regime this contribution is in general
overcome by thermal activation) \cite{bardas:12873}. Furthermore,
nonequilibrium effects in the N region as well in the S electrodes
may be a limitation. In the N region,  a suppression of the
cooling power  is expected by increasing the quasiparticle
relaxation time \cite{frank:281}, i.e., by driving the electron
gas far from equilibrium. In the superconductors, a
\textit{non-thermal} distribution can alter the cooling response
of the refrigeration process as well as the presence of hot
quasiparticle excitations (like with large-area junctions) may be
responsible of additional heat load into the N region. Yet,
quasiparticle states  within the superconducting gap represent a
crucial problem  at the lowest temperatures, yielding anomalous
heating in the N region and limiting the achievable minimum
temperature \cite{pekola:056804}. Such quasiparticle states, due
mainly to inelastic scattering in the superconductor
\cite{dynes:2437} or to inverse proximity effect from the nearby N
region, are generally taken into account by adding an imaginary
part ($\Gamma$) to the superconductor DOS, as  in Eq.
(\ref{eq:sdos}). Figure \ref{fig:traps}(c) shows a representative
set of cooling curves taken in an Al/Al$_2$O$_3$/Cu SINIS
refrigerator at different lattice temperatures, where at the
lowest temperature (red dash-dotted line) (this typically happens
around or below $T_0\approx 100$ mK) the electron gas first
undergoes \textit{heating}, then it is strongly cooled  at the
expected bias around $V_{refr}\simeq 2\Delta /e$. A similar
anomaly was reported in some experiments on SINIS refrigerators
\cite{pekola:056804,fisher:2705,savin:1471,pekola:485}. The
anomalous heating was attributed to the presence of such states
within the S gap that give rise to a sort of \textit{dissipative}
channel which dominates heat current in a certain bias range
\cite{pekola:056804}. \textcite{pekola:056804} theoretically
demonstrated that the minimum achievable electron temperature
($T_{min}$) in SINIS refrigerators is set by the amount of
quasiparticle states present within the superconducting gap,  and
is given by $T_{min}/T_c \simeq 2.5 (\Gamma/\Delta)^{2/3}$  at
$V_{refr}\simeq 2\Delta/e$ in quasiequilibrium (see Fig.
\ref{fig:traps}(d)). We note that a measure of the $\Gamma/\Delta$
value in real NIS junctions is approximately given  by the ratio
of the zero-bias to the normal state junction conductance at low
temperature. The existence of quasiparticle states within the
superconducting gap thus sets a \textit{fundamental} limit to the
minimum achievable temperature, and great care  has to be devoted
to get rid of their presence to optimize the refrigeration process
at the lowest temperatures.
\begin{figure}[tb!]
\begin{center}
\includegraphics[width=\columnwidth,clip]{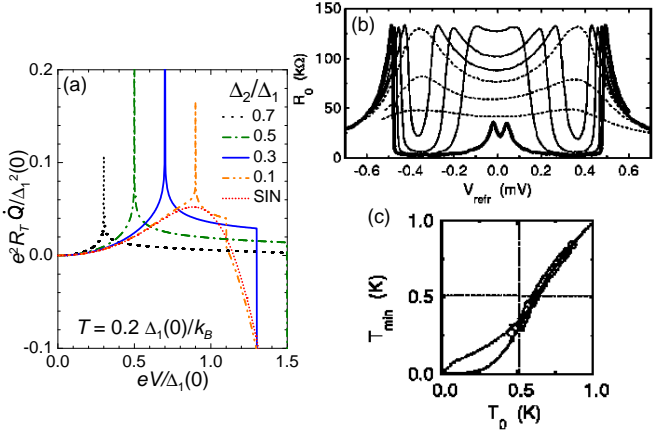}
\end{center}
\caption{S$_1$IS$_2$(IS$_1$) refrigerator. (a) Calculated cooling
power $\dot{Q}$ of a S$_1$IS$_2$ junction vs bias voltage $V$ at
$T=T_{e,S1}=T_{e,S2}=0.2~\Delta_{1}(0)/k_B$ for several
$\Delta_2/\Delta_1$ ratios. Red-dotted line represents $\dot{Q}$
when S$_2$ is in the normal state. (b) Measured cooling of a Ti
island to and in the superconducting state by quasiparticle
tunneling. $R_0$ is related to the electron temperature $T_{e,S2}$
in Ti (see text). Dashed lines: Ti is in the normal state. Thin
solid lines: Ti is cooled from the normal to the superconducting
state. Thick solid line: Ti is already superconducting at
$V_\textrm{refr}=0$. (c) Measured minimum electron temperature
($T_\textrm{min}$) of Ti versus bath temperature $T_0$. (b) and (c)
are adapted from \cite{manninen:3020}.} \label{fig:sisis}
\end{figure}

\subsubsection{S$_1$IS$_2$(IS$_1$) structures}
\label{sec:SIS}

The enhancement of superconductivity by quasiparticles extraction
was proposed in 1961 by \textcite{parmenter:274} in the context of
S$_1$IS$_2$ tunnel junctions, where S$_1$ and S$_2$ represent
superconductors with different energy gaps. Later on,
\textcite{melton:1858} theoretically discussed  the possibility to
exploit such a system to realize a refrigerator (addressing both
the basic features and performance). From an experimental point of
view, \textcite{chi:4465}  observed in Al films an enhancement of
the energy gap of the order of 40$\%$ due to quasiparticle
extraction. Then, \textcite{blamire:220}, using
Nb/AlO$_x$/Al/AlO$_x$/Nb structures, were able to obtain an
enhancement of the critical temperature of the Al layer by more
than $100\%$. The physical mechanism giving rise to this effect
was discussed by \textcite{heslinga:5157} and by
\textcite{zaitsev:45}. More recently, also
\textcite{nevirkovets:832} observed in similar structures an
enhancement of the Al gap by quasiparticle extraction.

Figure \ref{fig:sisis}(a) shows the calculated heat current versus
bias voltage for a S$_1$IS$_2$ tunnel junction at
$T=T_{e,S1}=T_{e,S2}=0.2~\Delta_{1}(0)/k_B$ for various
$\Delta_2/\Delta_1$ ratios (see Eq.~(\ref{eq:tunnelheatcurrent})).
The quantity $\dot{Q}(V)$ is an even function of $V$, thus
allowing connection of two junctions in a symmetric configuration
as in the NIS case, and is positive for
$|V|<(\Delta_1(T)+\Delta_2(T))/e$ where \textit{hot} quasiparticle
excitations are removed from S$_2$. Moreover, the heat current is
maximized at $V=\pm (\Delta_1(T)-\Delta_2(T))/e$, where it is
logarithmically diverging \cite{tinkham:96,harris:84} (note that
this is somewhat broadened by the smearing in the density of
states in a realistic situation
\cite{manninen:3020,pekola:056804,frank:281}). From Fig.
\ref{fig:sisis}(a) it follows that heat extraction from S$_2$ only
occurs if $\Delta_2(T)<\Delta_1(T)$ holds. Then, at $V=\pm
(\Delta_1(T)+\Delta_2(T))/e$ a sharp transition brings
$\dot{Q}(V)$ to negative values (more details about the heat
transport in S$_1$IS$_2$ junctions can be found in
\cite{frank:281}). The dotted line represents $\dot{Q}(V)$ when
the electrode S$_2$ is in the normal state (i.e., NIS case).
Notably, when both electrodes are in the superconducting state,
$\dot{Q}(V)$ can \textit{exceed} significantly that in the normal
state. This peculiar characteristic of heat transport in
S$_1$IS$_2$ junctions makes them promising in realizing a
"cascade" refrigerator that might operate at bath temperatures
higher than those for a SINIS cooler, where several combined
superconducting stages are used to efficiently cool a normal or a
superconducting region.

\begin{figure}[tb!]
\begin{center}
\includegraphics[width=\columnwidth,clip]{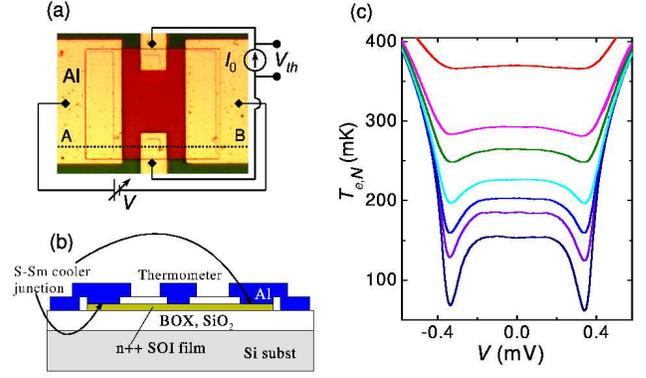}
\end{center}
\caption{Semiconductor-superconductor structures for electronic
cooling. (a) Optical micrograph of a typical Al/Si/Al
microrefrigerator and schematics of the measurement setup. The
current $I_{0}$ and voltage $V_{th}$ are used to determine the
electron temperature in the $n^{++}$-Si layer, and the bias voltage
$V$ is used to change its value. (b) Cross section of the structure
along the line AB in (a). (c) Measured electron temperature
$T_{e,N}$ in Si versus bias voltage across the Al/Si/Al cooler at
different substrate temperatures. Adapted from \cite{savin:1481}.}
\label{fig:ssms}
\end{figure}

The experimental observation of electron cooling in a
superconductor was reported by \textcite{manninen:3020} using
Al/insulator/Ti tunnel junctions. In this experiment, aluminum
(i.e, the larger-gap superconductor) was used to cool the electron
system of a Ti strip from $1.02~T_{c2}$ to below $0.7~T_{c2}$,
where $T_{c2}=0.51$ K was the Ti critical temperature. Figure
\ref{fig:sisis}(b) shows a representative set of measurements
taken at different bath temperatures $T_0$. In particular, it
shows the zero-bias resistance of the thermometer junctions $R_0$
against the bias voltage across the refrigerators
($V_\textrm{refr}$). Dashed lines indicate the  behavior at bath
temperatures where Ti is still in the normal state: the electron
temperature in Ti decreases (i.e., $R_0$ increases) by increasing
$V_\textrm{refr}$, reaching a minimum slightly below
$V_\textrm{refr}=2\Delta_1/e\simeq 420~\mu$eV, as expected for a
SINIS refrigerator ($\Delta_1$ is the energy gap of Al). Note that
dashed lines are the upside-down equivalent, for example, of the
measurements shown in Fig. \ref{fig:realsinis}(c). Thin solid
lines indicate the temperature behavior when Ti is cooled from the
normal to the superconducting state. This happens for $T_0$ larger
than $T_{c2}$ but below 625 mK, at  $V_\textrm{refr}$ values for
which $R_0$ has a deep minimum. Here, for instance, starting from
$T_0=520$ mK, the minimum electron temperature reached in Ti was
$T_\textrm{min}\approx 320\pm40$ mK, thus demonstrating the
effectiveness of the cooling mechanism also in all-superconducting
refrigerators. Thick solid line corresponds to the case where the
Ti strip is already superconducting, i.e., at a bath temperature
below $T_{c2}$. The results for the minimum electron temperature
reached in Ti against $T_0$ is represented by the open circles in
Fig. \ref{fig:sisis}(c). Comparison to the theory gave good
agreement with the experiment (lines in Fig. \ref{fig:sisis}(c))
with an electron-phonon interaction constant
$\Sigma_{\textrm{Ti}}\approx 1.3$ nW$/\mu$m$^3$K$^5$.

\subsubsection{SSmS structures}
\label{sec:SSmS}

Superconductor-semiconductor structures have recently attracted
interest  in the field of electron cooling
\cite{savin:1471,buonomo:7784,savin:1481}. The basic idea is to
exploit, for the active part of the cooler, heavily-doped
semiconductor layers instead of normal metals, with the natural
Schottky barrier forming at the contact to the metal electrodes
\cite{brennan:0,luth:0,rhoderick:0,sze:0,sze:1}. The peculiar
characteristic intrinsic to this scheme stems from the possibility
to alter up to a large extent the semiconductor electronic
properties (like, for example, the  charge density, etc.) and to
change the transmissivity of the Schottky barrier through proper
doping or choice of the semiconducting layer (see also Sec.
VI.F.2).

Heavily-doped silicon is a natural choice both from a practical
and technological point of view, mainly owing to the widespread
and well-developed Si technology. Up to now, the only evidence of
superconducting electron cooling with semiconductors comes from
microrefrigerators realized with the silicon-on-insulator (SOI)
technology \cite{savin:1471,buonomo:7784,savin:1481}. The first
demonstration of electron refrigeration in Si was reported by
\textcite{savin:1471}, where the authors obtained a maximum
temperature reduction under hot quasiparticle extraction of the
order of $30\%$ at the bath temperature $T_0=175$ mK. Figure
\ref{fig:ssms}(a) shows the optical micrograph of a typical
Al/Si/Al cooler. The two bigger Al electrodes are used for
cooling, while the two smaller for detecting the electronic
temperature in Si. A sketch of the structure cross-section is
displayed in Fig. \ref{fig:ssms}(b), where the $n^{++}$-Si region
appears just on top of the silicon dioxide insulating layer. The
structures were characterized by a doping level $N_D=4\times
10^{19}$ cm$^{-3}$. A set of cooling curves extracted from one of
such devices is shown in Fig. \ref{fig:ssms}(c) \cite{savin:1481}.
It displays the electron temperature $T_{e,N}$ in the $n^{++}$-SOI
layer versus bias voltage across the refrigerator at different
starting bath temperatures. The observed electronic temperature
reduction  is of the order of $60\%$ at $T_0\approx 150$ mK.
Although the device volume and the cooler resistances were rather
high, the authors attributed this significant cooling to a very
small electron-phonon coupling constant in Si: as a matter of
fact, the latter was determined to be $\Sigma_{\textrm{Si}}\simeq
0.1$ nW$/\mu$m$^3$K$^5$ \cite{savin:1471}, i.e., about one order
of magnitude smaller than in Cu. The maximum achieved cooling
power  was $\dot{Q}_{max}\approx 1.3$~pW at $T_0=175$ mK, mainly
limited by the high value of the specific contact resistances
($R_c\sim 67.5$ k$\Omega\mu$m$^2$). These authors also addressed
the effect of carrier concentration in Si on cooling performance
\cite{savin:1481}.

Buonomo et \textit{al.} \cite{buonomo:7784} also obtained a
cooling effect in Al/Si/Al refrigerators, with a device
configuration similar to that of Fig. \ref{fig:ssms}(a). Their
structures were characterized by a doping level $N_D=8\times
10^{18}$ cm$^{-3}$ and contact specific resistances $R_c\sim 100$
k$\Omega\mu$m$^2$. Notably, they confirmed the same value reported
by Savin \textit{et al.} (2001) for the electron-phonon coupling
constant in Si. In addition, these authors also explored  the
Nb/Si/Nb combination for electron cooling. Their results, however,
showed no cooling effect, owing probably to a too transmissive
interface at the contact with Si. This effect was ascribed both to
the slightly lower Schottky barrier height of Nb/Si contact ($\sim
0.5... 0.6$~eV \cite{heslinga:1048,chang:541}) with respect to the
Al/Si ($\sim 0.6... 0.7$~eV \cite{chang:541,singh:0}) and to
disorder-enhanced subgap conductance
\cite{badolato:9831,kastalsky:3026,vanwees:510,giazotto:1772a,giazotto:216808,nguyen:2847,nitta:3659,xiong:1907,poirier:2105,kutchinsky:931,magnee:4594,bakker:13275,giazotto:955}.
Further results on the electron-phonon coupling constant as well
as on the electronic thermal conductivity in $n$-type Si were
reported in \cite{kivinen:3201,heslinga:739}.

Because of the low electron-phonon coupling constant in heavily
doped Si, one may expect to cool islands with larger volumes than
in the normal metal case, and perhaps improve the high-temperature
limit of cooling by some factor. The issue of high $R_c$ values
that limit the maximum achievable cooling power still represents a
drawback of these refrigerators that has to be overcome. High
doping of the semiconductor (in the tunneling regime $R_c\propto
\textrm{exp}(N_D^{-1/2})$  \cite{sze:0,sze:1}, see also Sec.
VI.F.2) as well as \textit{engineering} of the Schottky barrier
height through incorporation of interface layers at the
metal-semiconductor junction
\cite{defranceschi:817,defranceschi:3890,defranceschi:1196,grant:1794,costa:382,koyanagi:502,cantile:988,cantile:2653,marinelli:2119}
can be exploited in order to tailor the interface transmissivity.

\subsubsection{SF systems}
\label{sec:FS}

As discussed at the beginning of Sec. \ref{sec:SINIS}, decreasing
the contact resistance ($R_T$) is not a viable route to enhance
heat current in NIS junctions. A possible  scheme to  surmount the
problem of Andreev reflection at the metal-superconductor
interface is to exploit, instead of an insulating  barrier, a thin
ferromagnetic (F) layer in good electric contact with S
\cite{giazotto:3784}. The physical origin of the SF cooler
operation stems from the spin-band splitting characteristic of a
ferromagnet. The electron involved in the Andreev reflection and
its phase-matched hole belong to opposite spin bands; as a
consequence, depending on the degree of spin polarization
$\mathcal{P}$ of the F layer, strong suppression of the Andreev
current may occur at the SF interface \cite{dejong:1657}. In the
limit of large $\mathcal{P}$, the subgap  current is thus strongly
suppressed while efficient hot-carrier transfer leads to a
sizeable heat current across the system. In the following we give
the main results of such a proposal.

\begin{figure}[tb!]
\begin{center}
\includegraphics[width=\columnwidth,clip]{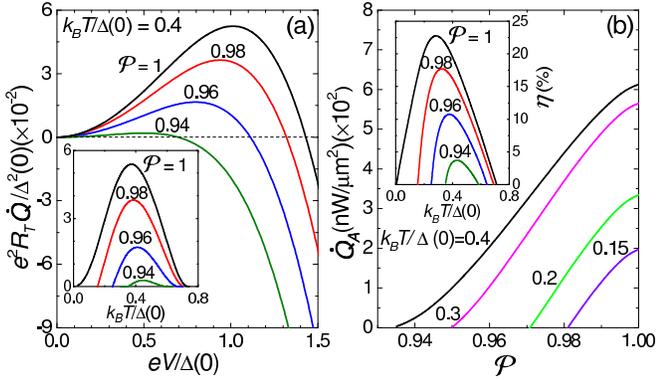}
\end{center}
\caption{Superconductor-ferromagnet refrigerator. (a) Calculated
heat current $\dot{Q}$ of a SF junction vs bias voltage $V$ at
$T=T_{e,F}=T_{e,S}=0.4~\Delta(0)/k_B$ for several  spin
polarizations $\mathcal{P}$. The inset shows the same quantity
calculated at the optimal bias voltage against temperature for some
values of $\mathcal{P}$. (b) Calculated maximum cooling power
surface density $\dot{Q}_{A}$ versus  $\mathcal{P}$ for various
temperatures. The inset shows the coefficient of performance $\eta$
calculated at the optimal bias voltage versus temperature for some
$\mathcal{P}$ values. Adapted from
\cite{giazotto:3784,giazotto:japan}.} \label{fig:fsjunction}
\end{figure}

The impact of partial spin polarization ($\mathcal{P}<100\%$) in
the F region is displayed in Fig.~\ref{fig:fsjunction}(a) where the
heat current $\dot{Q}$ versus bias voltage across the junction is
plotted for some values of $\mathcal{P}$ at
$T=T_{e,F}=T_{e,S}=0.4~\Delta(0)/k_B$ ($T_{e,F}$ is the electron
temperature in F). For each $\mathcal{P}$ value there exists an
optimum bias voltage ($V_{opt}$) which maximizes $\dot{Q}$. In the
limit of a half-metal ferromagnet (i.e., $\mathcal{P}=100\%$)
\cite{mazin:1427,coey:988} this value is around $V_{opt}\simeq
\Delta/e$. Moreover, for $\mathcal{P}=94\%$ there is still a
positive $\dot{Q}$ across the junction. The inset of Fig.
\ref{fig:fsjunction} (a) shows $\dot{Q}$ calculated at each
optimized bias voltage against temperature for various values of
$\mathcal{P}$. The heat current is maximized around
$T=T_{opt}\approx0.4~\Delta(0)/k_B$, rapidly decreasing both at
higher and lower temperatures.

The junction specific cooling power is shown in Fig.
\ref{fig:fsjunction}(b), where $\dot{Q}_{A}$ (evaluated at each
optimal bias voltage) is plotted versus $\mathcal{P}$ at different
bath temperatures. Notably, for a half-metallic ferromagnet at
$T=0.4~\Delta(0)/k_B$, cooling power surface densities  as high as
$600$ nW/$\mu$m$^2$ can be achieved, i.e., about a factor of 30
larger than those achievable in NIS junctions at the optimized
interface transmissivity (i.e., $\mathcal{T}\simeq 3\times
10^{-2}$ at $T\simeq 0.3\Delta(0)/k_B$, see Fig.
\ref{fig:coolprinciple}(b)). This marked difference stems from SF
specific normal-state contact resistances as low as some
$10^{-3}~\Omega\mu$m$^2$ that are currently achieved in
highly-transmissive junctions \cite{upadhyay:3247}. The inset of
Fig. \ref{fig:fsjunction}(b) shows the junction coefficient of
performance ($\eta$) calculated at the optimal bias voltage versus
temperature  for various $\mathcal{P}$ values. Notably, for
$\mathcal{P}=100\%$, $\eta$ reaches $\sim 23\%$ around
$T=0.3~\Delta(0)/k_B$ and exceeds $10\%$ for $\mathcal{P}=96\%$.
In addition, in light of a possible exploitation of this structure
in combination with a N region (i.e., a SFN refrigerator) it
turned out that cooling effects comparable to the SF case can be
achieved with a F-layer thickness of  a few nm (i.e., of the order
of the magnetic lenght). We note that the large $\dot{Q}_{A}$
typical of the  SF combination makes it promising as a possible
higher-temperature first stage in cascade cooling  (for instance,
over or around 1 K), where it could  dominate over the  large
thermal coupling to the lattice characteristic for such temperatures.

The results given above point out the necessity of strongly
spin-polarized ferromagnets for a proper operation of the SF
refrigerator. Among the predicted half-metallic candidates it is
possible to indicate CrO$_2$
\cite{kamper:2788,brener:16582,schwartz:L211}, for which values of
$\mathcal{P}$ in the range $85... 100\%$  have been reported
\cite{ji:5585,parker:196601,dedkov:4181,coey:3815},
(Co$_{1-x}$Fe$_{x})$S$_2$ \cite{mazin:3000}, NiMnSb
\cite{degroot:2024}, Sr$_2$FeMoO$_6$ \cite{kobayashi:677} and
NiMnV$_2$ \cite{weht:11006}. So far no experimental realizations
of SF structures for cooling applications have been reported.

\subsubsection{HT$_c$ NIS systems}
\label{sec:HTC}

Heat transport in high-critical temperature (HT$_c$) NIS junctions
was theoretically addressed by \textcite{devyatov:1050}. In these
systems the cooling power depends on interface transmissivity as
well as on orientation of the superconductor crystal axes and
temperature. In particular, these authors showed that the maximum
positive heat current in these structures can be achieved in
junctions with zero superconducting crystallographic angle, at
temperature $T=0.45~T_c$ and bias voltage $V\approx
0.8~\Delta(0)/e$. The behavior of $\dot{Q}(V)$ turns out to be
qualitatively similar to that of NIS junctions based on
low-critical temperature superconductors (see Fig.
\ref{fig:niscoolp}(a)), and with comparable values (in relative
units). From this it follows that the cooling power of electronic
refrigerators based on HT$_c$ materials is approximately two
orders of magnitude larger than in NIS junctions based on
low-critical temperature superconductors (at much lower
temperatures).

A somewhat different cooling effect in HT$_c$ superconductors was
predicted by \textcite{svidzinsky:144504}, who showed that an
adiabatic increase of the supercurrent  in a ring (or cylinder)
made from a HT$_c$ superconductor may lead to a cooling effect.
The maximum cooling occurs if the supercurrent is equal to its
critical value. For a clean HT$_c$ superconductor, the minimum
achievable temperature ($T_{min}$) was found to be around
$T_{min}=T_0^2/T_c$, with $T_0$ the initial temperature of the
ring, thus meaning that substantial cooling can be achieved using
large $T_c$ values.

Experimentally, \textcite{fee:1161} realized a Peltier
refrigerator junction exploiting a HT$_c$ superconductor and
operating around liquid nitrogen temperatures. In particular, his
device consisted of a BiSb  alloy and YBa$_2$Cu$_3$O$_{7-\delta}$
superconducting rods connected by a small copper plate which acted
as the cold junction of the device. The latter showed a maximum
cooling of $5.35$ K below the bath temperature $T_0=79$ K. The
figure of merit of the junction, $Z$, was estimated to be as large
as $2.0\times 10^{-3}$ K$^{-1}$.

\subsubsection{Application of (SI)NIS structures to lattice refrigeration}
\label{sec:extcool}

One  well established application of SINIS refrigerators concerns
lattice cooling \cite{manninen:1885,luukanen:281,clark:173508}. As
a matter of fact, while NIS tunneling directly cools  the electron
gas of the normal electrode, the phonon system can be refrigerated
through the electron-phonon coupling (see Eq.
(\ref{eq:phononpower})). The latter, however, is typically very small at the
lowest temperatures, thus limiting the heat transfer from the
surroundings to the electrons. This situation normally happens
whenever the metal to be cooled is in direct contact with a thick
substrate, for instance, oxidized Si \cite{nahum:3123,leivo:1996}:
only the electrons of the N region cool down while the metal
lattice presumably remains at the substrate  temperature. The
metal lattice can be refrigerated considerably if the thermal
resistance between the phonons and the substrate  is not
negligible compared to that between the electrons and the phonons.
One effective choice to meet this requirement, that was indeed
suggested at the beginning of cooling experiments
\cite{nahum:3123,fisher:561} as well as for detector applications
\cite{irwin:1945,nahum:3203,doriese:4762,ullom:4206,deiker:2137,pekola:41},
is to exploit a thermally isolated thin dielectric
\textit{membrane} on which the N region of the cooler is extended.
In this way, tunneling through the NIS junction will cool down the
electrons of the metal, then the phonons of the metal (via
electron-phonon coupling) that subsequently will refrigerate the
membrane phonons \cite{manninen:1885,luukanen:281,clark:173508}
according to
\begin{equation}
\label{eq:subphbaleq}
\dot{Q}_{\text{SINIS}}(V;T_{e,N},T_{e,S})+\dot{Q}_{ph-sub}(T_{ph},T_{sub})=0,
\end{equation}
where $T_{e,N}\simeq T_{ph}$, $T_{e,S}\simeq T_{sub}$, $T_{ph}$ is the lattice
temperature in the dielectric membrane, $T_{sub}$ is the lattice
temperature in the substrate, and $\dot{Q}_{ph-sub}$ is the rate
of exchanged energy between the membrane phonons and substrate
phonons. Eventually, if additional devices are standing on the
same dielectric platform (for instance, detectors, etc.), the
latter will cool down  first the phonons of the device and then
its electrons through the electron-phonon interaction.

\begin{figure}[tb!]
\begin{center}
\includegraphics[width=\columnwidth,clip]{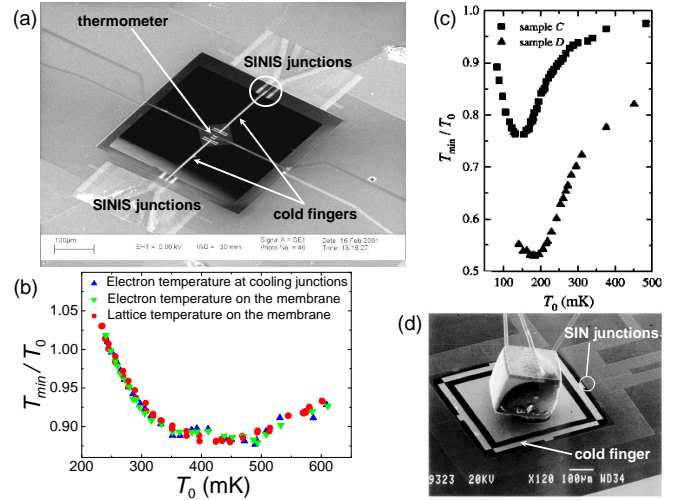}
\end{center}
\caption{Lattice refrigeration with SINIS. (a) SEM image of a
Si$_3$N$_4$ membrane (in the center) with self-suspended bridges.
Two normal-metal cold fingers extending onto the membrane are used
to cool down the dielectric platform. The Al/Al$_2$O$_3$/Cu SINIS
coolers are on the bulk (far down and top) and the thermometer
stands in the middle of the membrane. (b) Maximum temperature
decrease ($T_{min}/T_0$) versus bath temperature $T_0$ measured at
different positions in the cooler  shown in (a). (c) Maximum
\textit{lattice} temperature decrease on the membrane versus bath
temperature measured in two other samples similar to that shown in
(a). In this case, the refrigerators exploited many small-area
junctions arranged in parallel in a comb-like configuration. (d) SEM
micrograph of a NIS refrigerator device with a neutron transmutation
doped (NTD) Ge resistance thermometer attached on top of it. (c) is
adapted from \cite{luukanen:281}, (d) from \cite{clark:173508}.}
\label{fig:membrane}
\end{figure}

Dielectric membranes made of silicon nitride (Si$_3$N$_4$) have
proved to be attractive for this purpose in light of their
superior thermal isolation properties at low temperatures.
Low-temperature heat transport characterization as well as thermal
relaxation in low-stress Si$_3$N$_4$ membranes and films were
quite recently addressed \cite{leivo:1305,holmes:2250}. The first
demonstration reported of lattice cooling \cite{manninen:1885}
exploited such membranes  in combination with Al/Al$_2$O$_3$/Cu
SINIS refrigerators. In this experiment the authors were able to
achieve a $2\%$ temperature decrease in the membrane at bath
temperatures  $T_0\approx 200$ mK.

Figure \ref{fig:membrane}(a) shows a SEM image of a typical
new-generation lattice cooler fabricated on a Si$_3$N$_4$ membrane
with self-suspended bridges. The membrane consists of a low-stress
Si$_3$N$_4$ film deposited by low pressure chemical vapor
deposition (LPCVD) on Si,  and subsequently etched (with both wet
and dry etching)  in order to create the suspended bridge
structure. In such thin membranes phonon propagation is
essentially two-dimensional. The condensation of the phonon gas
into lower dimensions in ultrathin membranes was also
theoretically discussed
\cite{anghel:2958,anghel:9854,kuhn:125425}. The self-suspended
bridges improve thermal isolation of the dielectric platform from
the heat bath \cite{leivo:1305}. The image also shows the
Al/Al$_2$O$_3$/Cu refrigerators of the SINIS type that are placed
on the bulk substrate (i.e., outside the membrane) to ensure good
thermal contact with the bath. The N  cold fingers extend onto the
silicon nitride membrane, whose temperature is determined through
an additional SINIS thermometer placed in the middle of the
structure.

The lattice refrigeration effect achieved in this SINIS
refrigerator is shown in Fig.~\ref{fig:membrane}(b). Here the
maximum temperature decrease of the membrane  ($T_{min}/T_0$)
against bath temperature $T_0$ is displayed (red circles), and
shows that temperature reduction as high as about $12\%$  was
reached in the $400... 500$ mK range. The electron refrigeration
effect in the Cu region was also measured at two different
positions in the device, i.e., nearby the cooling junctions and on
the membrane. Notably, the $T_{min}/T_0$ behavior is almost the
same for the different sets of data; this basically means that: a)
good thermalization was achieved in the cold fingers; b) the
electron-phonon coupling was sufficiently large while Kapitza
resistance between Cu and Si$_3$N$_4$ was sufficiently small to
ensure the lattice temperature on the membrane to be nearly equal
to the Cu electron temperature on the membrane itself. The best
results of lattice refrigeration by SINIS coolers reported to date
are shown in Fig.~\ref{fig:membrane}(c) \cite{luukanen:281} for
two other devices (labeled C and D in the figure) similar to that
of Fig.~\ref{fig:membrane}(a). These devices exploited three Cu
cold fingers and several small-area junctions arranged in parallel
in a comb-like configuration for the SINIS cooling stage. The
junction specific resistances were $R_c=1.39$ k$\Omega\mu$m$^2$
and $R_c=220$ $\Omega\mu$m$^2$ for device C and D, respectively.
Lattice temperature reduction as high as $50\%$ at 200 mK was
achieved in the sample with lower $R_c$, thus confirming the
effectiveness of small-area junctions in yielding larger
temperature reductions.  The achieved cooling power in these
devices was estimated on the pW level. The reduction of the
refrigeration effect at the lowest temperatures can be explained
in terms of larger decoupling of electrons and phonons, but also
the effects discussed in Sec.~\ref{sec:SINIS} should play a role.

Figure \ref{fig:membrane} (d) demonstrates the realization of a
complete refrigerator device including a thermometer
\cite{clark:173508}, where four pairs of NIS junctions are used to
cool down a $450\times450\,\mu$m$^2$ suspended Si$_3$N$_4$
dielectric membrane. Each NIS junction area is $25\times
15\,\mu$m$^2$, and the N electrode consists of Al doped with Mn to
suppress superconductivity while Al is used for the
superconducting reservoirs. Also shown is a neutron transmutation
doped (NTD) Ge resistance thermometer (with volume $250\times 250
\times 250\,\mu$m$^3$) glued on the membrane. In such a
refrigerator the authors measured with the NTD Ge sensor a minimum
final temperature slightly below $240$ mK  starting from a bath
temperature $T_0=320$ mK, under optimal bias voltage across the
cooling junctions. This result is promising in light of a
realistic implementation of superconducting refrigerators, and
shows the possibility of cooling efficiently the whole content of
dielectric membranes through NIS junctions \cite{pekola:889}.

Possible strategies to attain enhanced lattice refrigeration
performance in the low temperature regime (i.e., below 500 mK)
could be a careful optimization in terms of the number and
specific resistance of the cooling junctions as well as to exploit
superconductors with the gap larger than in Al. Making the
dielectric platform thinner and reducing thermal conduction along
it should also increase the temperature drop across the membrane.
The exploitation of the described method around or above 1 K,
however, is still now not so straightforward, mainly due to the
strong temperature dependence of the electron-phonon interaction
that thermally shunts more effectively the N electrode portion
standing on the bulk substrate to the lattice (note that also the
thermal conduction along the membrane is strongly  temperature
dependent \cite{leivo:1305,kuhn:125425}). Toward this end, a
reduction of the N volume placed out of the membrane should help;
in addition, S$_1$IS$_2$(IS$_1$)  refrigerators (see Sec.
\ref{sec:SIS}) as well as SF junctions (see Sec. \ref{sec:FS})
might significantly improve the membrane cooling in the higher
temperature regime.

\subsubsection{Josephson transistors}
\label{sec:suptrans}

A further interesting field of application of SINIS structures
concerns superconducting weak links
\cite{likharev:101,golubov:411}. In particular, in
\textit{diffusive} SNS junctions, i.e.,  where the junction length
largely exceeds the elastic mean free path, coherent sequential
Andreev scattering between the superconducting electrodes may give
rise to a continuum spectrum of resonant levels
\cite{heikkila:184513,belzig:1251,yip:5803} responsible for
carrying the Josephson current across the structure. The
supercurrent turns out to be given by this spectrum weighted by
the occupation number of correlated electron-hole pairs,  the
latter being determined by the quasiparticle energy distribution
in the N region of the weak link. In \textit{controllable}
Josephson junctions, the supercurrent is  modulated by driving the
quasiparticle distribution out of equilibrium
\cite{vanwees:470,wilhelm:1682,yip:5803,heikkila:184513,volkov:4730,seviour:R9273,volkov:670,volkov:11184}
via dissipative current injection in the weak link from additional
normal-metal terminals. This operation principle leads to a
controlled supercurrent suppression and was successfully exploited
both in all-metal
\cite{morpurgo:966,baselmans:43,shaikhaidarov:R14649,huang:020507,baselmans:207002,baselmans:224513,baselmans:2940,baselmans:094504}
(where a  transition to a $\pi$-state was also reported) as well
as in hybrid semiconductor-superconductor weak links
\cite{kutchinsky:4856,schapers:2348,neurohr:11197,schapers:060502,schapers:014522,richter:321}.
The situation drastically changes if we allow current injection
from additional \textit{superconducting}  terminals arranged in a
SINIS fashion
\cite{baselmans:th,giazotto:2877,giazotto:137001,giazotto:435}. In
this way, thanks to the SINIS junctions, critical supercurrent can
be strongly increased  as well as steeply suppressed with respect
to equilibrium, leading to a tunable structure with large current
and power gain.

\begin{figure}[tb!]
\begin{center}
\includegraphics[width=\columnwidth,clip]{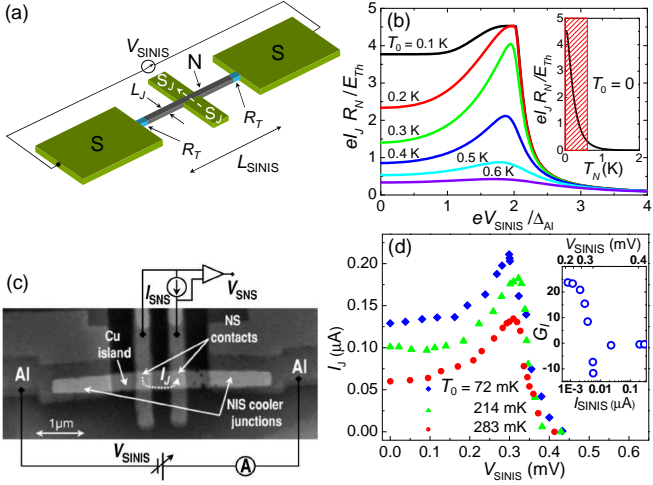}
\end{center}
\caption{Josephson transistor. (a) Simplified scheme of a
SINIS-controlled Josephson transistor. The Josephson current in the
S$_J$NS$_J$ weak link (along the white dashed line) is controlled by
applying a bias $V_{\text{SINIS}}$ across the SINIS line connected
to the weak link, allowing to increase or decrease its  magnitude
with respect to equilibrium. (b) Theoretical normalized critical
current $I_J$ versus $V_{\text{SINIS}}$ calculated in the
quasiequilibrium limit for several lattice temperatures $T_0$ for a
Nb/Cu/Nb \textit{long} junction. The inset shows the supercurrent vs
electron temperature characteristic calculated at $\phi=\pi/2$. (c)
SEM image of an Al/Cu/Al Josephson junction including the
Al/Al$_2$O$_3$/Cu symmetric SINIS electron cooler. The SNS long weak
link is placed in the middle of the structure. Also shown is a
scheme of the measurement circuit. (d) Measured critical current
$I_J$ versus $V_{\text{SINIS}}$ at three different bath temperatures
$T_0$ for the device shown in (c). The inset displays the measured
differential current gain $G_I$ versus $I_{\text{SINIS}}$ at
$T_0=72$ mK. (b) is adapted from \cite{giazotto:2877}, while (c) and
(d) from \cite{savin:4179}.} \label{fig:sinistrans}
\end{figure}
A simplified scheme of this class of transistors is displayed in
Fig. \ref{fig:sinistrans}(a). A diffusive S$_J$NS$_J$
\textit{long} Josephson junction of length $L_J$ (i.e., with
$L_J\gg \xi_0$, where $\xi_0$ is the $S_J$ coherence length)
shares  the N region of a SINIS control line of length
$L_{\text{SINIS}}$. The superconductors S$_J $ and S can be in
general different (i.e., with gaps $\Delta_J$ and $\Delta_S$,
respectively); in addition, the S$_J$ electrodes are kept at zero
potential, while the SINIS line is biased with $V_{\text{SINIS}}$.
The supercurrent ($I_J$) behavior  in response to a bias voltage
$V_{\text{SINIS}}$ stems from the degree of nonequilibrium the
SINIS line is able to generate in the weak link according to
\cite{wilhelm:1682,yip:5803,heikkila:184513}
\begin{equation}
\label{eq:supercurrent}
I_J(V_{\text{SINIS}})=\frac{1}{eR_N}\int_0^{\infty}dEj_S(E,\phi)(1-2f(E,V_{\text{SINIS}})),
\end{equation}
where $R_N$ is the weak link normal-state resistance, $\phi$ is
the phase difference across the superconductors, and $j_S(E,\phi)$
is the spectral supercurrent, obtainable from the solution of the
Usadel equations \cite{usadel:507}. The most straightforward
situation occurs at sufficiently low lattice temperatures, i.e.,
typically below $1$ K, when
$\ell_{e-e}<L_{\text{SINIS}}<\ell_{e-ph}$ (see Sec. II). In such a
case, strong electron-electron relaxation forces the electron
system in the N region to retain a local thermal quasiequilibrium
(see also Sec. \ref{sec:SINIS}), so that
$1-2f(E,V_{\text{SINIS}})=\text{tanh}[E/(2k_BT_{e,N}(V_{\text{SINIS}}))]$.
The transistor effect in the structure thus simply depends  on the
electron temperature $T_{e,N}(V_{\text{SINIS}})$ established in
the weak link upon biasing the SINIS line according to Eq.
(\ref{eq:ephbaleq}). The full calculation of the  behavior of a
prototype Nb/Cu/Nb long Josephson junction with integrated
Al/Al$_2$O$_3$/Cu SINIS electron cooler is displayed in Fig.
\ref{fig:sinistrans}(b) \cite{giazotto:2877}. Here the normalized
$I_J$ is plotted against  $V_{\text{SINIS}}$ for several bath
temperatures $T_0$. For all chosen values of $T_0$, the
supercurrent value increases monotonically up to about
$V_{\text{SINIS}}\approx 1.8~\Delta_{S}/e$, where the SINIS line
provides the largest cooling power and allows to attain the lowest
electron temperature. Then, by further increasing the voltage, an
efficient suppression of $I_J$ occurs due to electron heating: the
device behaves as a tunable superconducting junction. The above
given results can be easily understood recalling that for a
\textit{long} SNS junction at low lattice temperature (i.e.,
$k_BT_0\ll\Delta_J$) and for $k_BT_{e,N}\gg E_{Th}=\hbar D/L_J^2$
($D$ is the N-region diffusion coefficient), $I_J$ depends
exponentially on the effective electron temperature $T_{e,N}$
\cite{zaikin:184,wilhelm:305} and is almost independent of $T_0$.
Hence, long junctions are more appropriate for the device to
obtain large $I_J$ changes with small $T_{e,N}$ variations (as
indicated by the red hatched region in the inset of Fig.
\ref{fig:sinistrans}(b)). In the \textit{short} junction limit
($L_J\ll \xi_0$), conversely, the $I_J$ temperature dependence is
set by the energy gap $\Delta_J$, thus implying a much reduced
effect from the cooling line. Furthermore, it is expected that the
power dissipation values from two to four order of magnitude
smaller than with all-normal control lines can be obtained in
these structures, thus making them promising as mesoscopic
transistors for low dissipation cryogenic applications.

So far, the only successful demonstration of this transistor-like
operation was reported by \cite{savin:4179} in Al/Cu/Al SNS
junctions integrated with Al/Al$_2$O$_3$/Cu SINIS refrigerators.
The SEM image of one of these samples  along with a scheme of the
measurement setup is shown in Fig. \ref{fig:sinistrans}(c). The
structure parameters were the following: cooler junction
resistances $R_T\simeq 240~\Omega$, Josephson weak link normal
state resistance $R_N=11.5~\Omega$, and minimum SNS interelectrode
separation $L_J\simeq0.4~\mu$m that compared with the
superconducting coherence length ($\xi_0\approx62$ nm) provides
the frame of the \textit{long} junction regime. The critical
current $I_J$ versus $V_{\text{SINIS}}$ at three different $T_0$
is shown in Fig. \ref{fig:sinistrans}(d). For each bath
temperature, $I_J$ increases around $V_{\text{SINIS}}\approx
1.8\Delta_{\text{S}}/e$, as expected from the reduction of
$T_{e,N}$ by electron cooling, while it is steeply suppressed at
larger bias voltages. The resemblance of the experiment with the
curves of Fig. \ref{fig:sinistrans}(b) is evident. In the present
case, $I_J$ enhancement  under hot quasiparticle extraction by
more than a factor of two was observed at $T_0=283$ mK. The
transistor current gain $\textbf{G}_I=dI_J/dI_{\text{SINIS}}$,
shown in the inset of Fig. \ref{fig:sinistrans}(d), obtained
values in the range $-11... 20$  depending on the control bias. As
far as  power dissipation is concerned, these authors reported low
dissipated power on the $10^{-13}$ W level in the extraction
regime while of some tens of pW in the regime of $I_J$
suppression.

The transistor behavior for arbitrary inelastic scattering
strength in the SINIS line (or, equivalently, for arbitrary values
of $L_{\text{SINIS}}$) as well as for different SNS junction
lengths, was theoretically addressed in detail in
\cite{giazotto:137001,giazotto:435}. The role of geometry,
materials combination, phase dependence, and the input noise power
were also discussed. Notably, a marked supercurrent transition to
a $\pi$-state under nonequilibrium (about two times larger than
that achievable with an all-normal control channel
\cite{wilhelm:1682,yip:5803}) was predicted to occur for
negligible or moderate electron-electron interaction.
We recall that in the $\pi$-state \cite{bulaevskii:314} the supercurrent flows in  opposite direction with respect to the phase difference $\phi$ between the two superconductors, i.e., such a sign reversal is equivalent to the addition of a phase factor $\pi$ to the Josephson current-phase relation.
Furthermore,
current gain in the range $10^2...10^5$ and power gain up to
$10^3$ where predicted to occur depending on the control voltage.

Finally, transistor operation in a Josephson \textit{tunnel} junction integrated with a S$_1$IS$_2$IS$_1$ refrigerator was also theoretically addressed in the  quasiequilibrium regime \cite{giazotto:023908}.
In this case, the device benefits from the sharp characteristics due to the presence of superconductors with unequal energy gaps (see also \ref{sec:SIS}), that leads to improved overall characteristics as compared to the SINIS-controlled SNS junction in the same transport regime.

\subsection{Perspective types of refrigerators}

Any electric current that is accompanied by the extraction of hot
electrons (or holes) can be used, in principle, for refrigeration
purposes. For instance, this may happen in \emph{thermionic}
transport over a potential barrier as well as in energy-dependent
\emph{tunneling} through a barrier. Since an exhaustive analysis
of all possible predicted refrigeration methods  is far beyond the
limits set to this Review, in the following we give a brief
description of a few examples that we believe are relevant in the
present context. In such devices, several different effects may
contribute to the refrigeration process (e.g., thermionic
transport, quantum tunneling as well as  thermoelectric effects),
so that making a proper "classification" of the refrigeration
principle is, strictly speaking, rather difficult. As a
consequence, we shall try mainly to follow the definitions as they
were introduced in the original literature.

\subsubsection {Thermionic refrigerators}

A vacuum thermionic cooling device consists of two electrodes
separated by a vacuum gap. Cooling occurs when highly energetic
electrons overcome the vacuum barrier through thermionic emission,
thus reducing the electron temperature of one of the two electrodes.
In such a situation, the refrigerator operation is mainly affected
and limited  by radiative heat transfer between the electrodes. The
thermionic emission current density ($j_{R}$) is given by the
Richardson equation, $j_{R} = A_{0}T^{2}\exp(-\frac{\Phi}{k_BT})$,
where $\Phi$ and $T$ are the work function and the temperature of
the emitting electrode, respectively,  $A_{0}= 4 \pi e m k_B^2 /
h^{3}$ is the Richardson constant, and $m$ is the electron mass.
From the expression above, a strong reduction of $j_{R}$ is expected
upon lowering the temperature. \textcite{mahan:4362} developed a
simple model for thermionic refrigeration, and demonstrated that its
efficiency can be as high as 80\% of the Carnot value. Vacuum
thermionic refrigerators are generally characterized by a higher
efficiency as compared to thermoelectric coolers, and are considered
to be an attractive solution for future refrigeration devices
\cite{nolas:4066}. However, the high $\Phi$ values typical of
currently available materials make thermionic cooling efficient, at
present, only above 500 K.

Several ideas on how to increase the cathode emission current and to improve
the operation of these refrigerators have been
proposed \cite{korotkov:2491, hishinuma:4242, hishinuma:2572,
Purcell:17259}. Korotkov and Likharev suggested to cover the
emitter with a thin layer of a wide-gap semiconductor, and to exploit
the resonant emission  current to cool the emitter \cite{korotkov:2491}.
The analysis of such a thermionic cooler predicted an
efficient refrigeration down to 10 K.

\begin{figure}[tb!]
\begin{center}
\includegraphics[width=7cm,clip]{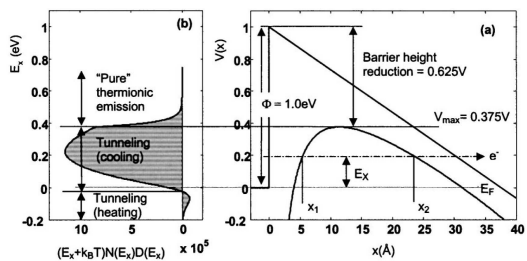}
\end{center}
\caption{Refrigeration by combined tunneling and thermionic
emission. (a) Schematic diagram of the  potential barrier profile
$V(x)$ for $\Phi = 1$ eV and with electrode separation of 60 \AA.
(b) Heat flow distribution at $T=300$ K. Adapted from
\cite{hishinuma:2572}.} \label{Hishinuma2572}
\end{figure}

The issue of high $\Phi$ values can be overcome by a
reduction of the distance between the electrodes (in the submicron range)
 and by the application of a strong electric field. Following this
scheme, \textcite{hishinuma:2572} theoretically analyzed a
thermionic cooler where the  two electrodes where separated by a
distance in the nanometer range. In such a situation, the
potential barrier is essentially lowered (see
Fig.~\ref{Hishinuma2572}(a)), allowing both thermionic emission
and energy-dependent tunneling. As a
consequence, rather small (if compared to vacuum thermionic
devices) external voltages ($\sim 1...3$ V) are required in order
to produce significant  electric currents. For suitable  values of
the applied voltage and  distances between the electrodes,
electrons above the Fermi level dominate the electric transport
(both thermionic emission over the barrier and tunneling through
the barrier), thus leading to cooling of the emitter. As it can be
inferred from Fig. \ref{Hishinuma2572}(b), the contribution of the
energy-dependent tunneling to total cooling is essential, and this
refrigerator could be classified as a vacuum tunneling device. The
cooling power surface density in this combined
thermionic-tunneling refrigerator was predicted to obtain values
as high as 100 W/cm$^{2}$ at room temperature. However, in the
only experimental demonstration of this device, a moderate
emission current (below 10 nA)  was reported at room temperature,
with an observed temperature reduction of about 1 mK
\cite{hishinuma:4690}. So far, no experimental demonstration of
vacuum thermionic refrigeration at cryogenic temperatures has been
reported.

\subsubsection{Application of low-dimensional systems to electronic
refrigeration}

The exploitation of low-dimensional systems gives additional
degrees of freedom in order to engineer materials that may lead to
enhanced operation of thermoelectric and thermionic devices
\cite{sales:1248, hicks:3230, hishinuma:4242, sofo:4565}. Some of
the limitations intrinsic to vacuum thermionic refrigerators can
be overcome with \emph{solid state} thermionic coolers
\cite{shakouri:1234, mahan:4016}. As a matter of fact, modern
growth techniques easily allow one to control both the barrier
height and its width within a wide range of values. One
disadvantage of solid state thermionic coolers stems from the
thermal conductivity of the barrier which is essentially absent in
vacuum devices. Nevertheless, a large temperature reduction can be
achieved by using a multilayered heterostructure
\cite{shakouri:1234, mahan:4016, zhou:1767}. According to theory
\cite{shakouri:1234, mahan:4016}, heterostructure-based thermionic
refrigerators perform somewhat better as compared to
thermoelectric coolers. The predicted temperature reduction of a
single-stage device at room temperature can be as high as 40 K,
and this value can be significantly increased in a multilayer
configuration. So far, however, the experimental implementations
of single barrier \cite{shakouri:88} and superlattice
\cite{fan:1580, zhang:374} refrigerators have reported temperature
reductions of a few degrees at room temperature.

\begin{figure}[tb!]
\begin{center}
\includegraphics[width=6cm,clip]{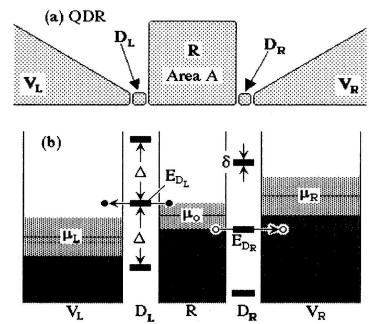}
\end{center}
\caption{Quantum-dot refrigerator. (a) Scheme of a quantum-dot
refrigerator. (b) Energy-level diagram of the structure. The
reservoir R is cooled as its quasiparticle distribution function is
sharpened by resonant tunneling through quantum dots D$_{\text{L}}$
and D$_{\text{R}}$ to the electrodes V$_{\text{L}}$ and
V$_{\text{R}}$. From \cite{edwards:5714}.} \label{QDR}
\end{figure}

Further improvement of thermionic refrigeration can, in principle,
be achieved by combining laser cooling and thermionic cooling
\cite{shakouri:1234, malshukov:5570}. In such an opto-thermionic
device, hot electrons and holes extracted through thermionic
emission lose their energy by emitting photons rather than by
heating the lattice. The theoretical investigation of a
GaAs/AlGaAs-based opto-thermionic refrigerator  predicted specific
cooling power densities of the order of several W/cm$^{2}$ at
$300$ K \cite{malshukov:5570}.

The presence of singularities in the energy spectrum of
low-dimensional systems can be used in solid state refrigeration.
For instance, the discrete energy spectrum in quantum dots can be
exploited  for  refrigeration at cryogenic temperatures
\cite{edwards:1815, edwards:5714}. In such a quantum-dot
refrigerator (QDR) (see Fig.~\ref{QDR}(a)) a reservoir (R) is
coupled to two electrodes via quantum dots (D$_{\text{L}}$ and
D$_{\text{R}}$) whose energy levels can be tuned through
capacitively-coupled electrodes. The QD energy levels can be
adjusted so that resonant tunneling to the electrode
V$_{\text{L}}$ is used to deplete the states in R above $\mu_0$
and, similarly, holes below $\mu_0$ in R tunnel to V$_{\text{R}}$
(see Fig.~\ref{QDR}(b)). As a consequence, the net result will be
to sharpen the quasiparticle distribution function in R, thus
leading to electron refrigeration. In spite of a rather moderate
achievable cooling power, the QDR was predicted to be effective
for cooling electrons of a micrometer-sized two-dimensional
electron gas reservoir at mK temperatures, and even of a
macroscopic reservoir at lower temperatures \cite{edwards:1815,
edwards:5714}.

\section{Device fabrication}
\label{sec:devicefabrication}

\subsection{Structure typologies and material considerations}
\label{sec:material}

This section is devoted to the description of the main techniques
and experimental procedures used for the fabrication of typical
superconducting electronic refrigerators and detectors. Owing to
the great advances reached in the last decades in micro- and
nanofabrication technology \cite{Timp:0,bhushan:0}, the amount of
information related to fabrication methods is too large to be
covered here and beyond the scope of the present review.
Therefore, we only briefly highlight all those issues that we
believe are strictly relevant for this research field. In
particular we first of all focus on the two typologies of existing
superconducting structures, namely all-metal and hybrid devices,
and on the main differences between them, both in terms of
materials and fabrication techniques. The former concern
structures where the active parts of the device, i.e., both the
superconducting elements and the normal regions are made of
metals; in hybrid structures, the non superconducting active part
of the device is made of doped semiconducting layers.

As far as all-metal-based structures are concerned, they are
 realized with low-critical-temperature thin-film
superconductors (normally Al, Nb, Ti and Mo), while the normal
regions usually consist of Cu, Ag or Au. Their fabrication
protocol includes patterning of a suitable radiation-sensitive
masking layer through electron-beam or optical lithography  in
combination with a shadow-mask (angle) evaporation technique
\cite{dolan:337}. The final device  is thus realized in a single
step in the deposition chamber, where additional  tunnel barriers
between different regions of the structure are \textit{in-situ}
created by suitable oxidation of the metallic layers. Although all
this leads to an efficient way for fabricating metallic
structures, however the electronic properties typical of metals
are only weakly dependent on their growth conditions and on the
specific employed technique. As a consequence, it is hard to
tailor the metallic properties in order to finally match some
specific requirements.

On the other hand, the situation is rather different with
semiconductors that offer some advantages in comparison to metals.
The large magnitude difference of Fermi wave-vector between metals
and semiconductors allows in general to observe quantum effects in
structures much bigger than with metals. In addition, the
availability of several techniques for growing high-purity
crystals (e.g., molecular beam epitaxy) yields the  capacity to
tailor the semiconductor electronic properties and, at the same
time, enables the fabrication of structures characterized by
coherent transport especially in reduced dimensionality
\cite{Capasso:1990}. The exploitation of low-dimensional electron
systems as active elements of electronic coolers was predicted to
be the next possible breakthrough in this research field
\cite{edwards:5714,edwards:1815,hicks:16631,hicks:12727,hicks:R10493,hicks:3230,koga:2950,koga:2438,koga:1490}.
Moreover, the realization of structures where charge carriers
experience arbitrarily chosen effective potentials is a further
advantage of modern engineered heterostructures \cite{Yu:0}. Last
but not least, the possibility of changing the carrier density
through electrostatic gating allows to strongly alter some
semiconductor  parameters (like the mobility as well as the
coherence length), thus enabling to have access  to different
electronic transport regimes in the same structure \cite{Ferry:0}.

Like all-metal devices, hybrid structures generally exploit  the
same superconducting materials. However, only heavily-doped Si
layers in combination with superconductors were exploited up to
now for the realization of superconducting refrigerators
\cite{savin:1471,buonomo:7784,savin:1481}. The fabrication
procedure of hybrid structures  involves typically two
lithographic steps, where the first allows to pattern the
semiconductor active layer through various techniques (such as wet
or dry etching). The second one the patterning of the
superconducting electrodes. Differently from all-metal devices, at
the metal-semiconductor interface forms the well-known Schottky
barrier  \cite{luth:0}. The latter can be tailored through
suitable doping of the semiconductor, allowing to control the
interface transmissivity (i.e., the junction specific resistance)
over several orders of magnitude.

Both all-metal and hybrid structures are normally fabricated on
semi-insulating semiconductor substrates which provide both the
mechanical support and rigidity as well as the thermalization of
the device at the bath temperature. It is noteworthy to mention
that undoped thermally-oxidized Si substrates are widely used
toward this end for metallic structures, while hybrid coolers
typically exploit the silicon-on-insulator (SOI) technique in
order to provide the Si active layer.

\subsection{Semiconductor growth techniques}
\label{sec:semgrow}

Depending on the physical principle exploited, semiconductor
growth technologies can be either physical or chemical, such as
gas and liquid-phase chemical processes. In the following we give
a brief survey of the most common semiconductor growth methods.

The \textcite{czochralski:219} crystal pulling method is probably
the most common technique for growing semiconductor bulk single
crystals and the largest amount of Si used in the
semiconductor industry is obtained using this technique. 
This technique allows to grow large semiconductor single crystals.
However, it is often sufficient and less demanding from an
economical point of view, to grow a thin ( $\sim 1~\mu$m) perfect
crystal layer on top of a bulk crystal of lower quality. This is
accomplished by growing \textit{epitaxially} the top layer, i.e.,
the atoms forming the latter build a crystal with the same
crystallographic structure and orientation as the starting
substrate (in contrast, non-epitaxial layers can be amorphous or
polycrystalline).

Gas-phase epitaxy is a widespread technique and represents
nowadays the most important process for the industrial fabrication
of Si and GaAs devices. In chemical vapour deposition (CVD)
\cite{adams:93,grove:0,sze:0}, the constituents of the vapor phase
chemically react at the substrate surface and the product of the
reaction is a film deposited on the substrate. Nowadays, CVD is
widely employed for growing several elemental semiconductors and
alloys and, depending on the specific application, it has been
implemented and adapted in a number of different configurations
\cite{jensen:199}. Among these we mention 1) metalorganic CVD
(MOCVD), exploits a thermally heated reactor and organometallic
gas sources; 2) plasma-enhanced CVD (PECVD), exploits a plasma
discharge to provide the energy for the reactions to occur; 3)
atmospheric pressure CVD (APCVD),  does not require a vacuum
environment.

While typical CVD processes are carried out in low vacuum (i.e.,
in the pressure range from $10^{-1}$ to several Torr), molecular
beam epitaxy (MBE) \cite{arthur:4032,cho:31,cho:157,chang:37} is
an ultrahigh vacuum (UHV) growth technique, performed at pressures
usually lower than $10^{-10}$ Torr. In such vacuum conditions the
amount of residual gas contaminants present in the growth chamber
is minimized, thus allowing to deposit epitaxial layers of high
purity and quality. The main features of this method are a precise
control of both chemical and stoichiometric composition, and a
perfect tuning of  doping profiles on the scale of a single atomic
layer \cite{cho:275}. The film growth concept is fundamentally an
evolution of UHV evaporation, where thermal or electron-beam
sources are used to create a flux of \textit{molecular} species.
Significant MBE work has been achieved with Si and Ge as well as
with IV-IV and II-VI semiconductor compounds and several metals,
but probably the largest amount of research has been devoted to
III-V semiconductor alloys.

\subsection{Thin-film metals deposition methods}
\label{sec:deposition}

\subsubsection{Thermal evaporation}

Vacuum evaporation is one of the oldest and simplest thin film
deposition techniques \cite{maissel:0,holland:1958}. It is a
physical vapor deposition (PVD) method widely used to deposit a
variety of materials, from elemental metals to alloys and
insulators. Evaporation is based on the boiling off or sublimation
of a heated source material onto a substrate surface. The result
of the evaporant condensation is the final film. The rate of atoms
or molecules ($\mathcal{N}_e$) lost under evaporation from a
source material per unit area per unit time can be expressed by
the Hertz-Knudsen relation $\mathcal{N}_e=a[P_V^{\ast}(T)-P][2\pi
Mk_BT]^{-1/2}$, where $a$ is the evaporation coefficient ($a=1$
for a clean evaporant surface), $P$ is the ambient hydrostatic
pressure acting on the evaporant in the condensed phase,
$P_V^{\ast}(T)$ is the equilibrium vapor pressure of the
evaporant, and $M$ its molecular weight. This expression  shows
that the evaporation rate strongly depends on the evaporant vapor
pressure. Most common metals typically deposited  by thermal
evaporation (such as Al, Au, Ga, and In) usually have vapor
pressures in the range between $10^{-2}$ to 1 Torr in the
temperature window of $600^{\circ}... 2000^{\circ}$~C; conversely
refractory metals (such as Nb, Mo, Ta, W and Pt) or ceramics (such
as BN, and Al$_2$O$_3$) reach such vapor pressures at much higher
temperatures, thus making more difficult the exploitation of this
technique for the deposition.

Usually evaporation is performed in high or ultrahigh vacuum (in
the range $10^{-5}... 10^{-10}$~Torr),  where the mean free path
$\ell$ for the evaporant species is much larger than the
substrate-source distance. This translates in an almost line of
sight evaporation which prevents  covering of edges perpendicular
to the source,  the latter also referred to as the lack of step
coverage \cite{madou:0} (note that this property is at the basis
of lift-off processes \cite{moreau:0} as well as of angle (shadow)
evaporation technique using suspended masks
\cite{dolan:337,dolan:7}). Furthermore, vacuum evaporation is a
low-energy process (implying a negligible damage to the substrate
surface), where the typical energy of the evaporant material
impinging on the substrate is of the order of 0.1~eV.
Nevertheless, radiative heating can be high.

\textit{Resistive} heating and \textit{electron-beam} deposition
are the two most common methods of evaporation. The former relies
on direct thermal heating to evaporate the source material. This
method is fairly simple, robust and economic but suffers from a
limited maximum achievable temperature (of the order of
$1800^{\circ}$~C), which prevents the evaporation of refractory
metals and several oxides. On the other side, electron-beam
evaporation represents a crucial improvement over resistive
heating. This method exploits a high-energy  electron beam that is
focused through a magnetic field on a localized region of the
source material. A wide range of materials (including refractory
metals and a wide choice of oxides) can be deposited owing to the
generation of high temperatures (in excess of $3000^{\circ}$~C)
over a restricted area. Its main drawback relies on the generation
of X-rays from the high-voltage electron beam which may damage
sensitive substrates (such as semiconductors)
\cite{sze:0,moreau:0}. Achievable deposition rates are up to
several hundreds \AA/sec (e.g., 0.5 $\mu$m/min for Al)
\cite{madou:0}.

\subsubsection{Sputter deposition}

The sputtering process has been known and used for over 150 years
\cite{chapman:0,chapman:291,wasa:0,rossnagel:609}. It is a PVD
method widely used nowadays for many applications, both in the
electronic and mechanic industry fields as well as in the pure
research environment. This process is based on the removal of
material from a solid target through its bombardment caused by
incident positive ions emitted from a (rare) gas glow discharge.
The transferred momentum of the ions leads to the expulsion of
atoms from the target material, thereby enabling the deposition
(condensation) of a film on the substrate surface. Sputter
deposition is generally performed at energies in the range of 0.4
to 3 keV. Furthermore, the average energy of emitted ions from the
target source is in the range $10... 100$ eV. At these energies
bombarding ions can penetrate up to two atomic layers in the
substrate thus leading to a great improvement of the adhesion of
the sputtered film \cite{wasa:0,maissel:0}. Normally, relatively
high pressures (from $10^{-4}$ to $10^{-1}$ Torr) are maintained
in the growth chamber during deposition. At these pressures the
mean free path is short (of the order of 1 mm at $10^{-1}$ Torr)
so that the material atoms reach the substrate surface with random
incident angles. As a consequence, a very good step coverage can
be achieved. Being essentially mechanical in nature, sputtering
successfully allows the deposition of refractory metals
(superconductors) like Nb, NbN, Ta, Mo and W at temperatures well
below their melting points.

\subsection{Thin film insulators}
\label{sec:dielectrics}

The roles of thin film insulators in solid state electronics are
various. In particular, deposited films are often used as
interlevel dielectrics for metals, to realize lithographic masking
for diffusion and implantation processes, as well as for
passivation and protective layers
\cite{sze:0,ghandhi:0,nicollian:0}. In addition they can be
exploited as thin amorphous membranes on which micro- and
nanostructured devices are realized
\cite{fisher:2705,irwin:1945,lanting:112511,nahum:3203,clark:173508,luukanen:281}
in light of their specific electric and thermal properties
\cite{leoni:3572,leivo:1305,manninen:1885}. In the following we
discuss those insulators which are considered particularly
relevant for microelectronic fabrication processing, i.e., silicon
dioxide and silicon nitride.

Silicon dioxide (SiO$_2$) is one of the most exploited insulators
in micro- and nanoelectronics based on Si, mainly due to the high
quality of the SiO$_2$/Si interface. SiO$_2$ films can be 
grown on Si substrates by thermal oxidation using oxygen or steam.
Thermal oxidation of Si is generally carried out in reactors at
temperatures between 900$^{\circ}$ C and 1200$^{\circ}$ C. The
resulting SiO$_2$ film is amorphous and characterized by good
uniformity, lack of porosity and very good adhesion to the
substrate. Some typical parameters of thermally grown silicon
dioxide at 1000$^{\circ}$ C are a refractive index of 1.46, a
breakdown strength larger than $10^7$ V/cm, and a density of 2.2
g/cm$^3$ \cite{nguyen:112,sze:0}. An alternative way to deposit
silicon dioxide layers is through CVD techniques
\cite{reif:260,nguyen:112}. In particular PECVD is considered an
effective technique \cite{hess:244,adams:111,kaganowicz:1233}
because of the low deposition temperature (SiO$_2$ is generally
deposited in the temperature range $200^{\circ}... 500^{\circ}$
C). Silicon oxide can be deposited from silane (SiH$_4$) with
O$_2$, CO$_2$, N$_2$O or CO \cite{adams:1545,hollahan:931}. In
addition, plasma oxide film properties are strongly dependent on
the growth conditions such as reactor configuration, RF power,
frequency, substrate temperature, pressure and gas fluxes. Typical
parameters for plasma-deposited silicon dioxide films at
$450^{\circ}$ C are  a refractive index of $1.44... 1.50$, a
breakdown strength of $2... 8\times 10^6$ V/cm, and a density of
2.1 g/cm$^3$.

Silicon nitride is another insulating film that forms good
interfaces with Si. It is nowadays successfully used as interlevel
dielectric \cite{swan:713}, in multilayer resist systems
\cite{suzuki:191}, as well as a protective coating as it provides
an efficient barrier against moisture and alkali ions (e.g., Na)
\cite{sze:0}. PECVD is commonly used \cite{hess:244,adams:111}
because of the low deposition temperature
($250^{\circ}...400^{\circ}$ C) implying low mechanical stress.
Silicon nitride is typically formed by reacting silane and ammonia
(NH$_3$) or nitrogen in the glow discharge. As for SiO$_2$, the
properties of the final film  strongly depend on the deposition
conditions \cite{dun:1556,nguyen:2348,adams:111,chow:5630}.
Typical parameters for plasma-deposited silicon nitride films at
$300^{\circ}$ C are a refractive index of $2.0... 2.1$, a density
of $2.5... 2.8$ g/cm$^3$, and a breakdown strength of $6\times
10^6$ V/cm.

\subsection{Lithography and etching techniques}
\label{sec:litho}

Fabrication of thin film metallic circuits as well as semiconductor micro- and nanodevices requires the generation of suitable patterns through lithographic processes \cite{campbell:0,jaeger:0,plummer:0}.
Lithography, indeed, is the method used to transfer such patterns onto a substrate (e.g., Si, GaAs, glass, etc.), thus defining those regions for subsequent etching removal or material addition.

In photolithography \cite{levinson:11}  a radiation-sensitive
polymeric material (called \textit{resist}) is spun on a substrate
as a thin film. The image exposure is then transferred to the
resist through a photomask, consisting normally of a glass plate
having the desired pattern of clear and opaque areas in the form
of a thin ($\sim1000$~\AA) Cr layer. Two types of resists can be
used in such a process, i.e., \textit{positive} and
\textit{negative} resists \cite{moreau:0,thompson:0,colclaser:0}.
In the former, the solubility of the exposed areas in a solvent
called developer is enhanced, while in the latter the solubility
is decreased. After exposure, the resist is developed and
reproduces the desired  pattern images for the subsequent
processing. The radiation source for photolithography depends on
the desired final resolution, although the latter is mainly
limited by effects due to light diffraction \cite{levinson:11}. In
particular, high pressure Hg lamps (with a wavelength $\lambda
=365$ nm i-line or $\lambda =436$ nm g-line) allow a line-width
larger than $0.250$ $\mu$m, while in the range $130... 250$ nm
deep UV (DUV) sources like excimer lasers are necessary, such as
KrF ($\lambda =248$ nm) and ArF ($\lambda =193$ nm).

Electron-beam lithography \cite{brewer:0,mccord:139} represents an
attractive technique for the fabrication of micro- and
nanostructures \cite{kern:87}. This method exploits a  focused
electron beam (with energy in the range $10... 100$~keV and
diameter of $0.2... 100$~nm) to expose a polymer-based
electron-sensitive resist (such as polymethyl-methacrylate
(PMMA)). Like in photolithography, the resist can be either
positive or negative. Even if the wavelength of the impinging
radiation beam can be smaller than $0.1$~nm, the maximum
achievable resolution is set by the electron scattering in the
resist and backscattering from substrate (known as the "proximity
effect"
\cite{kyser:1305,kyser:1391,jamoto:3855,howard:1101,jackel:698}),
so that the resolution is generally larger than $10$~nm (note,
however, that resolutions as high as $2$~nm have been achieved on
some materials \cite{mochel:38}).

In addition to lithographic procedures, etching of thin films or
bulk substrates represents an important step for the fabrication
of the final structure \cite{sze:0,madou:0}. Toward this end,
insulating or conducting thin films are exploited as masking
layers for  subsequent material removal. Two crucial parameters of
any etching process are \textit{directionality} and
\textit{selectivity}. The former refers to the etch profile under
the masking layer. In particular, for an \textit{isotropic} etch,
the etching rate is approximately the same in all directions,
leading to a spherical profile under the mask. With
\textit{anisotropic} etch, the etching rate depends on the
specific direction (e.g., a particular crystallographic plane)
thus leading to straight profiles and sidewalls. Selectivity
instead represents how well the etchant can differentiate between
the masking layer and the layer that has to be removed. Moreover,
etching techniques can be divided into wet
\cite{kendall:41,ghandhi:0} and dry \cite{wasa:0,madou:0}
categories. In wet etching, the substrate is placed in a liquid
solution, usually a strong base or acid. The advantage of wet
etching stems from its higher selectivity in comparison to dry
methods. Wet etching is in general isotropic for most substrates,
and various solutions that yield anisotropic etching are available
for some materials. In dry etching, the substrate is exposed to a
plasma in a reactor where ions can etch the substrate surface. The
great advantage of dry etching with respect to wet etching resides
in its higher anisotropy (that allows vertical etch walls), and
smaller undercut (that enables smaller lines to be patterned with
much higher resolution). Several materials (e.g., insulators,
semiconductors as well as refractory metals) can be successfully
dry etched, for which a variety of chemistries and recipes are
available \cite{cotler:38}.

\subsection{Tunnel barriers}
\label{sec:tunbar}

\subsubsection{Oxide barriers}
\label{subs:oxidebarriers}

Superconducting tunnel junctions represent key elements in a
number of electronic applications \cite{solymar:0} that span  from
single electron transistors \cite{grabert:0} to Josephson devices
\cite{barone:0}, just to mention two relevant examples. In
addition, they are crucial building blocks of  superconducting
microrefrigerators \cite{nahum:3123,leivo:1996} as well as of
ultrasensitive microbolometers \cite{nahum:3075,castellano:3251}.
The simplest picture of a tunnel  junction can be given assuming a
rectangular barrier of height $\phi_I$ and width $w$. The electron
transport across the barrier can be easily described within the
Wentzel-Kramers-Brillouin (WKB) approximation. The main results of
this analysis are \cite{simmons:2655} i) an exponential dependence
of the zero-bias junction conductance ($G_0$) on the barrier
width, $G_0\propto\textrm{exp}[-2w\sqrt{2m^{\ast}\phi_I}/\hbar]$,
where $m^{\ast}$ is the effective mass of electrons in the
barrier; ii) a quadratic voltage ($V$) dependence of the
conductance $G(V)$; iii) a weak insulating-like quadratic
dependence of $G$ on the temperature $T$. In general, image forces
acting on the electrons tunneling through the barrier will reduce
both its height and its effective thickness \cite{simmons:1793}.
Criteria i)-iii) require that the dominant process through the
barrier is direct tunneling and can be used to extract the
junction parameters in a realistic situation
\cite{simmons:1793,brinkman:1915}. Among the above given criteria,
i) seems a necessary, but not sufficient condition to establish
that tunneling is the dominant transport mechanism
\cite{rabson:2786,zhang:557} due to the possible presence of
pinholes (i.e., small regions where the insulator thickness
vanishes) in the barrier, and it is generally accepted that only
criterion iii) could be safely used  to rule out the presence of
such pinholes in the barrier \cite{jonson-akerman:1870} and assess
the junction quality (for instance, a reduction of junction
conductance of about $15\%$ on cooling from $295$~K to $4.2$~K is
believed to be a good practical indication of high-quality AlO$_x$
barriers \cite{gloos:1733,gloos:2915}; see also Secs. III and IV).
An important figure of merit of tunnel contacts is the junction
specific resistance $R_c =R_JA$, where $R_J\equiv G_0^{-1}$ is the
contact resistance and $A$ its area. A route to decrease $R_c$ is
to reduce $w$ or choose materials with lower effective barrier
height.

Among the available barrier materials, aluminum oxide (AlO$_x$) is
probably the most widespread insulator used to fabricate metallic
tunnel junctions because it can be easily and reliably grown
starting from an Al film. Its main parameters are a typical
barrier height $\phi_{I} \approx 2$ eV, although values in the
range $0.1... 8.6$ eV  have been reported
\cite{gundlach:125,kadlec:621,lau:2985,barner:2060} and a
dielectric constant smaller than that of bulk Al$_2$O$_3$ ($4.5...
8.9$ at 295~K \cite{bolz:0}). Several methods are currently
exploited to fabricate high-quality (i.e., highly uniform and
pinhole-free) aluminum oxide barriers such as \textit{in situ}
vacuum natural oxidation
\cite{sun:2424,parkin:5828,tsuge:3296,matsuda:5261,zhang:2219},
oxidation in air \cite{miyazaki:L231} and plasma oxidation
\cite{moodera:3273,sun:448,gallagher:3741} of a thin Al layer
(typically below 2 nm), just to mention  the most common
techniques.  The latter two methods lead in general to higher
$R_c$ values  (of the order of several k$\Omega\,\mu$m$^2$ or
larger) with respect to natural oxidation. The natural oxidation
allows to achieve the desired $R_c$ by simply changing the
oxidation pressure and time ($R_c$ also depends on the original
thickness of the Al layer), and almost any $R_c$ value can be
produced by following such a procedure. Aluminum oxide junctions
with $R_c$ values as low as some tens of $\Omega\,\mu$m$^2$ or
lower are currently realized with natural oxidation
\cite{parkin:5828,zhang:2219,deac:6792,fujikata:7558,liu:8385,sun:2424,childress:7353}.
However, low-$R_c$ junctions are more prone to  defects or
pinholes in the barrier that dramatically degrade their
performance. Promising methods and materials have been proposed
for the fabrication of oxidized barriers with $R_c$ as low as some
$\Omega\,\mu$m$^2$ among which we mention junctions made of
ZrAlO$_x$ \cite{wang:7463,liu:8385} and HfAlO$_x$
\cite{wang:8367}.

\subsubsection{Schottky barriers}
The metal-semiconductor (NSm) junction is an issue that dates back
more than 60 years \cite{schottky:367}, but still nowadays
represents a relevant topic both in the physics of semiconductors
\cite{rhoderick:0,sze:1,luth:0,sze:0,brennan:0} and in device
applications \cite{singh:0,millman:0}. 
The two most important types of NSm junctions are the Schottky
barrier (SB), showing a \textit{rectifying} diode-like I-V
characteristics, and the \textit{ohmic} contact, whose I-V is
almost linear. Most of NSm junctions, with a few exceptions such
as contacts with InAs, InSb and In$_{x}$Ga$_{1-x}$As (for $x\geq
70 \%$) \cite{kajiyama:458}, are affected by the presence of the
SB that drastically affect their electric behavior. For most
semiconductors (in the following we concentrate on $n$-type
semiconductors but similar conclusions can be given for $p$-type
ones), the SB height ($\phi_{SBn}$) has a rather weak dependence
upon the metal used for the contact; for instance, for
metal/$n$-Si contacts $\phi _{SBn}=0.7... 0.85$ eV, while for
metal/$n$-GaAs contacts $\phi _{SBn}=0.7... 0.9$ eV
\cite{singh:0,rhoderick:0}. This fact is explained in terms of the
Fermi-level pinning at the NSm interface
\cite{bardeen:717,heine:A1689,monch:221}.

The current across a Schottky junction depends on a number of
different mechanisms. In the limit where thermionic emission
dominates the electric transport, the rectifying action of a
biased NSm junction is described  as $I(V)=I_s
[\textrm{exp}(eV/k_B T)-1]$, where the detailed expression for the
saturation current $I_s$ depends on the assumptions made on
carrier transport \cite{brennan:0,sze:1,sze:0}. In such a case the
junction specific resistance is $R_c \propto
T^{-1}\textrm{exp}(e\phi _{SBn}/k_B T)$, thus meaning that it can
be lowered mainly by decreasing $\phi_{SBn}$ (typical $R_c$ values
at doping levels $N_D\leq 10^{17}$~cm$^{-3}$ for metal/$n$-Si
contacts are of the order of $10^{11}... 10^{13}$~$\Omega
\mu$m$^2$  almost independent of $N_D$). By contrast, tunneling
across the SB can be the dominating transport mechanism if the
semiconductor is heavily doped. In such a case $I(V)\propto
\textrm{exp}[-\alpha (\phi_{SBn}-V)/\sqrt{N_D}]$, with $\alpha
=2\hbar^{-1}\sqrt{m^{\ast}\epsilon}$ where $m^{\ast}$ is the
semiconductor effective mass and $\epsilon$ the dielectric
permittivity, i.e., the junction is not rectifying and the current
is proportional to $V$ for small voltages. The contact is thus
said to be ohmic and yields $R_c \propto \textrm{exp}(\alpha \phi
_{SBn}/\sqrt{N_D})$. This shows that $R_c$ can be reduced up to a
large extent by lowering the SB height and doping as heavily as
possible (again, for metal/$n$-Si contacts and $N_D\geq
10^{19}$~cm$^{-3}$, $R_c$ can be in the range $10^2...
10^8$~$\Omega \mu$m$^2$). All this shows the advantage of using
NSm contacts owing to the possibility of tuning the contact
specific resistance over several orders of magnitude (from
metallic-like to tunnel-like characteristics) through a careful
choice of metal-semiconductor combinations and proper doping
levels. This trick is commonly exploited to control the NSm
interface resistance in current semiconductor technology,
although heavy doping of the semiconductor just in proximity to
the metal is often preferred
\cite{shannon:537,kastalsky:3026,taboryski:656,giazotto:1772a,giazotto:216808}.

\section{Future prospects}
\label{sec:future} Low temperature solid-state cooling is still at
its infancy, although operation of a number of individual
principles and techniques have been demonstrated to work
successfully. Yet combinations of cascaded micro-refrigerators
over wider temperature ranges employing several stages, or
combinations of different refrigeration principles, e.g., fluidic
coolers together with electronic coolers, do not exist. In
principle, compact low-power refrigerators could be fabricated
using micro-machined helium-based fluidic refrigerators, for
instance based on Joule-Thomson process \cite{little:661}, and
these devices could then be directly precooling NIS-refrigerators
with niobium ($T_c= 9$ K) as a superconductor. A lot of
engineering effort is, however, needed to make this approach work
in practise as a targeted micro-refrigerator.

The solid-state micro-circuits have already proven to yield new
operation principles and previously unknown concepts have been
discovered in cryogenic devices, as demonstrated throughout this
review. We believe that what was demonstrated here is just a
presentation of a beginning of a new era in low temperature
physics and instrumentation. As possible new classes of devices we
could mention those utilizing thermodynamic Carnot cycles with
electrons. Brownian heat engines with electrons are predicted to
achieve efficiencies close to ideal \cite{humphrey:116801}. It may
be possible in the future to make use of other types of gated
cycles where energy selective extraction of electrons is producing
the refrigeration effect. As a conceivable example, a combination
of Coulomb effects and superconducting energy gap could form the
basis of operation of a refrigerator where cooling power would be
proportional to the operating frequency of the gate cycle. Such a
device would thus be principally different from the static
electronic refrigerators presented in this review, where a DC bias
is in charge of the redistribution of hot electrons.

At low temperatures, additional relaxation channels besides the
electron-phonon scattering, such as coupling between electrons and
photons, become important. More knowledge is needed on these
mechanisms.

In Subs.~\ref{sec:suptrans}, we describe how the non-equilibrium
shape of the distribution function sometimes leads to improved
characteristics of the device. It would be interesting to see if
such effects could be employed to improve also the properties of
the radiation detectors or other practical devices.

The presently obvious application fields of electronic
micro-refrigerators include astronomical detectors both in space
as well as those based on the earth, materials characterization
instrumentation, e.g., those devices employing ultra high
resolution x-ray micro-analysis, and security instrumentation,
e.g., concealed weapon search on the airports. It is, however,
evident that once realized in a user-friendly and economic way,
refrigeration becomes very important in high-tech based industry
in a much broader perspective. Low temperature electronics and
superconducting devices are often characterized by their
undeniably unique possibilities, but they are often superior to
the room temperature ones also in speed and power consumption.
Therefore, mesoscopic on-spot refrigerators to attain the low
temperatures forming the basis of these instruments are urgently
needed.

\acknowledgments We thank H. Courtois, R. Fazio, F. Hekking, M.
Paalanen, F. Taddei and P. Virtanen for their insightful comments
and for critically reading the manuscript. D. Anghel, A. Anthore,
F. Beltram, M. Feigelman, E. Grossman, K. Irwin, M. Meschke, A. J.
Miller, S. Nam, D. Schmidt, and J. Ullom are gratefully
acknowledged for enlightening discussions. This work was supported
by the Academy of Finland.

\bibliographystyle{apsrmp}
\bibliography{rmpnovember}

\end{document}